\documentclass[fleqn,usenatbib,useAMS]{mnras}
\usepackage{graphicx}	
\usepackage{amsmath}	
\usepackage{amssymb}	
\usepackage{multicol}
\usepackage{multirow}
\usepackage{mathtools}
\usepackage{caption}

\usepackage[T1]{fontenc}
\usepackage{ae,aecompl}
\usepackage{newtxtext,newtxmath}
\usepackage{threeparttable,tablefootnote}
\usepackage{url}
\usepackage{hyperref}
\usepackage{placeins}
\usepackage{gensymb}
\usepackage[dvipsnames]{xcolor}
\usepackage{etoolbox}
\usepackage{float}

\makeatletter
\patchcmd\@combinedblfloats{\box\@outputbox}{\unvbox\@outputbox}{}{\errmessage{\noexpand patch failed}}
\makeatother

\numberwithin{equation}{subsection}
\numberwithin{table}{section}
\numberwithin{figure}{section}

\title[]{Gaussian Processes, Median Statistics, Milky Way Rotation Curves}
\author[]{Hai Yu$^{1,2}$\thanks{E-mail:\href{mailto:yuhai@smail.nju.edu.cn}{yuhai@smail.nju.edu.cn}}, 
    Aman Singal$^{1,3}$\thanks{E-mail:\href{mailto:asingal14@gmail.com}{asingal14@gmail.com}},
	Jacob Peyton$^{1}$\thanks{E-mail:\href{mailto:JCPEYTON@phys.ksu.edu}{jcpeyton@phys.ksu.edu}}, 
	Sara Crandall$^{1,4}$\thanks{E-mail:\href{mailto:}{sacranda@ucsc.edu}}, 
	Bharat Ratra$^{1}$\thanks{E-mail:\href{mailto:ratra@phys.ksu.edu}{ratra@phys.ksu.edu}}\\
	$^{1}$Department of Physics, Kansas State University, 116 Cardwell Hall, Manhattan, KS 66506, USA\\
	$^{2}$School of Astronomy and Space Science, Nanjing University, Nanjing 210093, China\\
	$^{3}$Department of Physics, Indian Institute of Technology Delhi, Hauz Khas, Delhi 110016, India\\
	$^{4}$Department of Astronomy \& Astrophysics, UC Santa Cruz, 1156 High Street, Santa Cruz, CA 95064, USA}

\date{\today}
\pubyear{2018}
\begin{document}
	\label{firstpage}
	\pagerange{\pageref{firstpage}--\pageref{lastpage}}
	\maketitle
	\begin{abstract}
		We use the Iocco et al. (2015) compilation of 2,780 circular velocity measurements to analyze the Milky Way rotation curve. We find that the error bars for individual measurements are non-gaussian, and hence instead derive median statistics binned central circular velocity values and error bars from these data. We use these median statistics central values and error bars to fit the data to simple, few parameter, rotation curve functions. These simple functions are unable to adequately capture the significant small scale spatial structure in these data and so provide poor fits. We introduce and use the Gaussian Processes (GP) method to capture this small scale structure and use it to derive Milky Way rotation curves from the binned median statistics circular velocity data. The GP method rotation curves have significant small-scale spatial structure superimposed on a broad rise to galactocentric radius $R\approx7$ kpc and a decline at larger $R$. We use the GP method median statistics rotation curve to measure the Oort $A$ and $B$ constants and other characteristic rotation curve quantities. We study correlations in the residual circular velocities (relative to the GP method rotation curve). Along with other evidence for azimuthal asymmetry of the Milky Way circular rotation velocity field, we find that larger residual circular velocities seem to favor parts of spiral arms.
	\end{abstract}
	
	
	\begingroup
	\let\clearpage\relax
	\endgroup
	\newpage
	
	\section{Introduction}\label{sec:intro}
	The Milky Way rotation curve, the circular velocity of tracers in the disk as a function of the distance $R$ from the Galactic center, $v(R)$, is an important kinematical characteristic of the Milky Way and, in particular, is used to constrain the mass distribution of the Milky Way.\footnote{Dark matter \citep[for reviews see][and references therein]{Ratra08,Bertone2018,Drees2018} plays a prominent role in determining the Milky Way rotation curve. There is convincing evidence for dark matter from cosmic microwave background anisotropy measurements \citep[][and references therein]{Planck2015,Ooba2018,Park2018a}, from Hubble parameter measurements \citep[e.g.,][]{Farooq2013a, Moresco2016,Farooq2017}, and from other cosmological data \citep[e.g.,][]{Park2018b}.} For early discussions of the Milky Way rotation curve, see \cite{Kwee54}, \cite{Kerr62}, \cite{Burton66}, and \cite{Shane1966}. \cite{Sofue2013,Sofue2017} reviews the field. Recent work includes \cite{Russeil2017}, \cite{Sysoliatina2018}, \cite{Fernandez2018}, \cite{Mroz2018}, and \cite{Eilers2018}. 
	
	Implicit in most derivations of the rotation curve from observational data is the assumption of an axisymmetric circular velocity field. Under the assumption of azimuthal symmetry it is appropriate to combine circular velocity measurements in different galactocentric longitude bins at the same distance from the Galactic center. These combined, angle averaged, data allow for a more precise determination of the rotation curve. However, if the disk is azimuthally asymmetric then a rotation curve derived under the assumption of azimuthal symmetry is not an accurate characterization of the circular velocity field of the disk.
	
	\cite{Iocco2015} have compiled 2,780 Milky Way circular velocity measurements distributed over a significant fraction of the Milky Way disk. This useful large compilation includes velocities of many different tracers: gas tracers (such as CO and HI), star tracers (such as Cepheids), and maser tracers. This abundance of data provides an opportunity to examine the velocity field of the disk of the Milky Way, and through it the mass distribution, including that of the dark matter, of the Milky Way. This large data set should provide an opportunity to determine an accurate Milky Way rotation curve, as well as rotation curves in a few azimuthal sectors and so allow for a study of the azimuthal symmetry of the Milky Way circular velocity field. Unfortunately, as discussed next, this is not a straightforward task.
	
	Our first task is to examine whether the quoted velocity error bars are consistent with what is expected for a gaussian distribution. To study this we divide the data into small spatial bins and assume that the individual measurements in each bin are measurements of the same quantity. It is widely believed that a large enough number of independent measurements of the same quantity will have gaussian error bars, however many counter examples are known.\footnote{Examples include Hubble constant measurements \citep{Chen2003,Chen2011,Zhang2018}, $\rm ^7Li$ and D primordial abundance measurements \citep{Crandall2015,Zhang2017,Zavarygin,Penton2018}, LMC and SMC distance moduli measurements \citep{deGrijs2014,Crandall2015b}, and parameters of the Milky Way \citep{deGrijs2016,deGrijs2017,Rajan2018,Camarillo2018b,Camarillo2018}. For other examples see \cite{Bailey160600} and \cite{Rajan2018b}. A lot of effort is devoted to testing for intrinsic non-gaussianity \citep[e.g.,][]{Park2001,Planck2016b}, in contrast to non-gaussianity introduced by the measurement process, since gaussianity is usually assumed in parameter estimation \citep[e.g.,][]{Samushia2007,Chen2011b,Ooba2017}.}
	
	We construct error distributions using the \cite{Chen2003} procedure. We first assume azimuthal symmetry and bin the circular velocity data only in $R$. We find in this case that the circular velocity data are quite non-gaussian. It is possible that this result is a consequence of azimuthal asymmetry.\footnote{Because it bins together data for many different galactocentric longitude points that are very far apart in space.} So we next look at data in different angular sectors separately and bin in both galactocentric longitude and $R$. We find that the data binned in angle and $R$ are less non-gaussian than the data binned just in $R$, however, they are still significantly non-gaussian. We also analyze the two main tracer subsets of data, gas tracers and star tracers,\footnote{We do not have a sufficient number of maser measurements to test for non-gaussianity in maser tracer velocity error bars.} and find that they are slightly less non-gaussian than the complete data binned only in $R$, but still significantly non-gaussian.
	
	It is not known why these velocity errors are non-gaussian. It is possible that there are unaccounted-for systematic errors or correlations. Since they are non-gaussian, weighted comparisons between models and these data \citep{Iocco2015,Pato2015,Patoetal2015,Calore2015,Bozorgnia2016} are only illustrative, and it is incorrect to use the weighted mean technique to determine a Milky Way rotation curve from their data. More correctly, the error bars associated with the individual measurements are suspect, and so to make use of this large dataset we must determine more reliable error bars.\footnote{The other option is to simply discard these data and use more recent data such as that from Gaia. However, we shall find that when analyzed using medium statistics these older data provide results consistent with those derived from other datasets, as well as some new and interesting conclusions, and so we believe it is worthwhile to describe our analysis and results in detail in this paper.} Median statistics \citep{Gott2001} provides a technique to do this, and so allows one to make use of this circular rotation velocity dataset.
	
	Median statistics \citep{Gott2001} does not use the quoted error bars on the individual measurements (which in this case are non-gaussian). Instead the distribution of the binned data is used to compute median statistics error bars. We provide extensive tables of median statistics binned angular circular velocity and error bars, $\omega_i\pm\sigma_i\text{ at }R_i,\ i = 1,2,\dots,N$ bins, and recommend that these be used in model fitting and parameter constraints instead of the individual data points. Of course, since median statistics does not make use of the quoted error bars, the median statistics constraints will be less precise than the weighted mean ones, but probably more accurate.
	
	In summary, even though the individual measurement errors are non-gaussian, median statistics allows us to use this large circular rotation velocity dataset to determine Milky Way rotation curves, by binning together individual measurements and using the scatter in each bin to determine the error bars for that bin. In our analysis here, because of their non-gaussian error bars, the individual measurements are not used to determine rotation curves.
	
	We fit simple, few parameter, functional $\omega(R)$ forms to these median statistics binned $\omega_i\pm\sigma_i$ values, and determine the best-fit parameter values by minimizing $\chi^2$. We find that the reduced $\chi^2$ is large because the binned $\omega_i$ measurements are not a simple function of $R_i$, so a two or three parameter angular circular velocity function $\omega(R)$ is unable to adequately capture the finer scale spatial structure in the $\omega_i$ measurements as a function of $R_i$. This is one indication that there is significant smaller-scale structure in the Milky Way rotation curve.
	
	We make use of the Gaussian Processes (GP) method to better capture the smaller scale spatial structure in the median statistics binned, observed angular circular velocities. This is the first time that this useful regression method has been used to determine the Milky Way rotation curve. For the complete data set, the GP method results in a relatively robust rotation curve over the $2\text{ kpc}\lesssim R\lesssim 10$ kpc range, that is relatively insensitive to the binning strategy we use and to the GP method covariance function assumed. The linear circular velocity curve $v(R)$ rises from about $\rm 190\ km\ s^{-1}$ at $R\approx2$ kpc to about $\rm 230\ km\ s^{-1}$ at $R\approx7$ kpc, and then declines to about $\rm 200\ km\ s^{-1}$ at $R\approx10$ kpc. Superposed on the overall rise and fall are finer-scale spatial variations. We compute the slope of the GP method circular velocity rotation curve, $dv/dR$, and the index function $n(R) = d(\ln{v})/d(\ln{R})$, as well as the Oort functions $A(R)\text{ and }B(R)$. Our measured Oort constants $A(R_0)\text{ and }B(R_0)$ are consistent with other local measurements, but we have larger uncertainties (this is discussed in \S\ref{sec:FittingResults}). It is reassuring that our new GP method median statistics Milky Way rotation curve passes this consistency test. While we find that the rotation curve slope at the distance of the Sun, $dv/dR\ (R_0)$ is consistent with 0 and a flat rotation curve, $dv/dR$ assumes values between approximately $\pm40\rm\ km\ s^{-1}\ kpc^{-1}$ over $2\text{ kpc}\lesssim R\lesssim 9$ kpc, and the index $n(R)$ assumes values between approximately $-1.25$ and 0.75 over this range.\footnote{A flat rotation curve has $n = 0$ while Keplerian motion corresponds to $n = -1/2$.} These results show that the rotation curve we derive has more spatial structure than a simple flat one. 
	
	While it has long been known that the Milky Way rotation curve has significant finer-scale spatial structure superimposed on a broad rise to $R\approx7$ kpc and then a decline at larger $R$, most analyses to date are based on techniques that effectively obscure or average over this small-scale structure. The GP method we use here is better able to capture this small-scale spatial structure than most other techniques used to date, and we believe that most of this small-scale spatial structure we discover in the Milky Way rotation curve is physically significant. It is of interest to apply the GP method to more recent and more precise data, such as that from Gaia.
	
	Our GP method circular velocity rotation curve is derived using circular velocity data measured over a fairly large fraction of the Milky Way disk. It combines data over a large range of galactic longitude and if the circular velocity field is azimuthally anisotropic it is inappropriate to combine the data in such a manner. We find that the circular velocity field is indeed azimuthally anisotropic and circular velocity residuals (determined by subtracting the GP method determined rotation curve from the circular velocity measurements) show evidence of some characteristic spatial scales that might be related to dynamical processes in the Milky Way. Our results are qualitatively consistent with those found by \cite{McClureGriffiths2016} from a smaller data set over a smaller $R$ range. It is reassuring that our new GP method median statistics Milky Way rotation curve gives results that are consistent with those derived using the conventional technique.
	
	In \S\ref{sec:data} we summarize the \cite{Iocco2015} data, outline the data subsets we consider, and the binnings we use. Our gaussianity analyses are presented in \S\ref{sec:gaussianity}, where we show that the velocity error bars are non-gaussian, and so using weighted means of individual data points is inappropriate, and instead a median statistics analysis should be used to analyze these circular velocity data. In \S\ref{sec:gaussianity} we also present preliminary evidence for azimuthal anisotropy in the observed circular velocity field. In \S\ref{sec:functionalform} we fit the median statistics binned data to simple, few parameter $\omega(R)$ functions and show that such functions are unable to adequately describe the $\omega(R)$ measurements. In \S\ref{sec:GaussianProcessing} we use the GP method to compute the rotation curves from the $\omega(R)$ measurements. In this section we show that the circular velocity field is azimuthally anisotropic. We also measure a few characteristic spatial scales in the residual circular velocity field that are thought to be related to dynamical processes in the Milky Way. We conclude in \S\ref{sec:conclusions}.
	
	\section{Data}\label{sec:data}
	In this section we summarize the \cite{Iocco2015} data we use and describe how we bin these data.
	\begin{figure*}
	    \centering
	    \includegraphics[width = \linewidth]{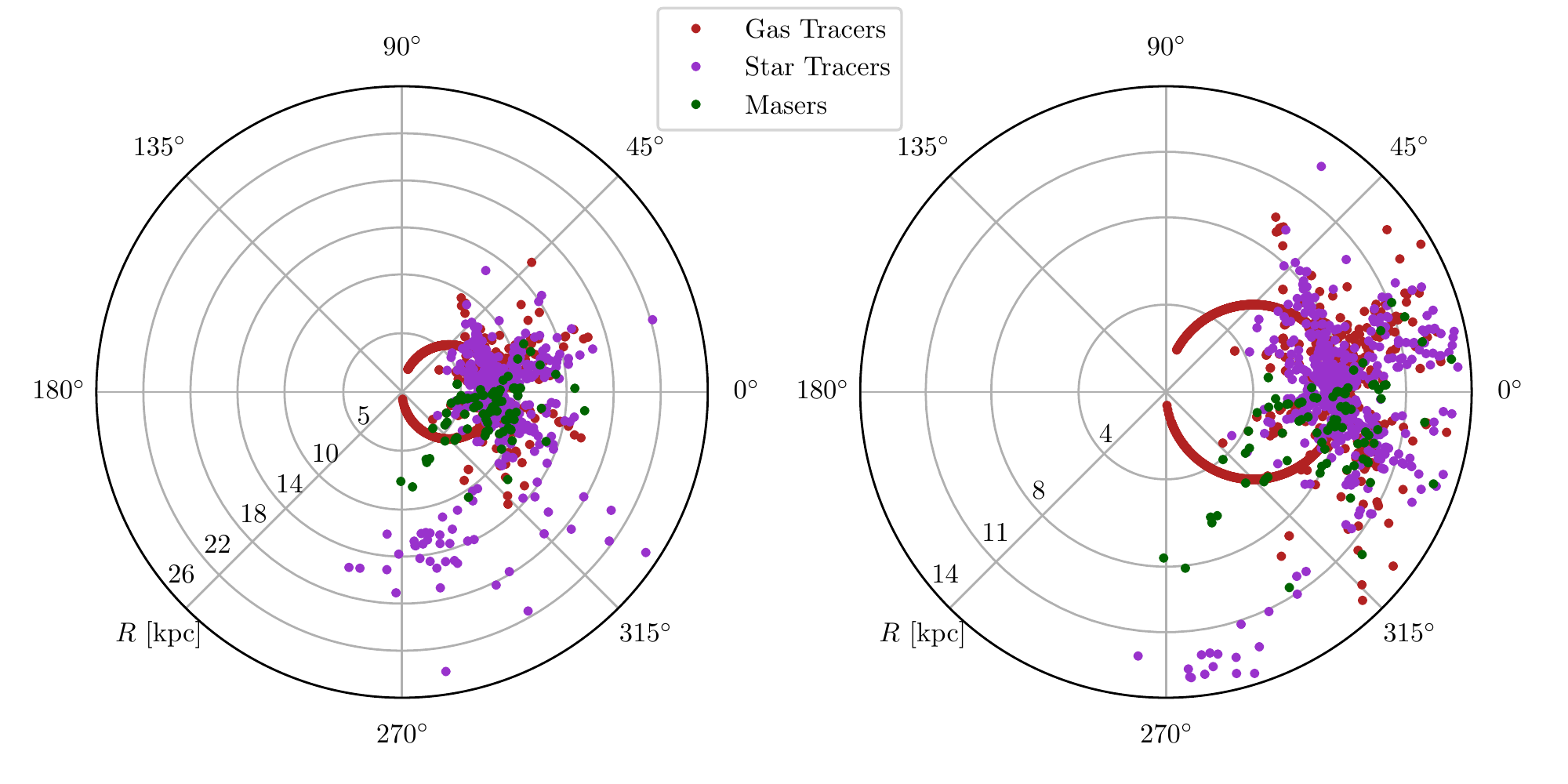}
	    \caption{The left panel shows the positions of 2,767 objects in the \protect\cite{Iocco2015} compilation. The Sun is at 0\degree and 8 kpc from the Galactic center which is at the origin. The density of measurement points fall rapidly beyond $R\sim13$ kpc. In our analyses we only consider measurements with $R\leq13.55$ kpc. In addition, we restrict our analyses to measurements close to the Galactic plane, retaining only those with $\left|z\right|\leq1.5$ kpc. The positions of the truncated set of 2,706 tracers we use in our analyses are shown in the right panel.}
	    \label{fig:position}
	\end{figure*}
	\subsection{Observational data}\label{subsec:observedata}
\begin{figure*}
		\centering
		\includegraphics[width = \linewidth]{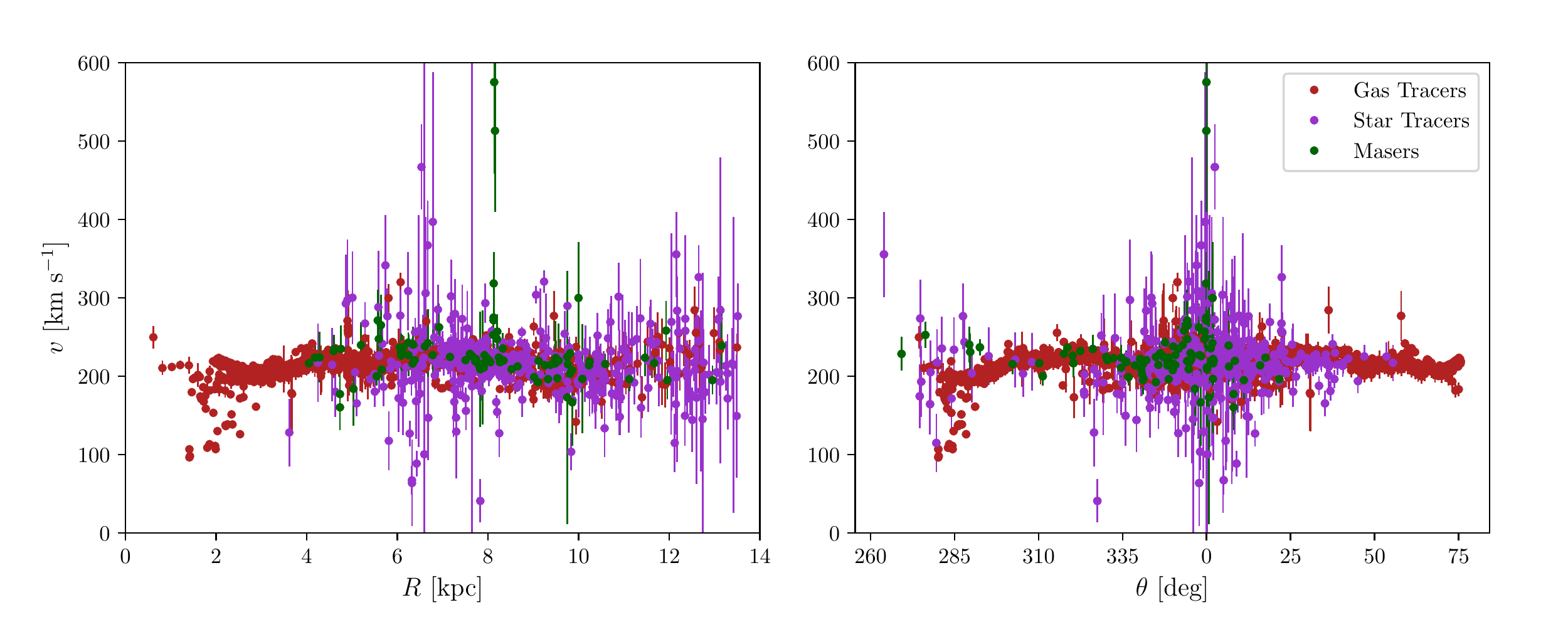}
		\caption{The left panel shows the 2,706 circular velocity $v$ measurements plotted at their radial distance $R$ from the Galactic center and the right panel shows them as a function of galactocentric longitude, $\theta$. All error bars are $\pm1\sigma$, and we do not show uncertainties in $R$.}
		\label{fig:velocity}
	\end{figure*}

	\cite{Iocco2015} compile 2,780 Milky Way circular velocity measurements of 2,274 gas tracers, 506 star tracers and 100 masers. They list the circular velocity and position relative to the Galactic center of these objects, $v(R)$, and in many cases they also list the uncertainties in $v$ and $R$ (uncertainties on $R$ are listed for only 1,083 measurements). Positional data is not provided for 13 objects, which we do not include in our analyses. The remaining 2,767 data points are shown in the left panels of Figs. \ref{fig:position} and \ref{fig:velocity}.
	
	We are interested in determining the rotation curve of the Milky Way disk and hence discard measurements more than 1.5 kpc above and below it. Due to the sharp decrease in density of measurements beyond 13.55 kpc from the Galactic center, we discard the measurements beyond that radial distance. The two constraints, $\left|z\right| \leq 1.5$ kpc and $R \leq 13.55$ kpc, reduce the number of measurements to 2,706. Of these 2,706 measurements, 2,146 are of gas tracers, 426 of star tracers, and 98 are of masers. These data are shown in the right panels of Figs. \ref{fig:position} and \ref{fig:velocity}.
	We calibrate the data we use here to $R_{0} = 8.0$ kpc and $v_0 = 220\ \rm km\ s^{-1}$ \citep[see][]{Camarillo2018b,Camarillo2018}. Here $R_0$ is the radial distance of the Sun from the Galactic center and $\Theta_0$ is the circular velocity about the Galactic center at the position of the Sun.\footnote{\cite[]{Camarillo2018b,Camarillo2018} have performed careful analyses of all recent measurements and conclude that at $2 \sigma$ the best summary values are $R_0 = 8.0 \pm 0.03$ kpc and $v_0 = 220 \pm 10\ \rm km\ s^{-1}$.} Because the angular velocity, $\omega$, is what is measured directly for most gas tracers (which make up by far the largest subgroup), we perform our analyses using this. However, we plot and present our results in terms of circular velocity $v$, as is more common.
	
	\subsection{\texttt{galkin}}\label{subsec:galkin}
	The \cite{Iocco2015} data can be accessed through the \texttt{galkin} Python program \citep{Pato2017}. \texttt{galkin} provides a graphical user interface where the user can select certain parameters including $R_0$ and $v_0$, and the tracers they wish to include from the compilation.
	
	The tool can be used to obtain three files,\footnote{All three files list a reference for each measurement, which can then be used to cross-reference the measurements between files.} which contain velocity data, position data, and raw data respectively:
	\begin{enumerate}
	    \item The velocity data file lists the tracer's distance from the Galactic center $R$, and the error $\Delta R$, in kpc ($\Delta R$ is provided for less than half the tracers). This file also lists linear circular velocity and angular circular velocity measurements with corresponding errors $(v, \Delta v, \omega\text{, and }\Delta \omega)$.
	    \item The position data file lists the distance from the Galactic center ($R$), and from the Sun ($d$), along with Galactic coordinates (longitude $l$ and latitude $b$), and cartesian coordinates with the Galactic center as the origin ($x, y\text{, and }z$).
	    \item The raw data file lists the line-of-sight velocity $v_{\rm los}$, the associated error $\Delta v_{\rm los}$, and the peculiar motion observations ($\mu_{l^{*}}, \Delta \mu_{l^{*}}, \mu_b\text{, and } \Delta \mu_b$).
	\end{enumerate}
	
	More detailed discussion of these data may be traced back through the papers of \cite{Iocco2015} and \cite{Pato2017}.
	
	\subsection{Binning, data subsets, and $\Delta R$'s}\label{subsec:bin}
	Given the reasonably large number of measurements, 2,706, distributed over a reasonably large area of the Milky Way disk, we may divide the full data set into different subsets and study consistency between results derived using the different subsets. Here we consider all 2,706 measurements together, as well as two different tracer subsets: the 2,146 gas tracers and the 426 star tracers.
	
	It is important to determine if the quoted error bars are consistent with what is expected for a gaussian distribution. To study this, we bin the data in small spatial bins and assume that individual measurements in a bin may be treated as measurements of the same quantity.
	
	We first assume that the circular velocity field is azimuthally symmetric and only consider radial bins. For $N$ individual measurements in a subset, we create $\lfloor\sqrt{N}\rfloor$ bins\footnote{Here, $\lfloor A\rfloor$ refers to the floor of the number $A$, i.e., the greatest integer less than $A$.} with approximately $\sqrt{N}$ measurements in each bin.\footnote{This is the best compromise between binwidth and number of measurements in each bin; it is of interest to have both narrow bins (to not smear out small-scale structure in the function) and many measurements in each bin (to better estimate the function at that point).} For example, for the full data compilation of 2,706 measurements, we create 52 radial bins with approximately 52 measurements in each bin.
	
	As an aside, we note that it is not straightforward to account for the uncertainty in $R$. The \cite{Iocco2015} compilation provides $\Delta R$ for only 966 of the 2,706 measurements. To examine the effect of uncertainties in $R$ on the results, we create a second set of bins by swapping the first third of the tracers in each bin, i.e., the third of the tracers with the smallest galactocentric radius, with the last third of the tracers in the previous bin. We then compare the results derived from the first set of bins with those derived from the second set of bins with interchanged $R$ values to get a qualitative indication of how the uncertainties in $R$ affect the results. Our procedure here is likely more dramatic than what the $R$ errors can cause. When we apply our procedure we find no significant change in our results. This is most probably due to the fact that we bin in $R$.
	
	It is also important to determine if the velocity field is azimuthally symmetric. To do this we first subdivide the data into 10\degree\ azimuthal angular sectors and then radially bin the data in each angular sector.
	
	\section{Gaussianity analyses}\label{sec:gaussianity}
	For multiple measurements $M_{i}$, $i = 1, 2, \dots, N$, of the same quantity $M$, with statistical errors $\sigma_{i}$, we ideally expect that the errors are distributed in a gaussian manner. If this is the case, the weighted mean technique can be used to derive a summary estimate of the data points with a smaller uncertainty than that of the individual measurements.
	
	In this section we examine if the angular velocity data described in the previous section have errors that are gaussianly distributed, and discover that they do not. In \S\ref{subsec:cestats}, \S\ref{subsec:errordist}, and \S\ref{subsec:kstest}, we summarize the tools required for such an analysis. We then present the results of the gaussianity analysis of the angular velocity data in \S\ref{subsec:results}, where we show that these angular velocity data have non-gaussianity distributed errors. We assume that the bins are small enough in space so that we may treat individual measurements in each bin as a measurement of the same quantity.

    In the Appendix we provide extensive tables to establish our conclusions. Among these tables are those for median statistics binned angular circular velocity $\omega(R)$ and errors. For model comparison and parameter estimation, we recommend that the data in these tables be used, rather than the individual measurements, which are non-gaussian.

	\subsection{Central estimate statistics}\label{subsec:cestats}
	To determine an error distribution of the measurements, we need to have a summary central value for the measurements. In our analyses we use two central estimates, the median and the weighted mean.
	
	\subsubsection{Median statistics}\label{subsubsec:median}
	The median is the central value in the sorted data set, the value that divides the data set into two halves each containing an equal number of elements. Median statistics assumes statistical independence of all measurements and an absence of systematic errors. It does not make use of a measurement's errors, which is an advantage if the errors are incorrectly estimated, however, as a result the median central estimate has a relatively large uncertainty.\footnote{For applications and discussion of median statistics see \cite{ChenRatra2003}, \cite{Mamajek2008}, \cite{Farooq2013b}, \cite{Crandall2014}, \cite{Zheng2016}, \cite{Leaf2017}, \cite{Sereno2017}, and \cite{Cao2018}.} To find the errors associated with the median we follow \cite{Gott2001}. For a data set of $N$ independent measurements, $M_i$, we define the probability of the median lying between $M_i$ and $M_{i+1}$ as a binomial distribution,
	\begin{equation}\label{eq:medianprobab}
	P = \frac{2^{-N}N!}{i!(N-i)!}\ .
	\end{equation}
	To determine the errors on the median $M_{\rm med}$, we start from $M_{\rm med}$, which has the highest probability, and integrate outwards from either side, stopping when the cumulative probability reaches 0.6827 of the total probability on either side. We take the difference between the median, ${M}_{\rm med}$, and the two $M$ values found at the end of the integrals to obtain the one standard deviation error bars, $1\sigma_{\pm}$. We repeat these integrals till the cumulative probability reaches 0.9545 of the total probability, to obtain the two standard deviation error bars, $2\sigma_{\pm}$. Note that the distribution need not be symmetric, and that the $2\sigma_{\pm}$ values are not necessarily twice the $1\sigma_{\pm}$ values.
	
	\subsubsection{Weighted mean statistics}\label{subsubsec:weighted}
	In contrast to the determination of the median, determination of the weighted mean makes use of the error of each measurement. Individual measurements with smaller errors are weighted more heavily in the weighted mean. For the median, each measurement carries the same weight. The weighted mean is defined for measurements $M_i$ each with non-zero error $\sigma_i$ as \citep{Podariu2001}
	\begin{equation}\label{eq:weightedmean}
	M_{\rm wm} = \frac{\sum_{i = 1}^{N} M_i/{\sigma_i}^2}{\sum_{i = 1}^{N} 1/{\sigma_i}^2}.
	\end{equation}
	The error associated with the weighted mean is symmetric ($\sigma_{+} = \sigma_{-}$), and given by
	\begin{equation}\label{eq:weightedmeanerr}
	\sigma_{\rm wm} = \frac{1}{\sqrt{\sum_{i = 1}^{N} 1/{\sigma_i}^2}}.
	\end{equation}
	Note that $\sigma_{\rm wm} < \sigma_i\ \forall\ i \in \{1, 2, 3, \dots, N\}$ if $N \neq 1$.
	
	\subsection{Error distributions}\label{subsec:errordist}
	An error distribution is a measure of how far each individual measurement is from the central estimate. It takes into account the error of the central estimate, $\sigma_{\rm CE}$, and of the individual measurement, $\sigma_i$. For a measurement $M_i$ and a central estimate $M_{\rm CE}$, we define
	\begin{equation}\label{eq:generrdist}
	{N_{\sigma_i}} = \frac{M_i - M_{\rm CE}}{\sqrt{{\sigma_i}^2 + {\sigma_{\rm CE}}^2}}.
	\end{equation}
	This assumes that $M_{\rm CE}$ is not determined by using the measurements $M_i$ and so it is not directly correlated with the data.
	
	For the median central estimate we have both an upper and a lower error. Hence we use the lower error for $M_i < M_{\rm med}$ and the upper error for $M_i \geq M_{\rm med}$. In this case, for the angular velocity measurements, $\omega_i$, with central estimate, $\omega_{\rm med}$, we have
	\begin{equation}\label{eq:mederrdist}
	{N_{\sigma_i}^{\rm med}} = 
	\begin{dcases*}
	\frac{{\omega}_i - {\omega_{\rm med}}}{\sqrt{{\sigma_i}^2  +{\sigma_{+}}^2}}& if ${\omega}_i \geq {\omega_{\rm med}}$\\
	\frac{{\omega}_i - {\omega_{\rm med}}}{\sqrt{{\sigma_i}^2  +{\sigma_{-}}^2}}& if  ${\omega}_i < {\omega_{\rm med}}$.
	\end{dcases*}
	\end{equation}
	
	For the weighted mean case, when the central estimate, $\omega_{\rm wm}$, is derived from the individual measurements, $\omega_i$, accounting for the correlations under the assumption that the measurements are gaussianly distributed \citep[see the appendix of][]{Camarillo2018}, we have
	\begin{equation}\label{eq:wmerrdist}
	{N_{\sigma_i}^{\rm wm}} = \frac{{\omega}_i - {\omega_{\rm wm}}}{\sqrt{{\sigma_i}^2 - {\sigma_{\rm wm}}^2}}.
	\end{equation}
	
	Ideally, we expect the error distribution $N_{\sigma_i}$ to follow a gaussian distribution of unit standard deviation ($\sigma = 1$), given by
	\begin{equation}\label{eq:Gauss}
	{\rm P}(x) = \frac{1}{\sqrt{2\pi{\sigma}^2}}{\exp}\left\{-\frac{x^2}{2{\sigma}^2}\right\} = \frac{1}{\sqrt{2\pi}}\exp\left\{-\frac{x^2}{2}\right\}.
	\end{equation}
	Here, 68.27\% of the data points lie within the 1$\sigma$ range of $x = 0$, 95.45\% within 2$\sigma$. Typically, real measurements are not gaussianly distributed \citep{Chen2003,Bailey160600}. We shall find that this is also the case for the angular velocity data we study here. It is of interest to see if the angular velocity error distributions can be fit by other functional forms. In addition to the gaussian distribution, we consider the Cauchy, Student's $t$, and Laplace distributions.
	
	Compared to the gaussian distribution, the Cauchy distribution has a smaller FWHM, but a much higher probability in the tails, with probability distribution function
	\begin{equation}\label{eq:Cauchy}
	{\rm P}(x) = \frac{1}{\pi}\frac{1}{1 + x^2}.
	\end{equation}
	For this distribution, 68.27\% and 95.45\% of the probability lie in the range $\left|x\right| \leq 1.8$ and $\left|x\right| \leq 14$ respectively.
	
	The Student's $t$ distribution is a generalized distribution, with an additional parameter $\nu$. The Student's $t$, qualitatively speaking, lies between the gaussian and the Cauchy, reducing to the former in the limit $\nu \rightarrow \infty$, and the latter in the limit $\nu = 1$. The limits containing 68.27\% and 95.45\% of the data vary with $\nu$, with the kurtosis reducing when $\nu$ is large. The Student's $t$ probability distribution function is
	\begin{equation}\label{eq:t_dist}
	{\rm P}(x) = \frac{\Gamma\left[(\nu + 1)/2\right]}{\sqrt{\pi\nu}\Gamma(\nu/2)} \frac{1}{\left(1 + x^2/\nu\right)^{(\nu + 1)/2}}
	\end{equation}
	where $\Gamma(x)$ is the gamma function. In our analyses we limit $\nu$ to lie between $[1, 100]$. Qualitatively, at $\nu = 100$ the $t$ distribution is very similar to the gaussian. 
	
	The Laplace distribution is an exponential distribution that has a relatively lower kurtosis. It does, however, have a longer tail than the gaussian. 68.27\% and 95.45\% of the probability lie in the range $\left|x\right| \leq 1.2$ and $\left|x\right| \leq 3.1$ respectively. The Laplace distribution is
	\begin{equation}\label{eq:Laplace}
	{\rm P}(x) = \frac{1}{2}\exp\left\{-\left|x\right|\right\}.
	\end{equation}
	
	Data with underestimated or overestimated errors would not have an error distribution consistent with the gaussian distribution of unit standard deviation, eq. \eqref{eq:Gauss}. So in addition to comparing the angular velocity error distributions to the functional forms described above, it is useful to also compare them to scaled versions of these distributions. We accomplish this by using a scale factor $S$ to form scaled error distributions,
	\begin{equation*}
	N_{\sigma_i}^S = N_{\sigma_i}/S,
	\end{equation*}
	which we then fit to the above functional forms.
	
	In a gaussian distribution of unit standard deviation, 68.27\% and 95.45\% of the probability lies in the ranges [-1, 1] and [-2, 2] respectively. To study the gaussianity of the angular velocity error distributions, we first determine the ranges of the error distributions that contain 68.27\% and 95.45\% of the error distribution probability on each side. To eliminate the need to have to consider a possible non-zero mean value, we then symmetrize the error distribution by appending its additive inverse to the original error distribution (i.e. if we define $^{\rm inv}N_\sigma \equiv -1 \times N_\sigma$ where the multiplication happens to the error of each individual measurement in the error distribution $N_\sigma$, then we use the error distribution formed by appending $^{\rm inv}N_\sigma\text{ to } N_\sigma$)\footnote{Now $0$ lies at the middle of the error distribution, and the error distribution is symmetric.}. We then determine its parameter ranges that contain 68.27\% and 95.45\% of the error distribution probability and the percentage of probability contained within the ranges $[-1, 1]$ and $[-2, 2]$ for $\left|N_\sigma\right|$. We call the latter two the expected fraction (abbreviated as EF in the tables) of the error distribution.
	
	\subsection{Kolmogorov-Smirnov (KS) test}\label{subsec:kstest}
	We next fit the symmetrized error distributions to the functional forms described in the previous subsection to determine which functional form best fits the error distribution. To accomplish this we employ the KS test, a well-understood goodness of fit test, which finds the maximum difference between the expected cumulative distribution from a hypothesis, say $\mathcal{H}_{\rm o}$, and the cumulative distribution function from the error distribution. This maximum difference is called the $D$ statistic, and the KS test uses it, along with the number of data points in the symmetrized error distribution, $N$, as factors for obtaining the probability that $\mathcal{H}_{\rm o}$ cannot be rejected.
	
	To do this, we use a modified statistic, $z = {f}(N, D)$, to obtain the corresponding probability, or $p$--value, of the data arising from our hypothesis. There are multiple such functions, with common ones including $z = \sqrt{N}D$. For our work, we use the $z$ statistic mentioned in \cite{NumRep}
	\begin{equation}
	z = \left(\sqrt{N}+ 0.12 + \frac{0.11}{\sqrt{N}}\right)D.
	\end{equation}
	We use this $z$ to find the $p$--value from its formula
	\begin{equation}
	p(z) = 2\sum_{j = 1}^{\infty}(-1)^{j - 1}\exp\left\{-2j^2z^2\right\}.
	\end{equation}
	From these equations, it can be shown that as $D\rightarrow 0$, $p \rightarrow 1$, and as $D \rightarrow 1$, $p \rightarrow 0$ for a large $N$. Non-trivially, it can also be shown that $p$ is a strictly decreasing function of $D$ for a constant $N$.
	
	We use this test to quantitatively find the best-fit functional form and scale factor $S$ by computing $p$ for the probability distributions described in \S\ref{subsec:errordist} by optimizing $S$ to provide the lowest $D$ statistic.
	
	The KS test makes no prior assumption about the data. However, it has an important shortcoming: it is more sensitive to the peak of the corresponding PDF than it is to the edges.
	
	\subsection{Gaussianity results}\label{subsec:results}
	
	Here we present and discuss the results of the analyses described in \S\ref{subsec:cestats}, \S\ref{subsec:errordist}, and \S\ref{subsec:kstest}. The results are given in tables in the Appendix. We first discuss the results obtained for the complete data set of 2,706 measurements. Next, we present the results of the subsets formed by splitting the data into tracer types, and finally, we examine the azimuthal symmetry of the data by subdividing them into angular sectors.
	
    Splitting the data into tracer types shows that velocity measurements of gas and star tracers are mildly inconsistent at $R\sim7$ kpc. The gas tracer error distributions are slightly less non-gaussian than the complete and star tracer error distributions.
	
	Assuming azimuthal isotropy, there is significant non-gaussianity in the error distributions. However, assessing the data in azimuthal sectors we find less non-gaussianity in the error distributions, indicating a mild deviation from azimuthal symmetry in these data.	
	
	All error distributions we study are non-gaussian, thus weighted mean analyses of these data are inappropriate.
	
	\subsubsection{The complete data set}\label{subsubsec:alldata}
	Here we present results from our analysis of the complete velocity data set. Table \ref{tab:cvall} lists the angular and linear circular velocity median central values for the radially binned data without the $1/3\ R$ bin exchanges.\footnote{Results obtained by binning the data with the $1/3\ R$ bin exchanges are only used to qualitatively asses how much the $R$ uncertainties could affect the results}
	
	Table \ref{tab:nsall} shows results from the $N_\sigma$ analysis of the complete data, using both the median and the weighted mean estimates of the central angular circular velocities. From it we see that most bin error distributions have high kurtosis and high skewness. The high kurtosis is especially important, as it is a sign of non-gaussianity for a symmetrized distribution. We expect 95.45\% of the error distribution to lie within the 2$\sigma$ range of the center of the gaussian distribution if 68.27\% lies within $\sigma$ of it. From this table we qualitatively see that in most bins the kurtosis is more consistent with distributions which have higher probability in the tails than a gaussian distribution, with 95.45\% of the error distribution lying in a range much larger than should be expected from the range containing 68.27\% of the error distribution for most bins. In addition, for a gaussian error distribution 68.27\% of the data ought to be contained in the $\pm1\sigma$ range. This is not consistent with the error distributions of most bins, possibly implying erroneous uncertainties, correlations, or intrinsic non-gaussianity of the system. This improves for bins closer to $R_0 = 8.0$ kpc, however, the high kurtosis does not drop, still indicating that a higher fraction of the error distribution lies in the tails compared to that of a gaussian distribution.
	
	The smaller error bars for the weighted mean central estimate results in a broader error distribution than its median central estimate counterpart. This further escalates the non-gaussianity of the error distributions derived from the weighted mean central estimates. The observed non-gaussianity invalidates usage of the weighted mean technique, thus we do not list weighted mean circular velocity central values in Table \ref{tab:cvall}.
	
	Comparing the results without and with the $1/3\ R$ bin exchanges (not listed), we see that they are quite similar. This indicates that the $R$ uncertainties cannot significantly change the conclusions. This is probably mostly because we bin in $R$.
	
	Table \ref{tab:ksall} gives the best-fit functional forms for the error distributions obtained for the median and weighted mean central estimates. These results largely corroborate the results observed in Table \ref{tab:nsall}, with best-fit distributions generally being the high kurtosis Cauchy, Laplace, and low to mid $\nu$ Student's $t$ distributions. There is an increase in the incidence of bins with gaussian distributions being the best fitting functional form for error distributions built using the weighted mean central estimate, but the corresponding scale factors $S$ are large, indicating that the error distributions are highly non-gaussian with standard deviations far from 1.
	
	\subsubsection{Tracer type subsets}\label{subsubsec:tracertypes}
	The gas tracer and star tracer data subsets, with 2,146 and 426 measurements respectively, are large enough for us to determine rotation curves. Tables \ref{tab:cvgas} and \ref{tab:cvstar} list the angular and linear circular velocities median central values for the radially binned gas and star tracer data without the $1/3\ R$ bin exchanges.
	
	Comparing the median central values in Tables \ref{tab:cvall}, \ref{tab:cvgas}, and \ref{tab:cvstar} we see that the $\omega$ values are largely mutually consistent for the complete data and the gas and star tracer data. However, for $R\sim7$ kpc, the star tracer binned velocities are somewhat lower.
	
	The $1/3\ R$ exchange affects the median central values of $\omega$ a little more significantly in the star tracer data subset. This is probably a consequence of the lower radial density of star tracers. Overall, the $R$ uncertainties are unlikely to significantly affect our conclusions.
	
	Tables \ref{tab:nsgas} and \ref{tab:nsstar} show the results of the $N_\sigma$ analyses for the gas and star tracer data subsets. We observe that the kurtosis and skewness of the error distributions reduce (and in some bins, significantly) for gas tracers data subset compared to the complete data set. Even the ranges containing 68.27\% and 95.45\% of the symmetrized gas tracer error distributions get closer to $\pm 1\sigma$ and $\pm 2\sigma$ respectively, indicating less non-gaussianity. For star tracers, the skewness reduces compared to the results for the complete data set, but the kurtosis remains high. This indicates that the star tracer measurement error distributions are more non-gaussian than the gas tracer measurement error distributions.
	
	These differences between the results of analyses of the complete data and the gas and star tracer data sets might be indicating that there is a systematic difference between the different data subsets. This is consistent with the discussion above about the inconsistencies between the median central angular velocity for the gas and star tracers near $R\sim 7$ kpc. We return to this issue in \S\ref{sec:FittingResults}.
	
	Tables \ref{tab:ksgas} and \ref{tab:ksstar} list the best-fit functional forms for the error distributions of the binned gas tracer and star tracer data. For the without $R$ error cases for median central values, we see from Tables \ref{tab:ksall}, \ref{tab:ksgas}, and \ref{tab:ksstar}, that 13\% (7 of 52), 33\% (15 of 46), and 19\% (4 of 21) of the bin error distributions for the complete data set and the gas and star data subsets are most consistent with a gaussian distribution respectively. However, most of the gaussian error distributions for the three data sets have scale factors significantly different from 1, with the closest two $S$'s being 1.24 and 1.41 for the complete data set, 0.9 and 1.28 for the gas tracers, and 0.83 and 1.32 for the star tracer subset. 
	
	For the error distributions formed using the weighted mean, there is an overall increase in the incidence of best-fit gaussian distributions, especially for the complete data, however, the scale factors $S$ are much larger. This pattern holds for all tables, though for a few bins, $S$ is pretty close to 1. These are mostly bins where the corresponding error distributions from median statistics have $S<1$.
	
	\subsubsection{Angular sector data subsets}\label{subsubsec:angularslices}
	
	To investigate the azimuthal symmetry of the Milky Way disk, we study the rotation curves in a number of angular sectors. It is also important to do this to see if the non-gaussianity we have found in the complete data set and in the gas and star tracer data subsets are a consequence of combining together measurements that are far apart in azimuthal separation. The radially binned data in each angular sector consists of measurements taken at points that are spatially adjacent.
	
	Dividing the 2,706 measurements among 10\degree\ sectors, the 4 sectors between 290\degree\ and 330\degree, and the 9 sectors between 340\degree\ and 70\degree, contain enough measurements to allow reasonable gaussianity studies.
	
	Tables \ref{tab:cvang1} through \ref{tab:cvang4} list the median central estimates of angular and linear circular velocity measurements for radially binned data without the $1/3\ R$ exchange for each 10\degree\ sector. By comparing the results for the original $R$ binning and the $1/3\ R$ exchange binning, we find that the $R$ uncertainties cannot significantly affect the conclusions. Similarly, although we haven't explicitly checked, it is likely that azimuthal position uncertainties also cannot significantly alter the conclusions. 
	
	There are hints of deviation from azimuthal isotropy in the central values listed in these tables. For example, for the sector between 290\degree\ and 300\degree, we observe for the radial bin $\sim2.80$ kpc from the Galactic center that the median angular circular velocity is $70.4_{-1.0}^{+0.6}\rm\ km\ s^{-1}\ kpc^{-1}$, compared to $75.2_{-0.5}^{+0.2}\rm\ km\ s^{-1}\ kpc^{-1}$ for the same galactic radius for the section between 60\degree and 70\degree. This might be caused by the bar at the center of the Milky Way, or might be an effect due to difference in peculiar motions of tracers at the two locations.
	
	Tables \ref{tab:nsang1} through \ref{tab:nsang4} list parameters of the error distributions obtained from the angular sector data. The kurtosis of the error distributions is high, but the range containing 68.27\% and 95.45\% of the error distributions is typically closer to 1 and 2 respectively than for the complete and gas and star tracers data sets. The skewness is also smaller for the angular sector data. These results generally indicate that the angular sector data are less non-gaussian than the complete and gas and star tracers data, although the angular sector data are still non-gaussian. This reduction in non-gaussianity is consistent with the velocity field being slightly azimuthally asymmetric.
	
	Tables \ref{tab:ksang1} through \ref{tab:ksang4} show the best-fit functional forms for the error distributions formed for all 13 angular sector subsets of data. The results show that about 27\% (46 of 171) of all bins have median error distributions best-fit by gaussian distributions, when binned using the original $R$ values. For binning with the $1/3\ R$ exchanges, the fraction increases to about 36\% (61 out of 171). For these error distributions best-fit by a gaussian, a fair fraction of the scale factors are close to 1, however, a number are significantly larger. Overall the sector data are less non-gaussian, although it is fair to conclude that even these error distributions are mostly non-gaussian.
	
	The weighted mean error distributions are best-fit by gaussian distributions more frequently than are their median counterparts. However, they have higher non-gaussianity, as their scales $S$ are further from 1. For median error distributions best-fit by gaussian distribution, about 13\% (6 of 46) and 20\% (12 of 61) of the best-fit scales $S$ lie in the range $[0.95, 1.05]$ for binning without and with the $1/3\ R$ exchanges respectively. These fractions are 8\% (5 of 65) and 8\% (6 of 76) for the weighted mean error distributions best-fit by gaussian distributions, and formed without and with the $1/3\ R$ exchanges respectively.
	
	Partially motivated by the gaussianity analyses of tracer-type data subsets, we check for non-gaussianity in measurements for gas tracers without the assumption of anisotropy. The error distributions hence formed are more gaussian than our previous results, albeit still non-gaussian. Therefore, the various angular sectors within the data subset are ill-suited for a weighted mean analysis.\footnote{We do not list tabulated results for this finding.}
	
	\section{Functional form fitting}\label{sec:functionalform}
	It is of interest to try to condense the information in the large number of circular velocity measurements by fitting these data to simple functions of $R$ characterized by a few parameters. 
	
	Previous efforts have been made to fit rotation curve measurements to simple functions of $R$ such as linear functions and power laws \citep[see e.g.][]{Clemens85,Fich1989,Brand93,McClure-G07,Reid2014}. These have typically involved using many fewer (and less heterogeneous) measurements than the \cite{Iocco2015} compilation we consider here. 
	
	We have shown in the previous section that the error bars of these velocity measurements are non-gaussian. Consequently we ignore the quoted error bars and instead use median statistics to determine central value and error bars for the spatially binned \cite{Iocco2015} data. In this section we fit functional form to these median statistics binned data measurements.
	
	We find that the simple two and three parameter functions we fit to these data have a very large reduced $\chi^2$. Interestingly, the three parameter functions fit better, having a smaller reduced $\chi^2$. We argue that the simple functional forms we use are not flexible enough to fit the small-scale spatial structure in the circular rotational velocity data.
	
	\begin{figure}
		\begin{center}
			\centering
			\includegraphics[height=66mm,width=84mm]{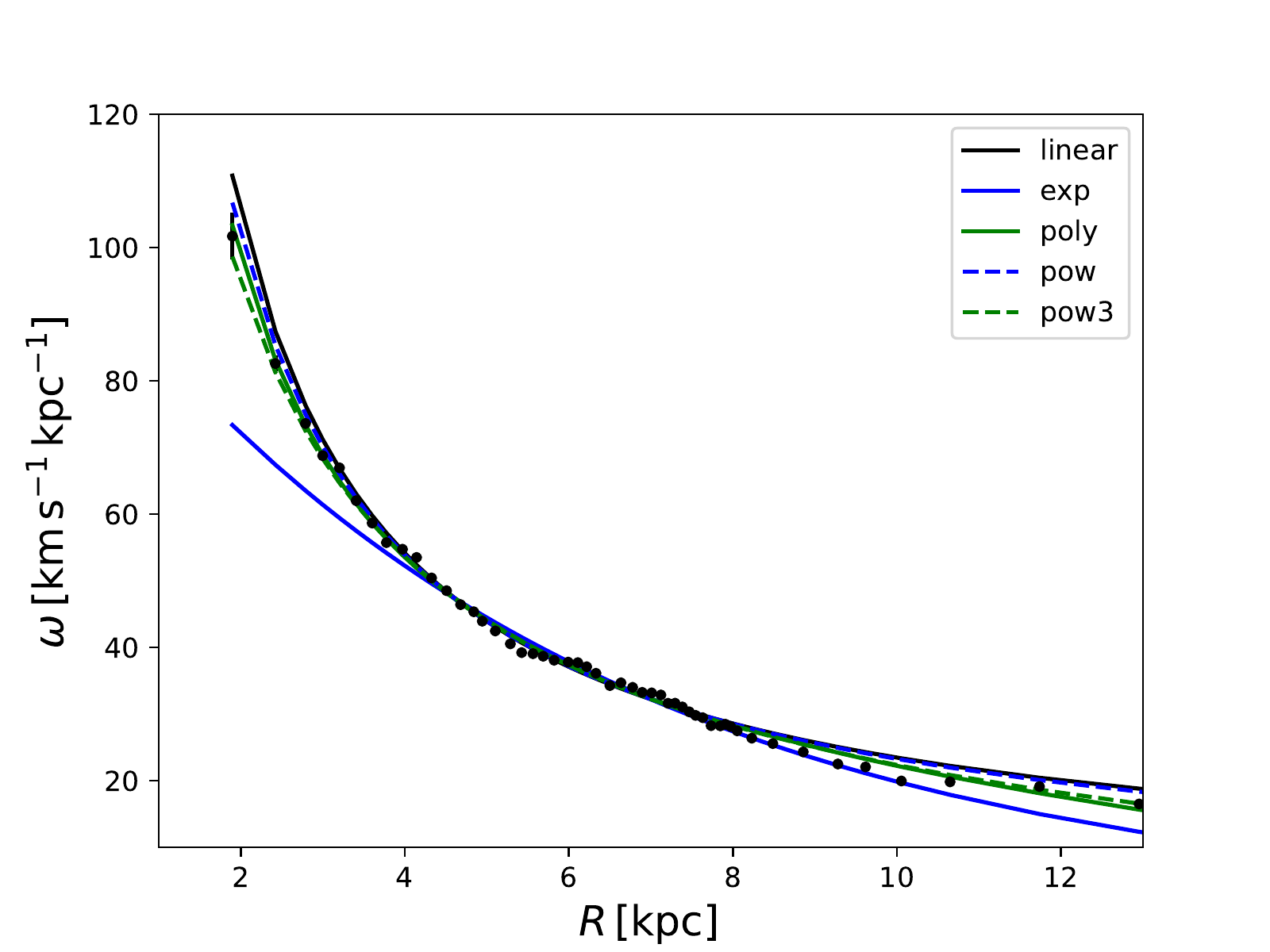}
			\caption{Best fits to the median statistics measurements obtained from the complete data binned in 52 bins for all five functional forms for the angular circular velocity $\omega(R)$. These correspond to the cases $\rm AelB_{52}$.}
			\label{fig:linear_ff_fit}
		\end{center}
	\end{figure}
	
	\subsection{Functional forms}\label{subsec:ff}
    
	We fit the angular circular velocity data to five different functional forms. This is partly motivated by the desire to try to determine model independent summary descriptions for these data. By fitting to different models we can determine which results are independent of the models used. We find that the best-fitting five different $v(R)$ functions agree reasonably well over $3 \text{ kpc}\lesssim R\lesssim 9\rm\ kpc$, although there are discernible differences between the five rotation curves.
	
	The first function we consider is the linear function
    \begin{equation}
    \omega(R) = a_{0}\frac{v_{0}}{R}+ a_{1}\omega_{0},
    \label{eq:lin_fit}
    \end{equation}
    where $a_{0}$ and $a_{1}$ are fitting parameters to be determined from the data and $v_0$ and $R_0$ are the local linear circular velocity and distance of the Sun from the Galactic center respectively, and $\omega_0 = v_0/R_0$ is the local angular circular velocity. We use $v_0 = 220\rm\ km\ s^{-1}$ and $R_0 = 8.0\rm\ kpc$ \citep[see][]{Camarillo2018b,Camarillo2018}.
    
    The second and third functional forms we consider are the two parameter power law
    \begin{equation}
    \omega(R) = a_0\omega_0\left(\frac{R_0}{R}\right)^{a_1},
    \label{eq:pow}
    \end{equation}
    and the three parameter power law
    \begin{equation}
      \omega(R) = a_0\omega_0\left[\left(\frac{R_0}{R}\right)^{a_1}+a_2\right].
      \label{eq:power3}
    \end{equation}
    Here $a_0$, $a_1$, $a_2$ are the fitting parameters.
    
    We also study two other functional forms from \cite{Fich1989}. The first is an exponential function
    \begin{equation}
    \omega(R) = \omega_{0}a_{0} \exp{\left[a_{1}\frac{R}{R_{0}}\right]},
    \label{eq:exp}
    \end{equation}
    and the second is the second-order polynomial function
    \begin{equation}
    \omega(R) = \frac{v_{0}}{R}\left[a_{0}+ a_{1}\left(\frac{R}{R_{0}}-1\right)+ a_{2}\left(\frac{R}{R_{0}}-1\right)^2\right].
    \label{eq:poly}
    \end{equation}
    
    \subsection{Fitting procedure}\label{subsec:fitprocedure}
    \begin{figure}
    \begin{center}
        \centering
        \includegraphics[height=66mm,width=84mm]{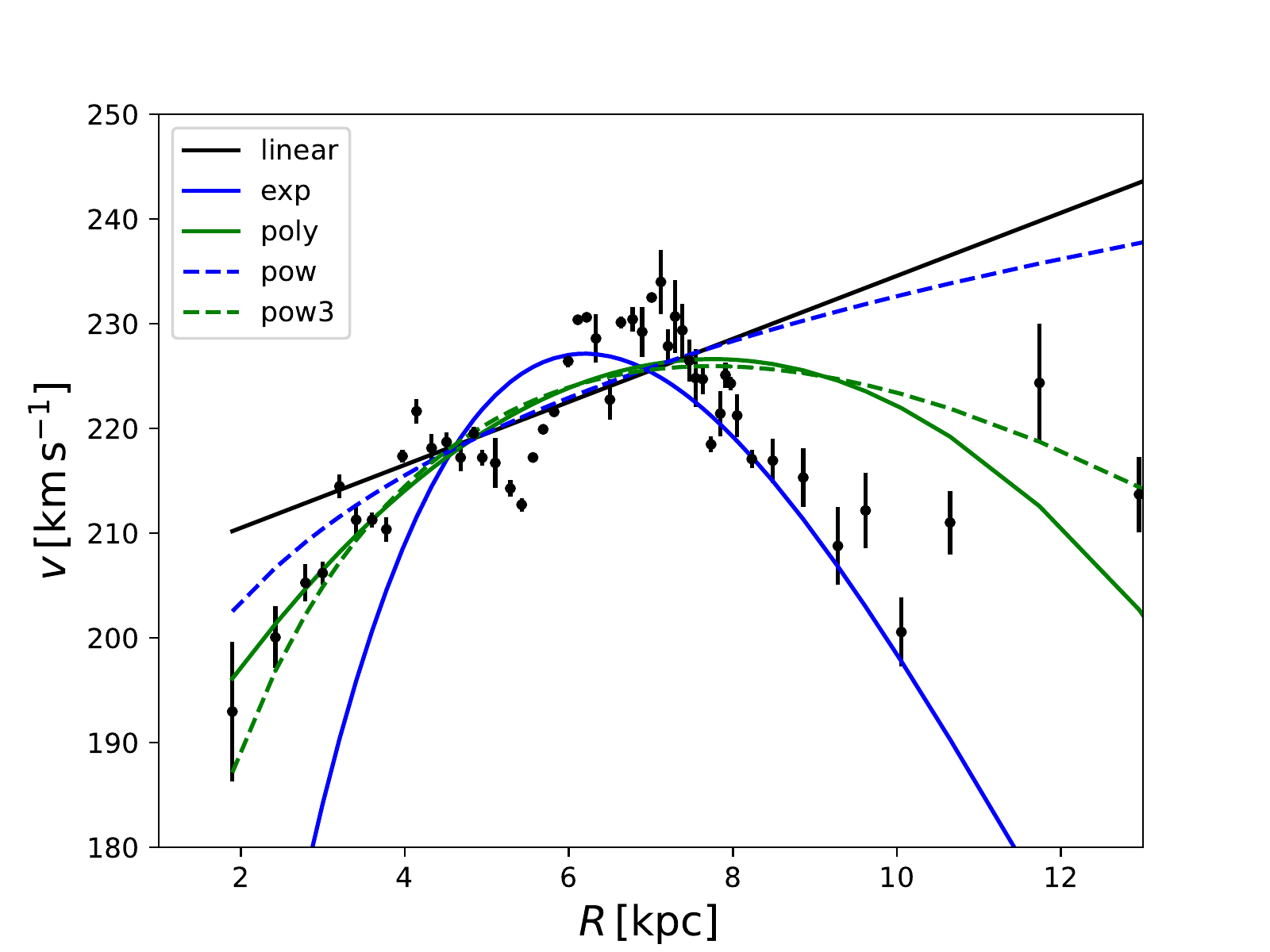}
        \caption{Linear circular velocity $v(R) = \omega(R)\times R$ corresponding to the best-fit angular circular velocity $\omega(R)$ shown in Fig. \ref{fig:linear_ff_fit} determined using median statistics measurements for the complete data set binned in 52 bins, for all five $\omega(R)$ functional forms. These correspond to the cases $\rm AelB_{52}$.}
        \label{figure:fits1}
    \end{center}
    \end{figure}
    \begin{figure}
	\begin{center}
		\centering
		\includegraphics[height=66mm,width=84mm]{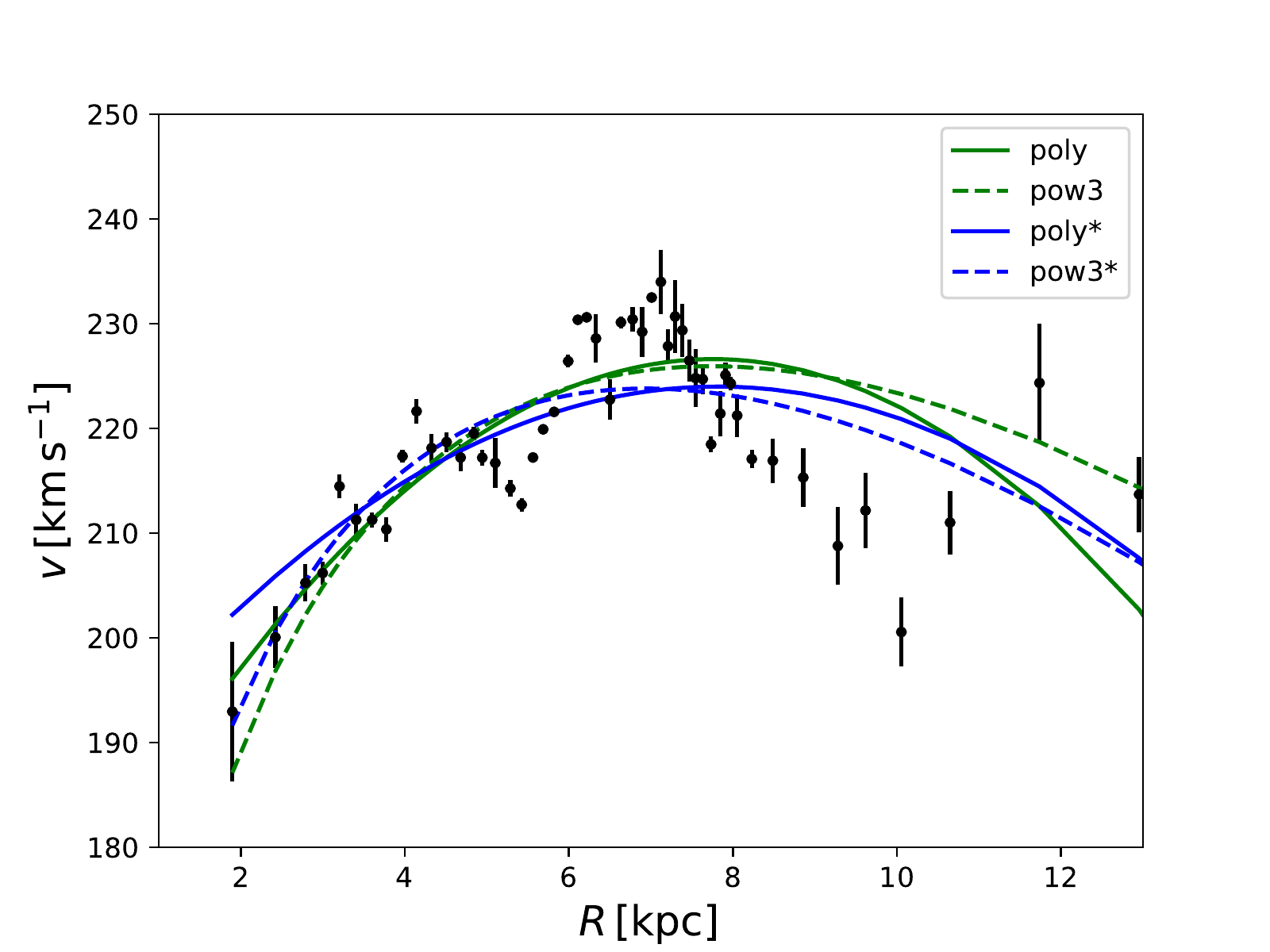}
		\caption{Linear circular velocity $v(R)$'s corresponding to the best-fit $\omega(R)$'s determined using median statistics measurements for complete data binned in 52 bins, excluding [$\rm AelB_{52}M_2\text{ and }AelB_{52}M_4$, un-starred in top-right linestyle legend] and including [$\rm AeLB_{52}M_2\text{ and }AeLB_{52}M_4$, starred] the additional error $\sigma_{\rm int}$. The solid lines are fits to the polynomial functional form of eq. \eqref{eq:poly} and the dashed lines are fits to the three-parameter power-law functional form of eq. \eqref{eq:power3}.} 
		\label{figure:fits2}
	\end{center}
    \end{figure}
    \begin{figure}
	\begin{center}
		\centering
		\includegraphics[height=66mm,width=84mm]{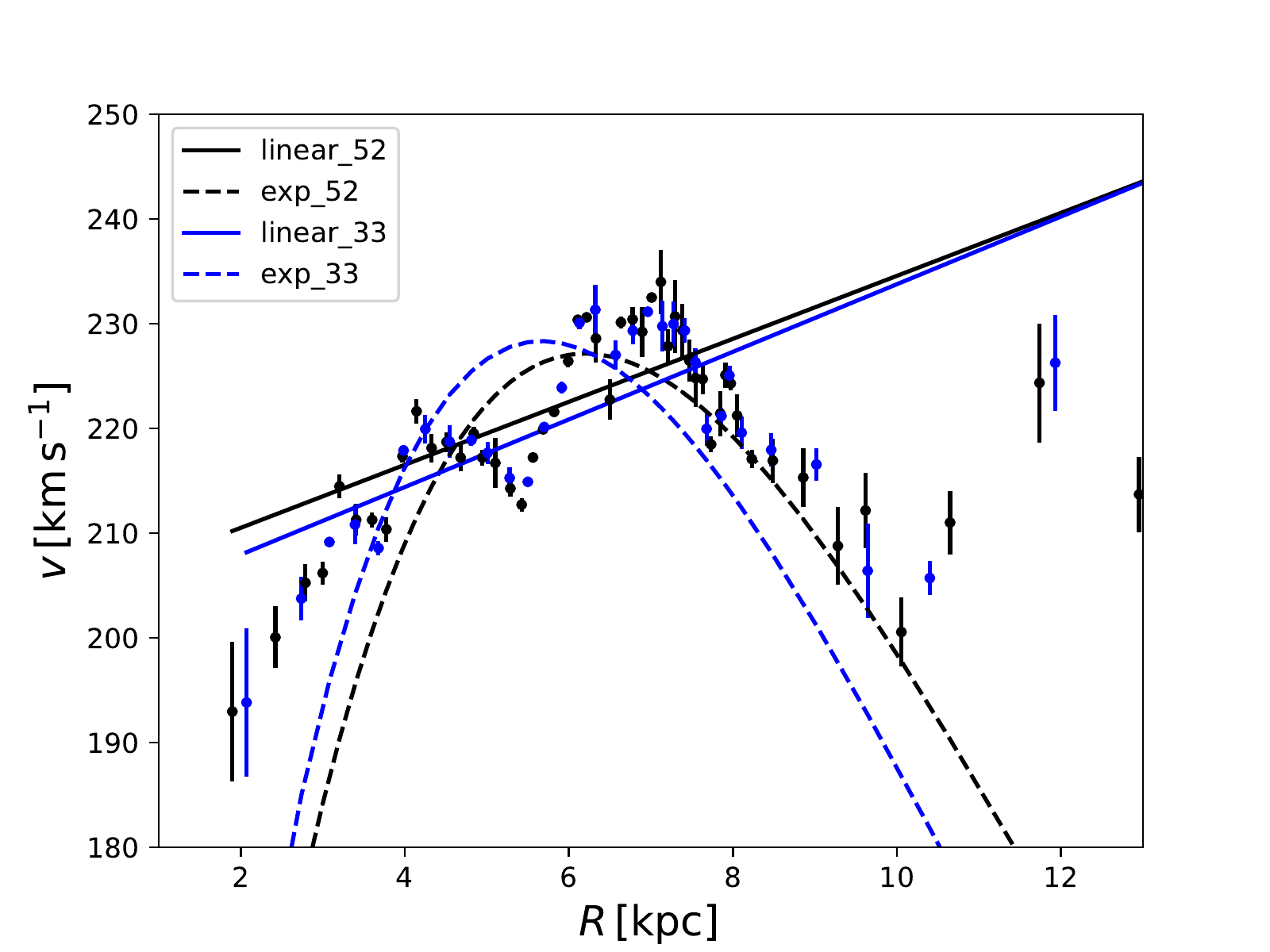}
		\caption{Linear circular velocity $v(R)$'s corresponding to best-fit $\omega(R)$'s determined using median statistics measurements for the complete data binned in 52 (black) and 33 (blue) bins. Solid lines are fits to the linear function of eq. \eqref{eq:lin_fit} and dashed lines are fits to the exponential form in eq. \eqref{eq:exp}. The top left linestyle legend entries correspond to cases $\rm AelB_{52}M_1, AelB_{52}M_5, AelB_{33}M_1\text{, and } AelB_{33}M_5$ respectively.} 
		\label{figure:fits3}
	\end{center}
    \end{figure}
	\begin{figure}
    \begin{center}
		\centering
		\includegraphics[height=66mm,width=84mm]{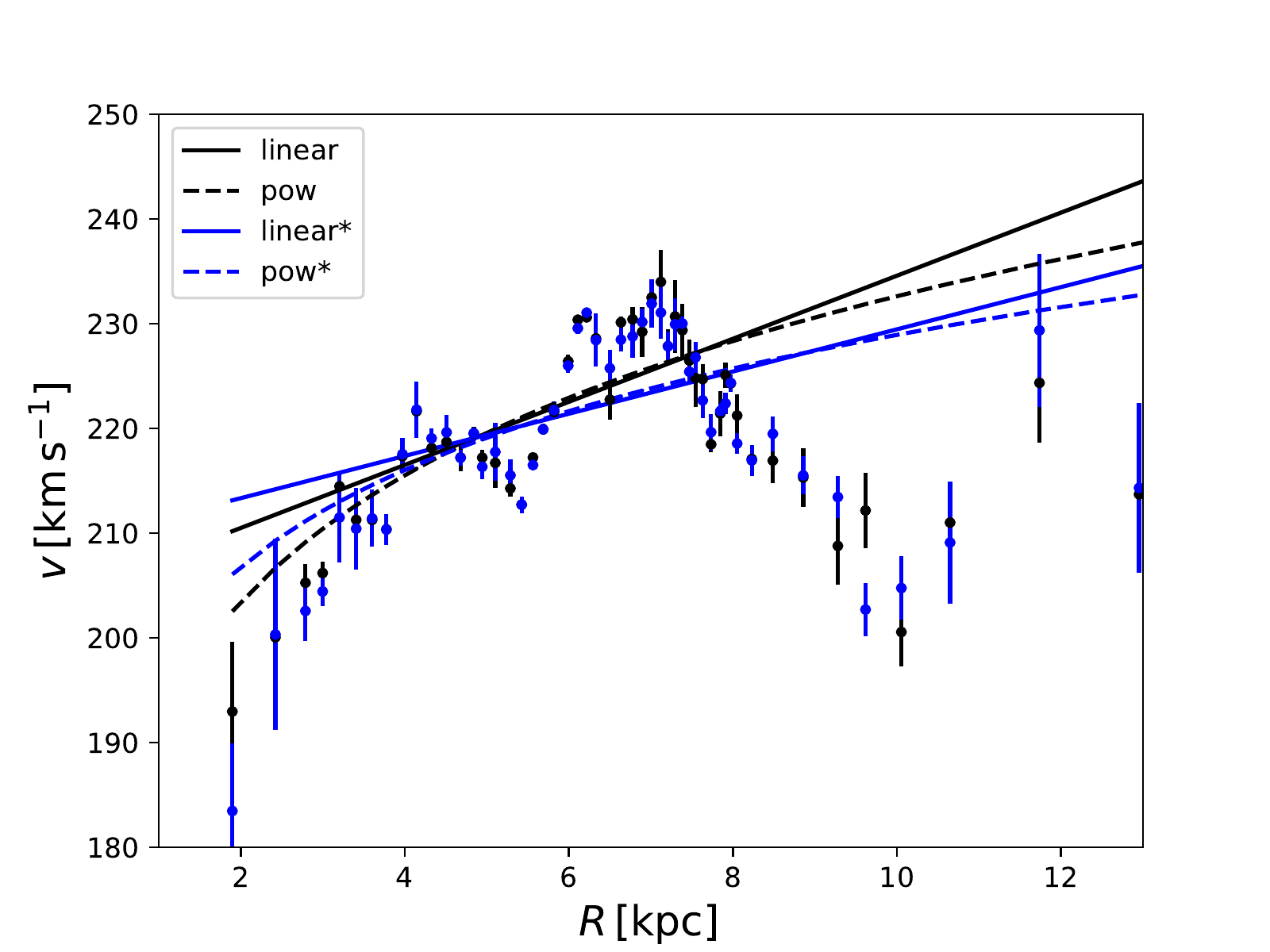}
		\caption{Linear circular velocity $v(R)$'s corresponding to best-fit $\omega(R)$'s determined using median statistics measurements for the complete data binned in 52 bins with (black) and without (blue) the $1/3 R$ exchange operation. The solid lines are fits to the linear form in eq. \eqref{eq:lin_fit}, and the dashed lines are fits to the two-parameter power law model in eq. \eqref{eq:pow}.  The top left linestyle legend entries correspond to cases $\rm AelB_{52}M_1, AElB_{52}M_3, AElB_{52}M_1\text{, and } AElB_{52}M_3$ respectively.} 
		\label{figure:fits4}
	\end{center}
    \end{figure}
	\begin{figure}
	\begin{center}
		\centering
    	\includegraphics[height=66mm,width=84mm]{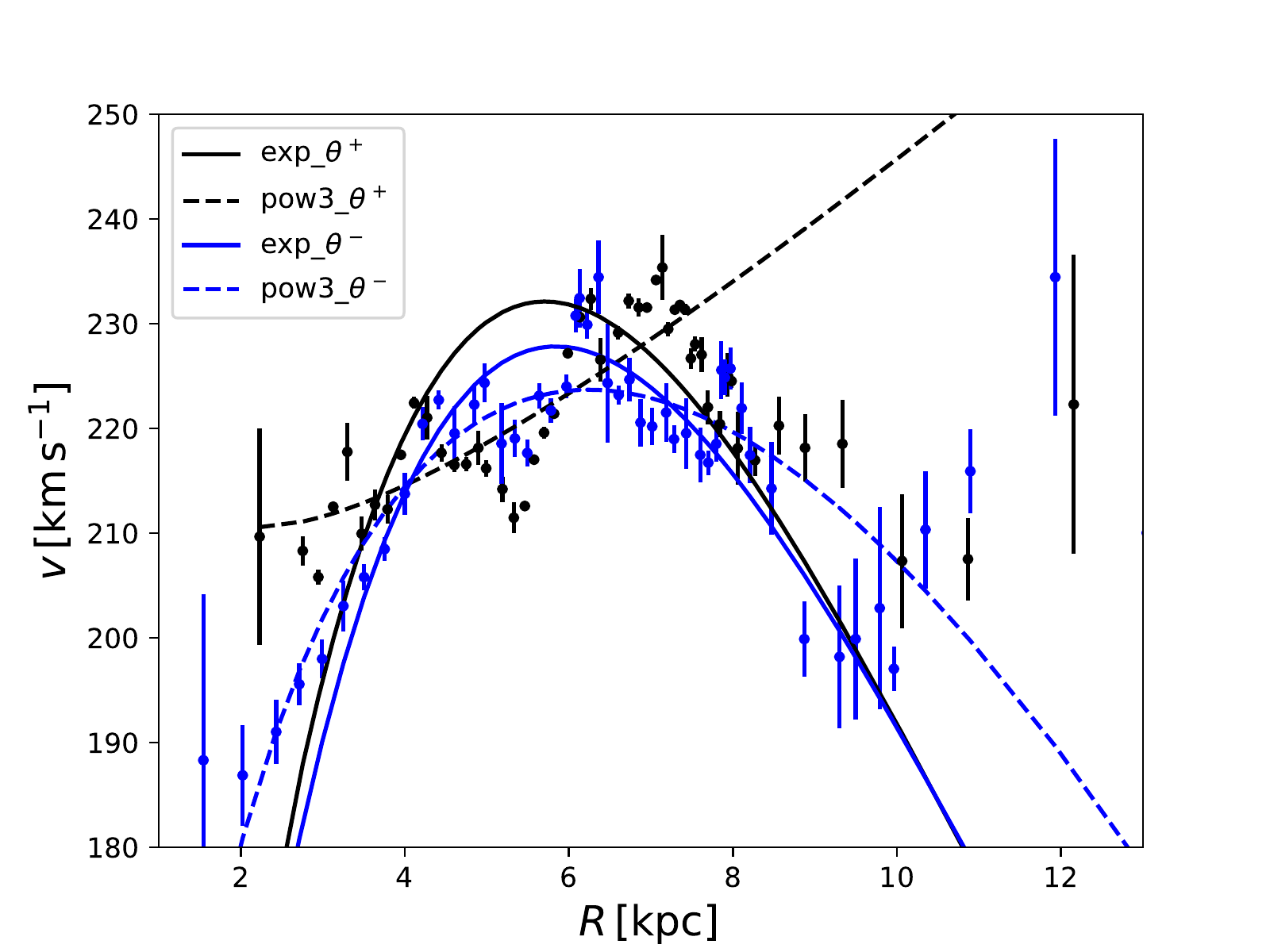}
		\caption{Linear circular velocity $v(R)$'s corresponding to best-fit $\omega(R)$'s determined using median statistics measurements for the positive (black) and negative azimuthal angle (blue) data. The solid lines are fits to the exponential form in eq. \eqref{eq:exp}, and the dashed lines are fits to the three-parameter power law model in eq. \eqref{eq:power3}.  The top left linestyle legend entries correspond to cases $\rm PelB_{52}M_5, PelB_{52}M_4, NelB_{52}M_5\text{, and } NelB_{52}M_4$ respectively.} 
		\label{figure:fits5}
	\end{center}
    \end{figure}
    As discussed in \cite{Fich1989}, fitting should be performed using angular circular velocities as a function of distance, as the errors of $\omega$ are not correlated with those of $R$, unlike the case for the $v$ errors.
    
    While \cite{Fich1989} use a 2D $\chi^{2}$ fitting technique that makes use of the information in both the $\omega$ and $R$ uncertainties, we cannot use this procedure. This is because the data we use have non-gaussian error bars. So instead we discard the quoted errors, bin the data, and use median statistics to derive angular velocity central values and error bars for the binned data. We fit the above functional forms to the median statistics determined $\omega_{{\rm med}, i}\pm \sigma_i$, of the bins centered at $R_i$.\footnote{We have shown in the previous section that the $R$ uncertainties do not significantly affect our results.}
    
    To fit the above functional forms to the median statistics data we use the Markov chain Monte Carlo (MCMC) procedure to maximize a likelihood and determine the best-fit parameter values. Here we utilize the Python based software \texttt{emcee} \citep[][]{Mackey2013}.
    
    The likelihood we use is $\mathcal{L}\propto\exp\left\{-\chi^2/2\right\}$ where $\chi^2$ is
    \begin{equation}\label{eq:chi2}
    \chi^2 = \sum_{i=1}^{N}\frac{(\omega_{i}-\omega_{{\rm med},i})^2}{\sigma_i^2},
    \end{equation}
    where $N$ is the number of measurements (bins). The goodness of fit is determined from the reduced $\chi^2$ value
    \begin{equation}\label{eq:red_chi2}
    \chi^{2}_{\nu} = \frac{\chi^{2}}{\nu},
    \end{equation}
    where $\nu$ is the number of degrees of freedom.
    
    We see that the functional forms of \S\ref{subsec:ff} are poor fits to the velocity data. This is reflected in the large $\chi^2_\nu$ values we find. This is almost certainly a consequence of significant spatial substructure in the $\omega(R)$ data that is not adequately captured by the simple, few parameter, functional forms of \S\ref{subsec:ff}
    
    While not likely, it is possible that this substructure is not physical but rather a consequence of unobserved systematic errors or correlations in the velocity measurements. To study the possible consequences of this we introduce an additional unknown uncertainty $\sigma_{\rm int}$ and regard it as an additional free parameter to be determined by the MCMC fitting procedure. In this case, we use the modified $\chi^2$
    \begin{equation}\label{eq:mod_chi2}
    \chi^2 = \sum_{i}\ln(\sigma_i^2+\sigma_{\rm int}^2)+\sum_{i}\frac{(\omega_{i}-\omega_{{\rm med},i})^2}{\sigma_i^2+\sigma_{\rm int}^2},
    \end{equation}
    and the modified reduced $\chi^2$ is 
    \begin{equation}\label{eq:mod_redchi2}
    \chi^{2}_{\nu} = \frac{1}{\nu}\sum_{i}\frac{(\omega_{i}-\omega_{{\rm med},i})^2}{\sigma_i^2+\sigma_{\rm int}^2}.
    \end{equation}
    We use these two definitions of $\chi^2$ in our fitting procedure and study the effects of including $\sigma_{\rm int}$.
	
	\subsection{Fitting results}\label{subsec:ffresults}
    \setlength\tabcolsep{4.5pt}
    \begin{table}
    \tiny 
      	\caption{Best-fit parameter values and error bars.}
      	\label{table:fitPars}
      	\begin{center}
      	\begin{tabular}{lrrrrl}\hline\hline
      	Case name & $a_0$ & $a_1$ & $a_2$ & $\sigma_{\rm int}^{\rm a}$& $\chi^2_\nu$ \\ \hline
	    AelB$_{33}$M$_{1}$	&	$0.916 \pm 0.002$	&	$0.117 \pm 0.003$	&	...	&	...	&	50.47	\\
		AelB$_{33}$M$_{2}$	&	$1.021 \pm 0.001$	&	$-0.027 \pm 0.007$	&	$-0.241 \pm 0.011$	&	...	&	37.32	\\
		AelB$_{33}$M$_{3}$	&	$1.031 \pm 0.001$	&	$0.919 \pm 0.002$	&	...	&	...	&	44.20	\\
		AelB$_{33}$M$_{4}$	&	$1.337 \pm 0.029$	&	$0.764 \pm 0.012$	&	$-0.237 \pm 0.017$	&	...	&	40.43	\\
		AelB$_{33}$M$_{5}$	&	$3.982 \pm 0.009$	&	$-1.411 \pm 0.003$	&	...	&	...	&	176.6	\\
		AeLB$_{33}$M$_{1}$	&	$0.962 \pm 0.019$	&	$0.045 \pm 0.023$	&	...	&	$7.939 \pm 1.155$	&	1.037	\\
		AeLB$_{33}$M$_{2}$	&	$1.020 \pm 0.007$	&	$-0.006 \pm 0.022$	&	$-0.187 \pm 0.044$	&	$5.936 \pm 0.960$	&	1.124	\\
		AeLB$_{33}$M$_{3}$	&	$1.012 \pm 0.008$	&	$0.951 \pm 0.016$	&		&	$7.378 \pm 1.068$	&	1.022	\\
		AeLB$_{33}$M$_{4}$	&	$1.485 \pm 0.142$	&	$0.710 \pm 0.056$	&	$-0.316 \pm 0.064$	&	$5.419 \pm 0.860$	&	1.084	\\
		AeLB$_{33}$M$_{5}$	&	$3.525 \pm 0.155$	&	$-1.220 \pm 0.051$	&	...	&	$16.17 \pm 2.45$	&	1.133	\\
		AElB$_{33}$M$_{1}$	&	$0.918 \pm 0.004$	&	$0.125 \pm 0.005$	&	...	&	...	&	23.19	\\
		AElB$_{33}$M$_{2}$	&	$1.029 \pm 0.002$	&	$-0.063 \pm 0.012$	&	$-0.367 \pm 0.021$	&	...	&	14.05	\\
		AElB$_{33}$M$_{3}$	&	$1.042 \pm 0.001$	&	$0.906 \pm 0.004$	&	...	&	...	&	20.00	\\
		AElB$_{33}$M$_{4}$	&	$1.931 \pm 0.108$	&	$0.559 \pm 0.026$	&	$-0.468 \pm 0.030$	&	...	&	14.93	\\
		AElB$_{33}$M$_{5}$	&	$3.628 \pm 0.016$	&	$-1.284 \pm 0.006$	&	...	&	...	&	31.79	\\
		AELB$_{33}$M$_{1}$	&	$0.961 \pm 0.021$	&	$0.043 \pm 0.025$	&	...	&	$8.090 \pm 1.275$	&	1.099	\\
		AELB$_{33}$M$_{2}$	&	$1.019 \pm 0.006$	&	$-0.034 \pm 0.023$	&	$-0.254 \pm 0.045$	&	$4.984 \pm 0.887$	&	1.204	\\
		AELB$_{33}$M$_{3}$	&	$1.011 \pm 0.008$	&	$0.949 \pm 0.017$	&	...	&	$7.394 \pm 1.190$	&	1.089	\\
		AELB$_{33}$M$_{4}$	&	$1.716 \pm 0.180$	&	$0.630 \pm 0.056$	&	$-0.409 \pm 0.061$	&	$4.625 \pm 0.762$	&	1.059	\\
		AELB$_{33}$M$_{5}$	&	$3.576 \pm 0.130$	&	$-1.246 \pm 0.043$	&	...	&	$12.69 \pm 2.12$	&	1.257	\\
		AelB$_{52}$M$_{1}$	&	$0.929 \pm 0.002$	&	$0.109 \pm 0.003$	&	...	&	...	&	50.00	\\
		AelB$_{52}$M$_{2}$	&	$1.030 \pm 0.001$	&	$-0.015 \pm 0.006$	&	$-0.258 \pm 0.010$	&	...	&	38.31	\\
		AelB$_{52}$M$_{3}$	&	$1.038 \pm 0.001$	&	$0.917 \pm 0.002$	&	...	&	...	&	45.22	\\
		AelB$_{52}$M$_{4}$	&	$1.497 \pm 0.036$	&	$0.693 \pm 0.013$	&	$-0.314 \pm 0.017$	&	...	&	40.24	\\
		AelB$_{52}$M$_{5}$	&	$3.623 \pm 0.010$	&	$-1.291 \pm 0.004$	&	...	&	...	&	98.36	\\
		AeLB$_{52}$M$_{1}$	&	$0.979 \pm 0.015$	&	$0.021 \pm 0.018$	&	...	&	$8.365 \pm 0.958$	&	1.061	\\
		AeLB$_{52}$M$_{2}$	&	$1.018 \pm 0.005$	&	$-0.008 \pm 0.014$	&	$-0.181 \pm 0.027$	&	$5.771 \pm 0.714$	&	1.076	\\
		AeLB$_{52}$M$_{3}$	&	$1.006 \pm 0.006$	&	$0.965 \pm 0.013$	&	...	&	$7.818 \pm 0.911$	&	1.073	\\
		AeLB$_{52}$M$_{4}$	&	$1.471 \pm 0.118$	&	$0.719 \pm 0.050$	&	$-0.311 \pm 0.056$	&	$5.355 \pm 0.932$	&	1.009	\\
		AeLB$_{52}$M$_{5}$	&	$3.538 \pm 0.126$	&	$-1.224 \pm 0.041$	&	...	&	$16.59 \pm 1.81$	&	1.044	\\
		AElB$_{52}$M$_{1}$	&	$0.951 \pm 0.003$	&	$0.073 \pm 0.004$	&	...	&	...	&	33.23	\\
		AElB$_{52}$M$_{2}$	&	$1.018 \pm 0.001$	&	$-0.086 \pm 0.008$	&	$-0.386 \pm 0.017$	&	...	&	23.57	\\
		AElB$_{52}$M$_{3}$	&	$1.026 \pm 0.001$	&	$0.937 \pm 0.003$	&	...	&	...	&	30.90	\\
		AElB$_{52}$M$_{4}$	&	$1.939 \pm 0.085$	&	$0.558 \pm 0.021$	&	$-0.476 \pm 0.023$	&	...	&	24.65	\\
		AElB$_{52}$M$_{5}$	&	$3.457 \pm 0.012$	&	$-1.236 \pm 0.004$	&	...	&	...	&	41.24	\\
		AELB$_{52}$M$_{1}$	&	$0.959 \pm 0.017$	&	$0.047 \pm 0.020$	&	...	&	$8.100 \pm 0.963$	&	1.057	\\
		AELB$_{52}$M$_{2}$	&	$1.017 \pm 0.005$	&	$-0.023 \pm 0.021$	&	$-0.220 \pm 0.040$	&	$5.705 \pm 0.783$	&	1.206	\\
		AELB$_{52}$M$_{3}$	&	$1.012 \pm 0.006$	&	$0.949 \pm 0.014$	&	...	&	$7.564 \pm 0.902$	&	1.032	\\
		AELB$_{52}$M$_{4}$	&	$1.628 \pm 0.143$	&	$0.660 \pm 0.048$	&	$-0.378 \pm 0.054$	&	$5.093 \pm 0.677$	&	1.129	\\
		AELB$_{52}$M$_{5}$	&	$3.566 \pm 0.113$	&	$-1.240 \pm 0.037$	&	...	&	$13.63 \pm 1.81$	&	1.304	\\\hline\hline
		\multicolumn{6}{l}{$^{\rm a}$: In units of $\rm km\ s^{-1}\ kpc^{-1}$.}
      	\end{tabular}
    \end{center}
    \end{table}
    \begin{figure}
    \begin{center}
    	\centering
		\includegraphics[height=80mm,width=80mm]{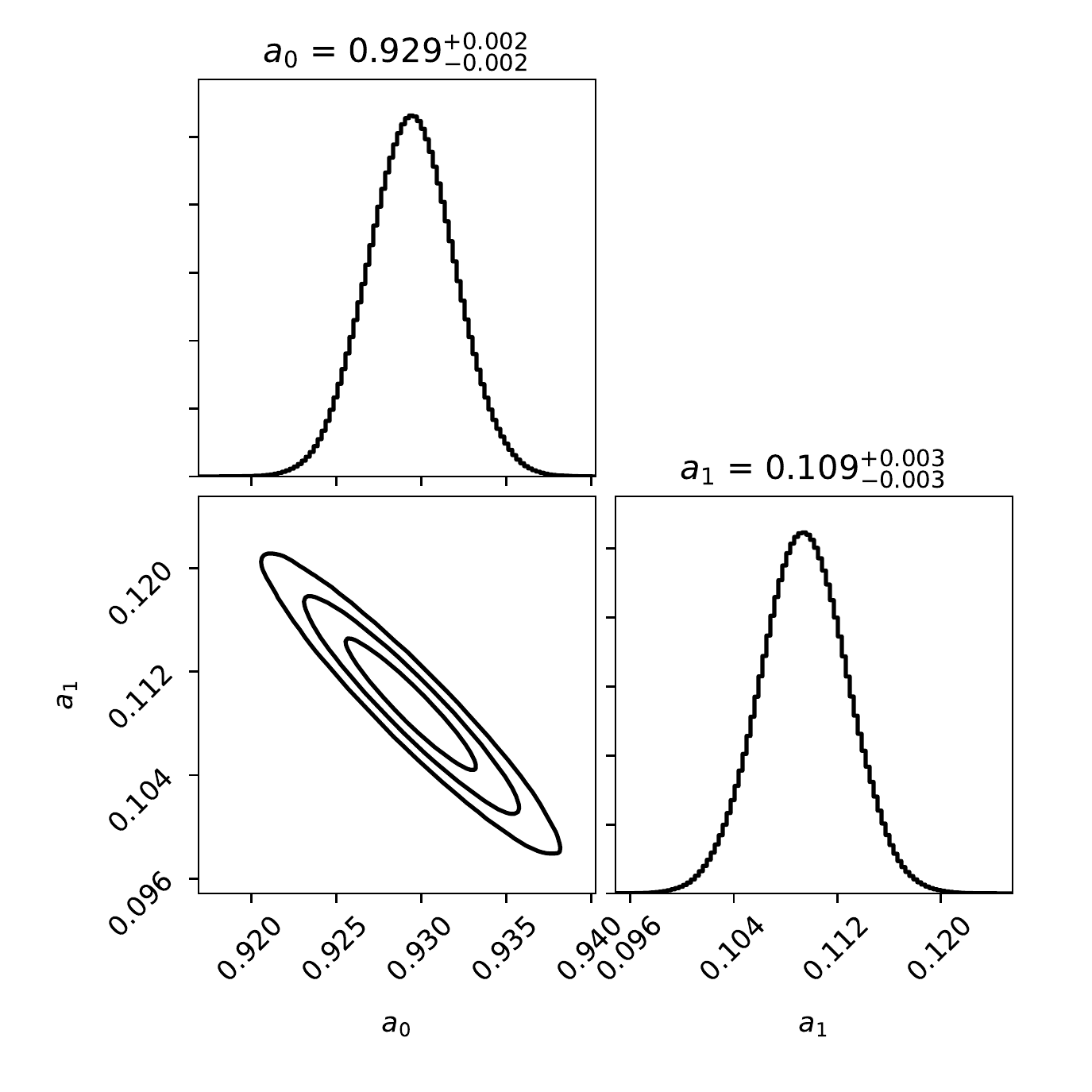}
		\caption{Two-dimensional likelihood contours and one-dimensional likelihoods for the linear function of eq. \eqref{eq:lin_fit} fit to the complete data binned in 52 bins, $\rm AelB_{52}M_1$.} 
		\label{figure:c1}
	\end{center}
    \end{figure}
    \begin{figure}
 	\begin{center}
 		\centering
 		\includegraphics[height=80mm,width=80mm]{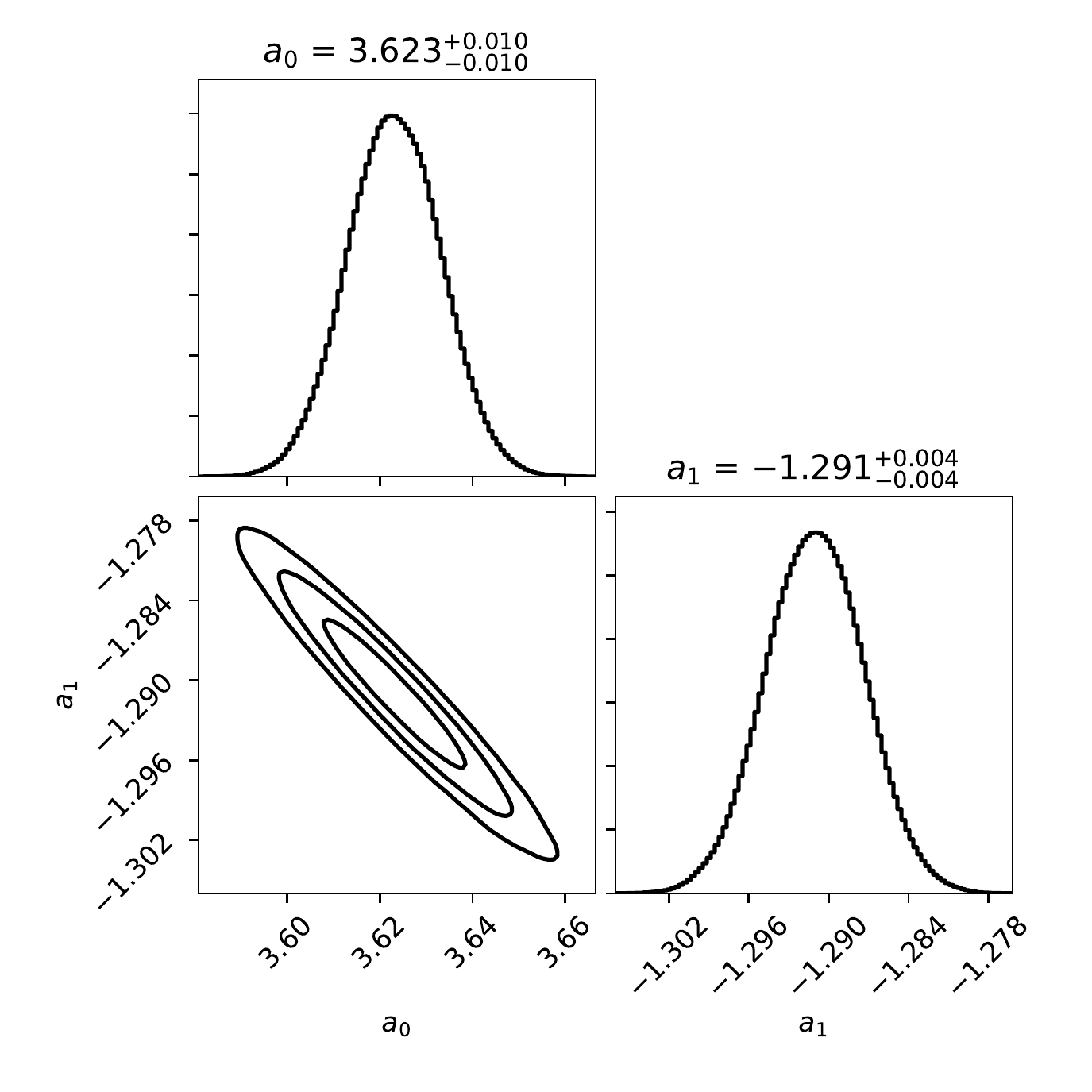}
 		\caption{Two-dimensional likelihood contours and one-dimensional likelihoods for the exponential function of eq. \eqref{eq:exp} fit to the complete data binned in 52 bins, $\rm AelB_{52}M_5$.} 
 		\label{figure:c2}
 	\end{center}
    \end{figure}
    \begin{figure}
 	\begin{center}
 		\centering
 		\includegraphics[height=80mm,width=80mm]{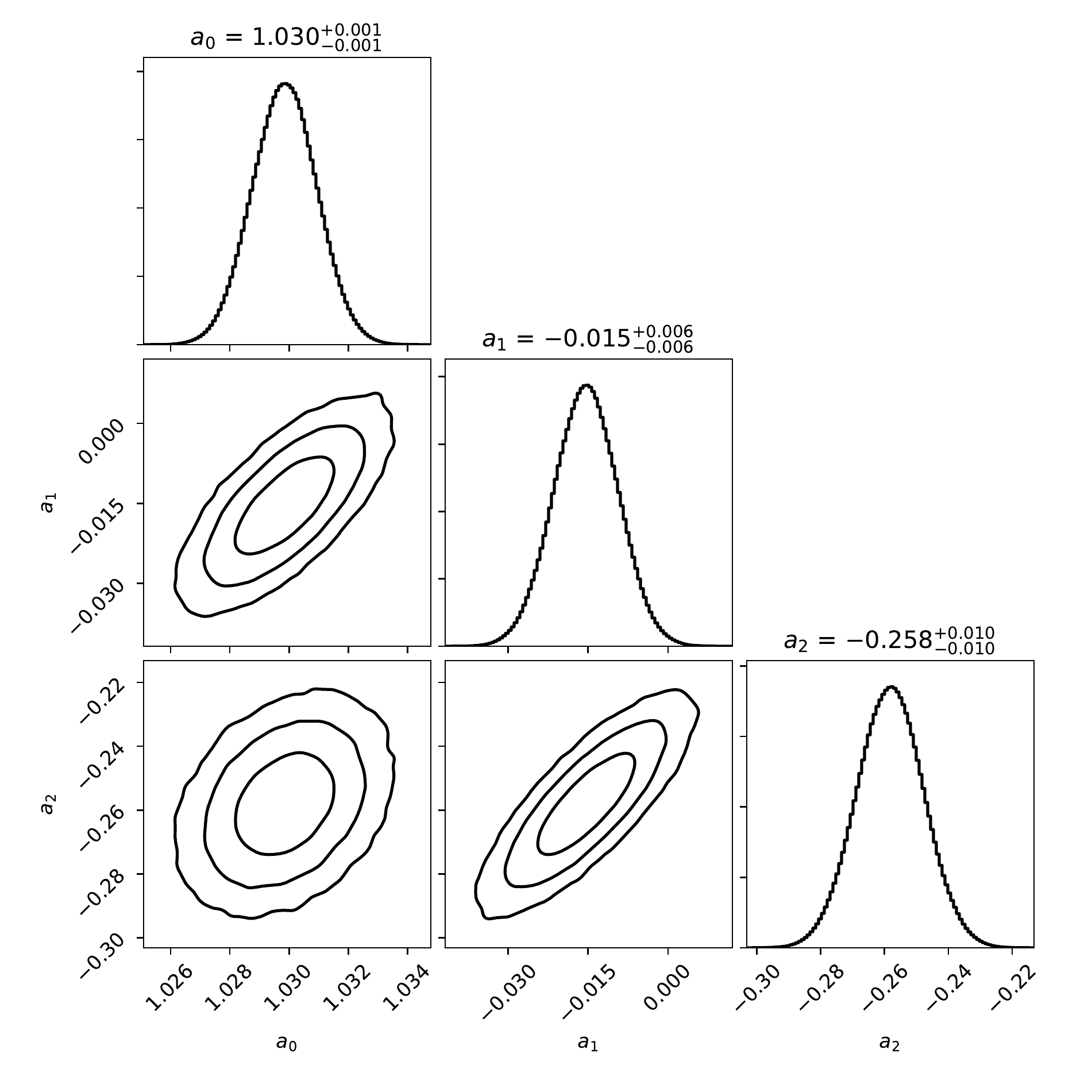}
 		\caption{Two-dimensional likelihood contours and one-dimensional likelihoods for the polynomial function of eq. \eqref{eq:poly} fit to the complete data binned in 52 bins, $\rm AelB_{52}M_2$.} 
 		\label{figure:c3}
     \end{center}
     \end{figure}
    \begin{figure}
 	\begin{center}
 		\centering
 		\includegraphics[height=80mm,width=80mm]{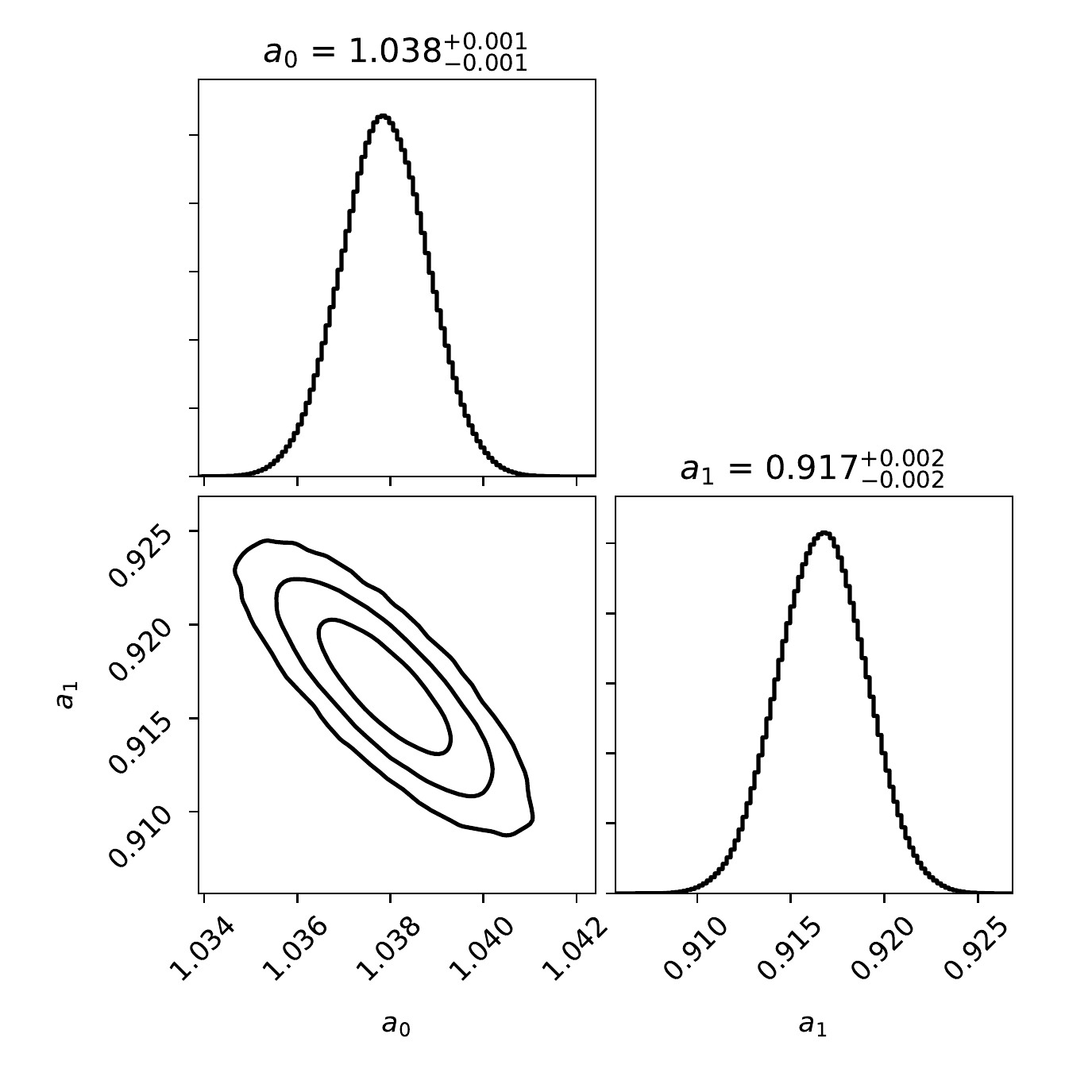}
 		\caption{Two-dimensional likelihood contours and one-dimensional likelihoods for the two-parameter power-law function of eq. \eqref{eq:pow} fit to the complete data binned in 52 bins, $\rm AelB_{52}M_3$.} 
 		\label{figure:c4}
 	\end{center}
 	\end{figure}
    \begin{figure}
    \begin{center}
 		\centering
 		\includegraphics[height=80mm,width=80mm]{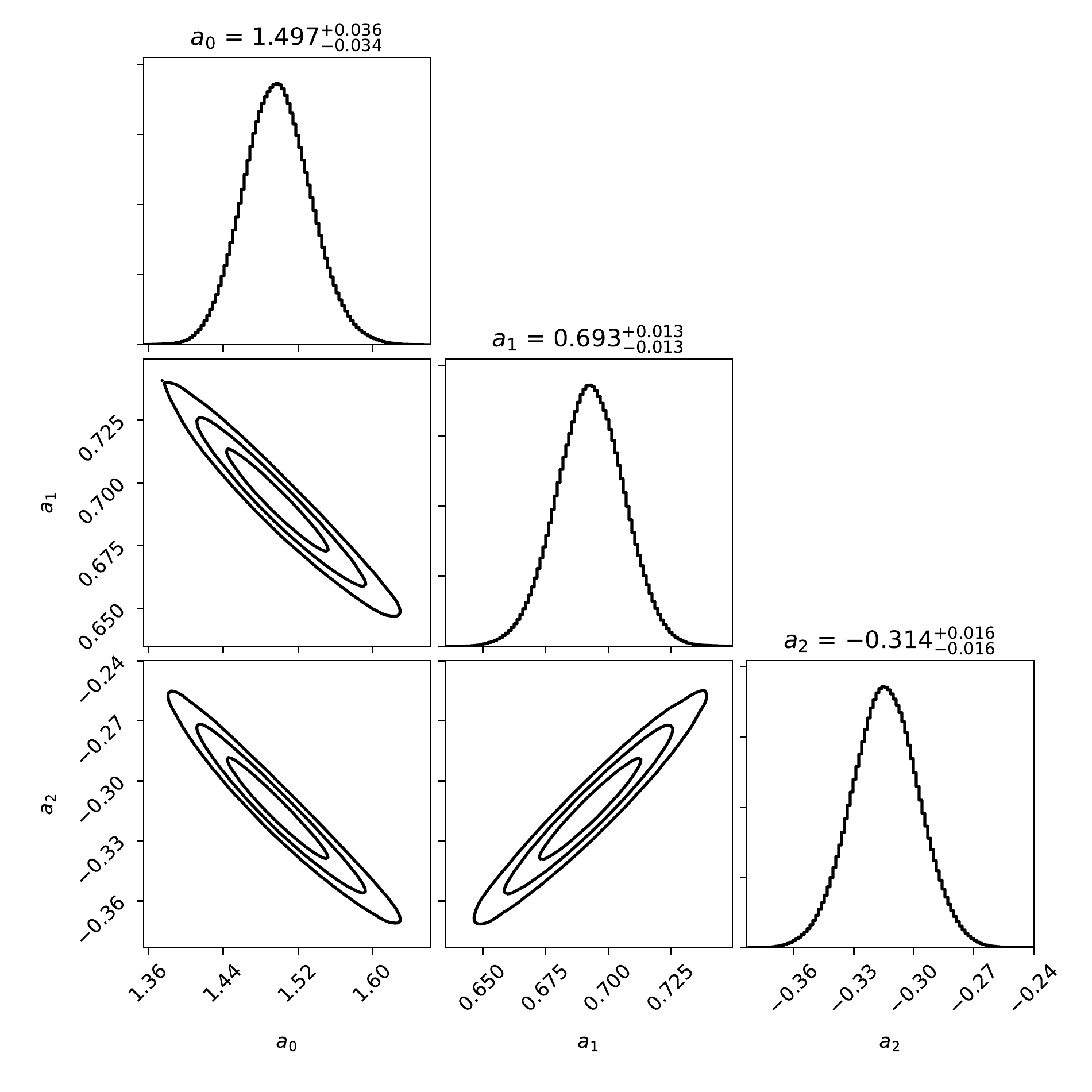}
 		\caption{Two-dimensional likelihood contours and one-dimensional likelihoods for the three-parameter power-law function of eq. \eqref{eq:power3} fit to the complete data binned in 52 bins, $\rm AelB_{52}M_4$.} 
 		\label{figure:c5}
    \end{center}
 	\end{figure}
    \setlength\tabcolsep{4pt}
    In order to determine what model independent conclusions can be drawn from these data, we fit to a variety of different data and functional form combinations. In what follows, we introduce our labelling conventions for these combinations.
    
    $\rm M_1$ through $\rm M_5$ refer to the linear, polynomial, two-parameter power law, three-parameter power law, and exponential functions described in eqs. \eqref{eq:lin_fit}, \eqref{eq:poly}, and \eqref{eq:pow}-\eqref{eq:exp} above. $\rm B_{52} \text{ and } B_{33}$ refer to the median statistics measurements for the data binned in 52 and 33 bins. P ($\theta^+$) and N ($\theta^-$) refer to positive ($0\degree\leq\theta<180\degree$, with 1,641 measurements) and negative azimuthal angle ($180\degree\leq\theta<360\degree$, with 1,065 measurements) data (we don't tabulate best-fit parameter values for the P and N data subsets). A refers to the complete data set of 2,706 measurements. e and E refer to the cases without and with the $1/3\ R$ error exchange, while l and L refer to the standard $\chi^2$, eq. \eqref{eq:chi2}, and the modified $\chi^2$ with $\sigma_{\rm int}$ as an additional parameter to be determined from the fit, eq. \eqref{eq:mod_chi2}. For example, a case with name AELB$_{33}$M$_1$ means that we bin the complete data (A) into 33 bins ($\rm B_{33}$) with $1/3\ R$ error exchange (E) and use the modified $\chi^2$ (L) to fit the median angular circular velocities of each bin to the linear model ($\rm M_1$).
    
    Figure \ref{fig:linear_ff_fit} shows the five angular circular velocity functions that best fit the complete data set binned in 52 bins. Since $\omega(R)$ drops rapidly as $R$ increases, the $y$ axis extends over a large range of $\omega$ values, resulting in relatively small error bars on the measurements shown in this plot. Figure \ref{figure:fits1} shows the same five models, but now we plot the linear circular velocity $v(R)$ which does not change as significantly as a function of $R$. This figure clearly shows that the simple functional forms we assume don't adequately capture the spatial substructure in the velocity data. It is interesting that the best-fit functions at $R>6\rm\ kpc$ range from a mildly rising linear function to a fairly steeply dropping exponential function. All five best-fit functions are reasonably mutually consistent in the $3\text{ kpc}\lesssim R\lesssim 9$ kpc range. However, even over this range of $R$ there are discernible differences between the five rotation curves. Consequently there will be significant differences between velocity residuals derived using each of these five rotation curves, making conclusions drawn from velocity residuals very sensitive to the rotation curve model assumed.
    
    Figures \ref{figure:fits2} through \ref{figure:fits5} show some of the various combinations of models and data we study. Table \ref{table:fitPars} lists best-fit parameter values and $1\sigma$ error bars as well as $\chi^2_\nu$ values for forty different cases (we don't list these values for the P and N data subsets). This table shows that when we do not include the additional error $\sigma_{\rm int}$, the reduced $\chi^2_\nu$ values are very large. Figures \ref{figure:fits1}-\ref{figure:fits5} clearly show that the simple functional forms do not adequately capture the spatial substructure seen in the velocity data. This is reinforced by the fact that Table \ref{table:fitPars} shows that when we do not included the additional error $\sigma_{\rm int}$, the reduced $\chi^2$'s are lower for the three-parameter models compared to the two-parameter cases.
    
    Allowing for the additional uncertainty $\sigma_{\rm int}$ results in shifts in the best-fit parameter values and curves -- see Table \ref{table:fitPars} and Fig. \ref{figure:fits2}. Figure \ref{figure:fits2} also shows that the best-fit curves do not drastically change when $\sigma_{\rm int}$ is introduced as an additional fitting parameter, compared to the case when it is not included. Figure \ref{figure:fits3} and Table \ref{table:fitPars} show that changes in the binning do not significantly alter the curves and the best-fit parameter values. Table \ref{table:fitPars} and Fig. \ref{figure:fits4} show that the $1/3\ R$ exchange does not significantly alter the results: since we have binned the data in $R$ it is relatively safe to ignore the $R$ uncertainties.
    
    Figures \ref{figure:c1} through \ref{figure:c5} show two-dimensional likelihood contours and one-dimensional likelihoods for the parameters of the five functional forms of \S\ref{subsec:ff}, for the complete data binned in 52 bins; these plots show that all the adjustable parameters we use are reasonably well constrained. The corresponding $\omega(R)$ and $v(R)$ best-fit functions are shown in Figs. \ref{fig:linear_ff_fit} and \ref{figure:fits1}.
    
    We emphasize that none of the models we discuss here adequately fit the data, as they do not capture the substructure, and some don't even capture the general shape of the rotation curve beyond some range.
    
	\section{Gaussian Processes}\label{sec:GaussianProcessing}
	
	In the previous section we found that the median statistics binned circular velocity data has significant substructure as a function of $R$. It is not clear how much of this is physical and how much, if any, of it is caused by unaccounted-for systematic errors or correlations. In the previous section we saw that simple, few parameter, functions of $R$ are not able to capture this substructure and so provide a bad fit to the circular velocity data.
	
	In this section we use the Gaussian Processes (GP) method to determine the rotation curve from the median statistics binned circular velocity data. The GP method is better able to capture the spatial substructure in measurements. It determines a continuous function, here $\omega (R)$, that best represents discrete data, here $\omega_i\pm \sigma_i$ at $R_i$, the binned median statistics angular circular velocity data. The GP method was first used cosmologically by \cite{Holsclaw2010a,Holscalw2010b,Holsclaw2011}, \cite{Shafieloo2012}, and \cite{Seikel2012JCAP...06..036S,Seikel2012PhRvD..86h3001S}. Here we present a first application of the GP method to the study of the Milky Way rotation curve.
	
	The GP method median statistics Milky Way rotation curves we derive have significant small-scale spatial structure superimposed on a broad rise to $R \approx 7$ kpc and a decline at larger $R$. We believe most of this small-scale spatial structure is physical and reinforces a revised picture of the Milky Way rotation curve.
	
	\subsection{Gaussian Processes method}\label{ssec:gpintro}
	\begin{figure}
    	\centering
        \begin{center}
        \includegraphics[width=84mm]{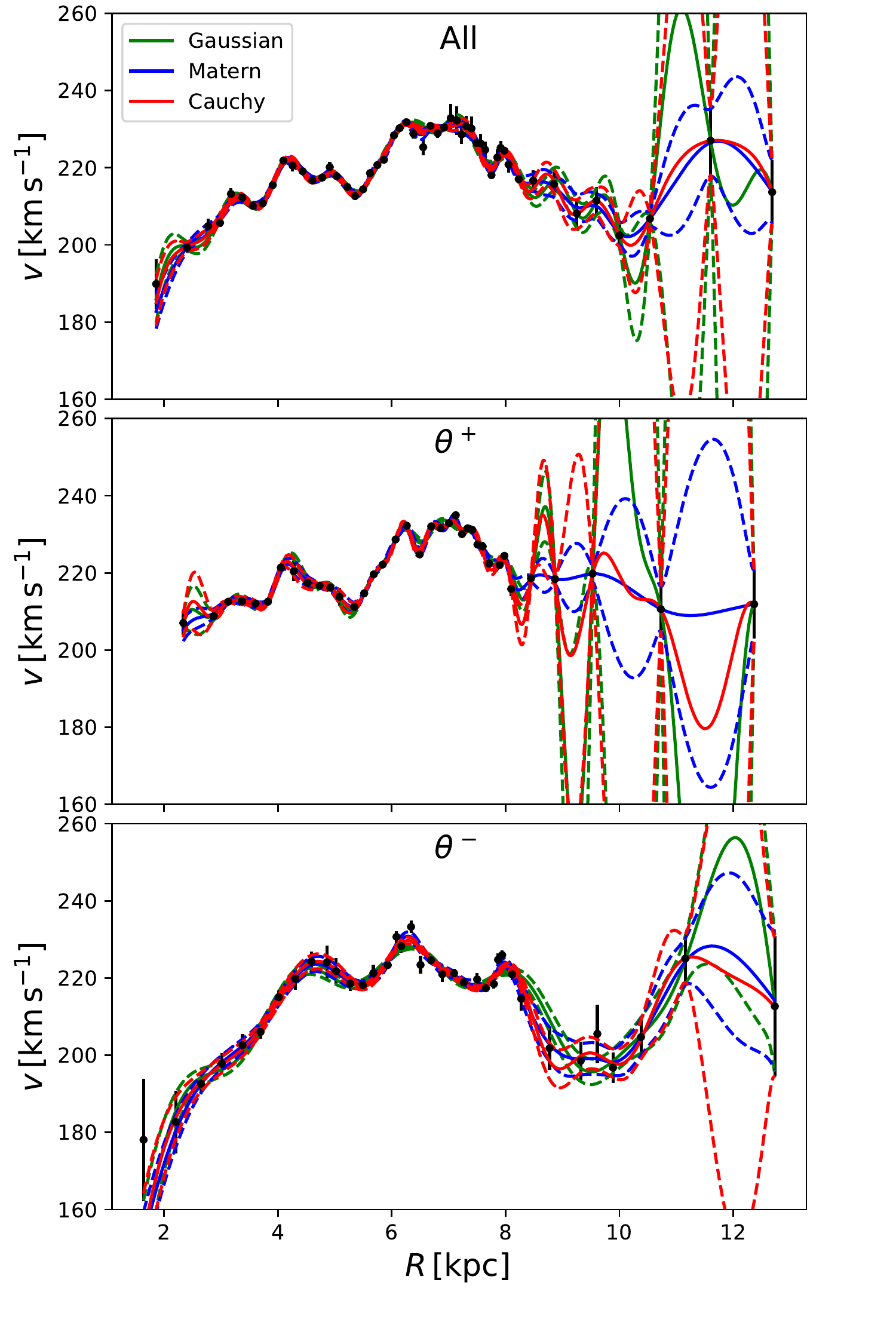}
        \end{center}
    	\caption{GP method $v(R)$ fits using gaussian, Mat\'ern, and Cauchy covariance functions. $v(R)$ central values (solid lines) and $\pm1\sigma$ limits (dashed lines) are shown. Top panel shows results for the complete data set (52 bins), middle panel shows results for the positive azimuthal sector (37 bins), and bottom panel shows results for the negative azimuthal sector (37 bins).} 
    	\label{figure:gp1}
	\end{figure}
    \begin{figure}
		\centering
        \begin{center}
		\includegraphics[width=84mm]{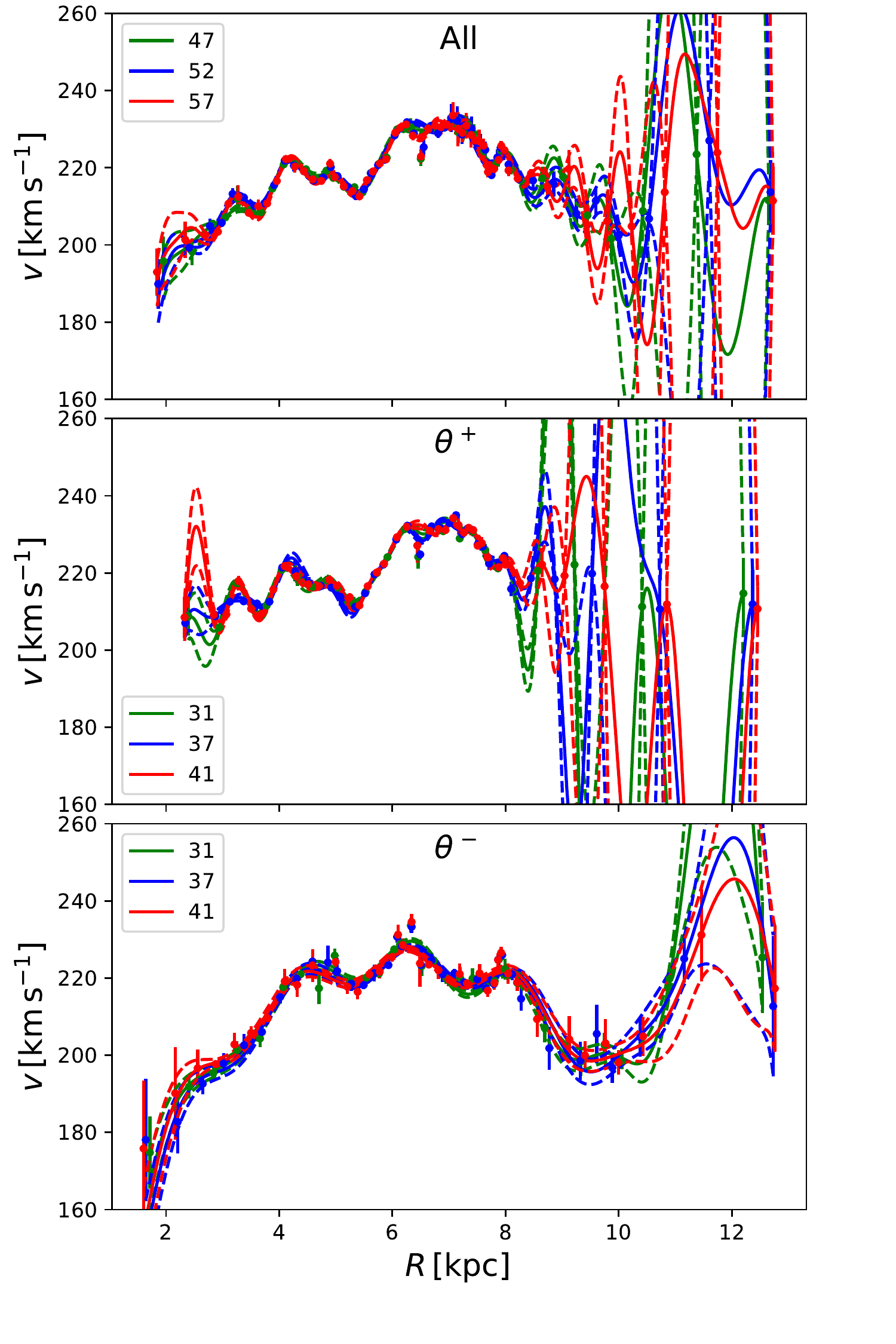}
        \end{center}
		\caption{Top, middle and bottom panels show GP method $v(R)$ fits for the complete data set, the positive azimuthal sector, and the negative azimuthal sector respectively, for variable number of bins, as indicated in the left top linestyle legend, for the gaussian covariance function. $v(R)$ central values (solid lines) and $\pm1\sigma$ limits (dashed lines) are shown. Over the range of $R$ where the rotation curve is well determined, altering the binning results in small but non-zero changes to the rotation curve.} 
		\label{figure:gp2}
    \end{figure}
    \begin{figure}
		\centering
		\begin{center}
		\includegraphics[width=84mm]{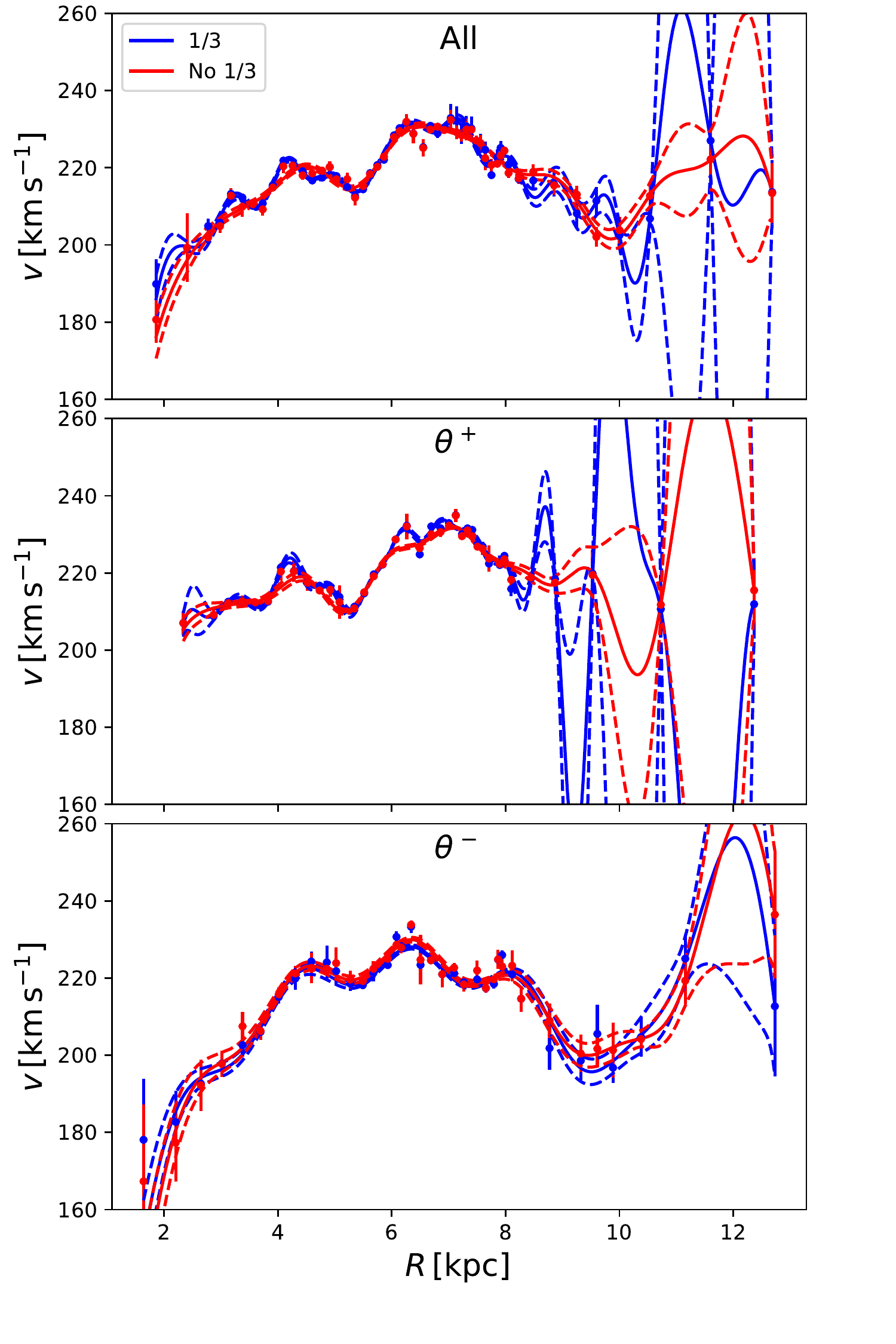}
        \end{center}
		\caption{Top, middle and bottom panels show GP method $v(R)$ fits for the complete data (52 bins), the positive azimuthal sector, and the negative azimuthal sector (both 37 bins), with and without the $1/3\ R$ exchange, as indicated in the left top linestyle legend, for the gaussian covariance function. $v(R)$ central values (solid lines) and $\pm1\sigma$ limits (dashed lines) are shown.} 
		\label{figure:gp3}
    \end{figure}
    \begin{figure}
	    \begin{center}
		\centering
		\includegraphics[width=84mm]{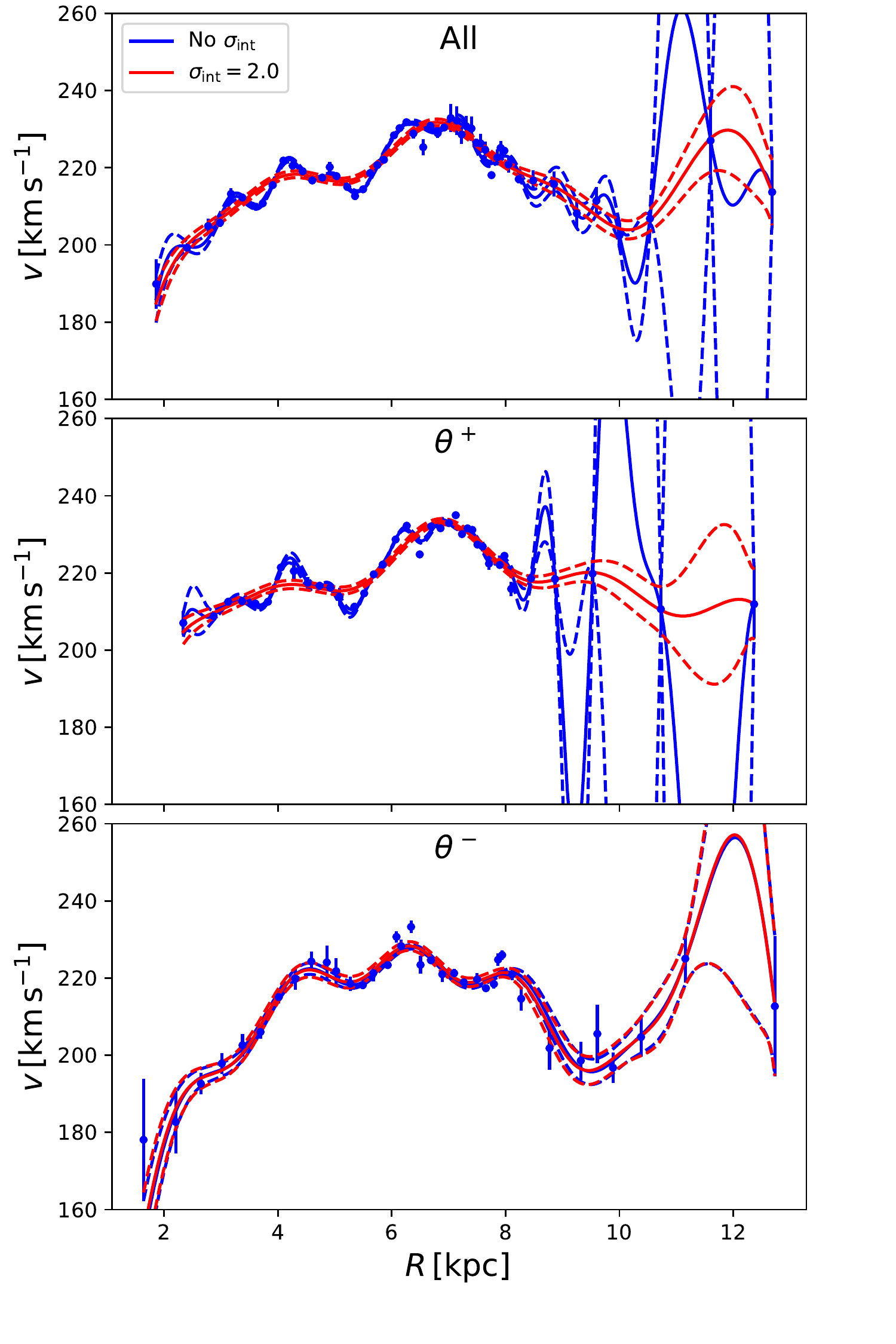}
	    \end{center}
    	\caption{Top, middle and bottom panels show the GP method $v(R)$ fits for the complete data set (52 bins), the positive azimuthal sector, and the negative azimuthal sector (both 37 bins), for $\sigma_{\rm int} = 0$ and $\sigma_{\rm int} = 2.0\rm\ km\ s^{-1}\ kpc^{-1}$, for the gaussian covariance function. $v(R)$ central values (solid lines) and $\pm1\sigma$ limits (dashed lines) are shown.}
		\label{figure:gp4}
	\end{figure}
    \begin{figure}
        \centering
    	\begin{center}
		\includegraphics[width=84mm]{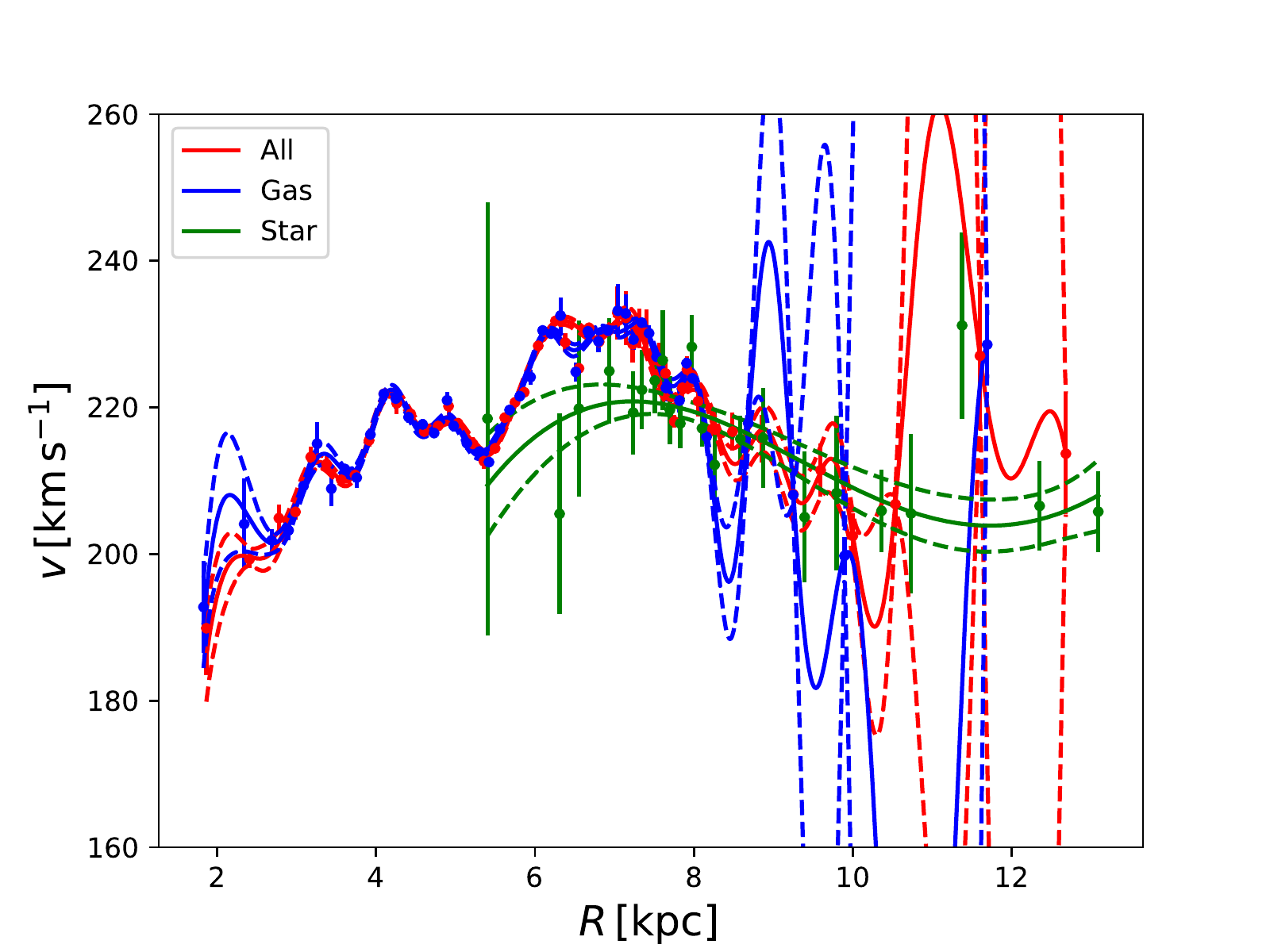}
        \end{center}
    	\caption{GP method $v(R)$ fits for the complete data set as well as for the gas and star tracers data subsets, for the gaussian covariance function. $v(R)$ central values (solid lines) and $\pm1\sigma$ limits (dashed lines) are shown.} 
    	\label{figure:gp5}
    \end{figure}
    \begin{figure*}
    	\centering
    	\begin{center}
    		\includegraphics[width=84mm]{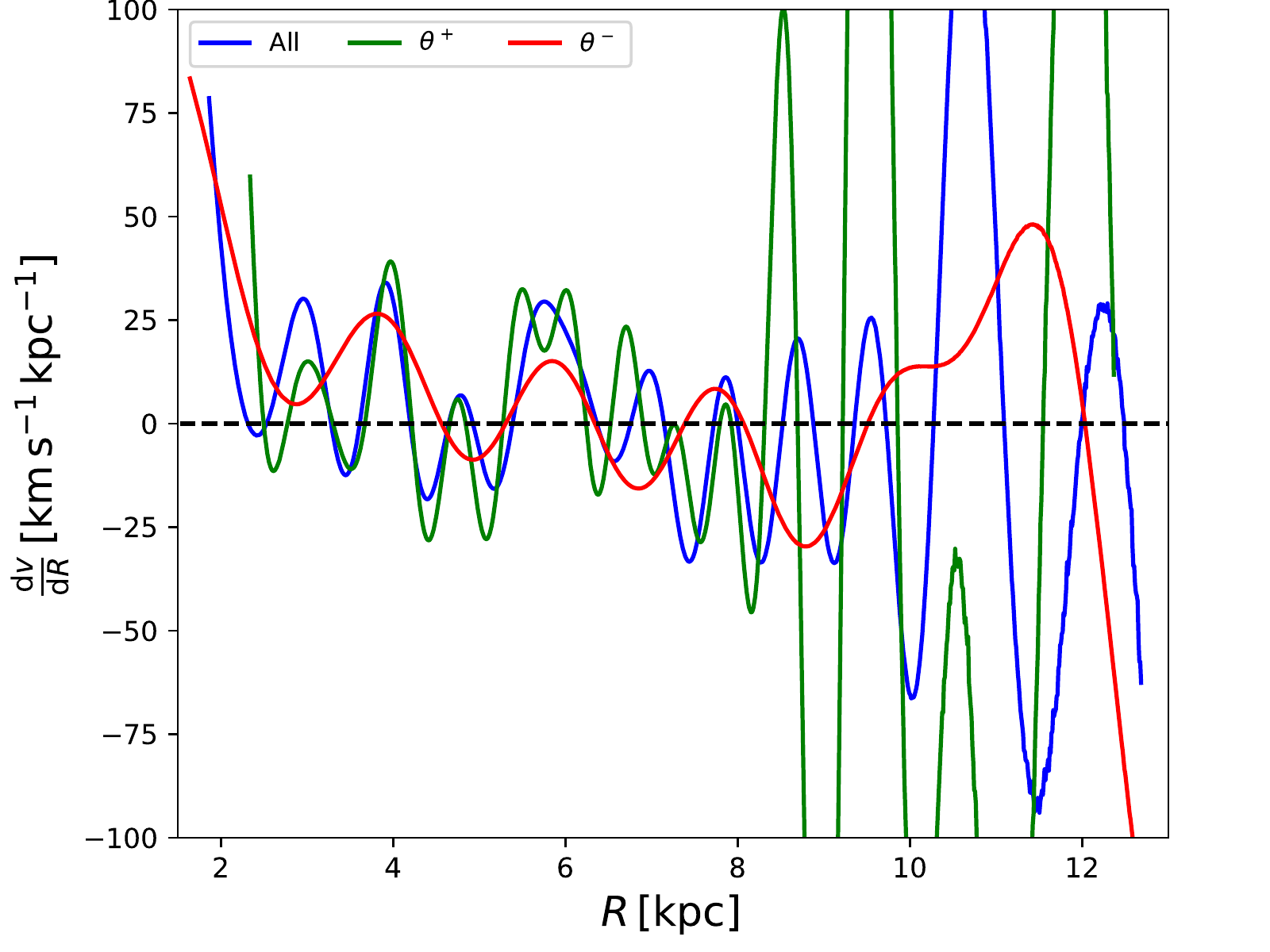}
    		\includegraphics[width=84mm]{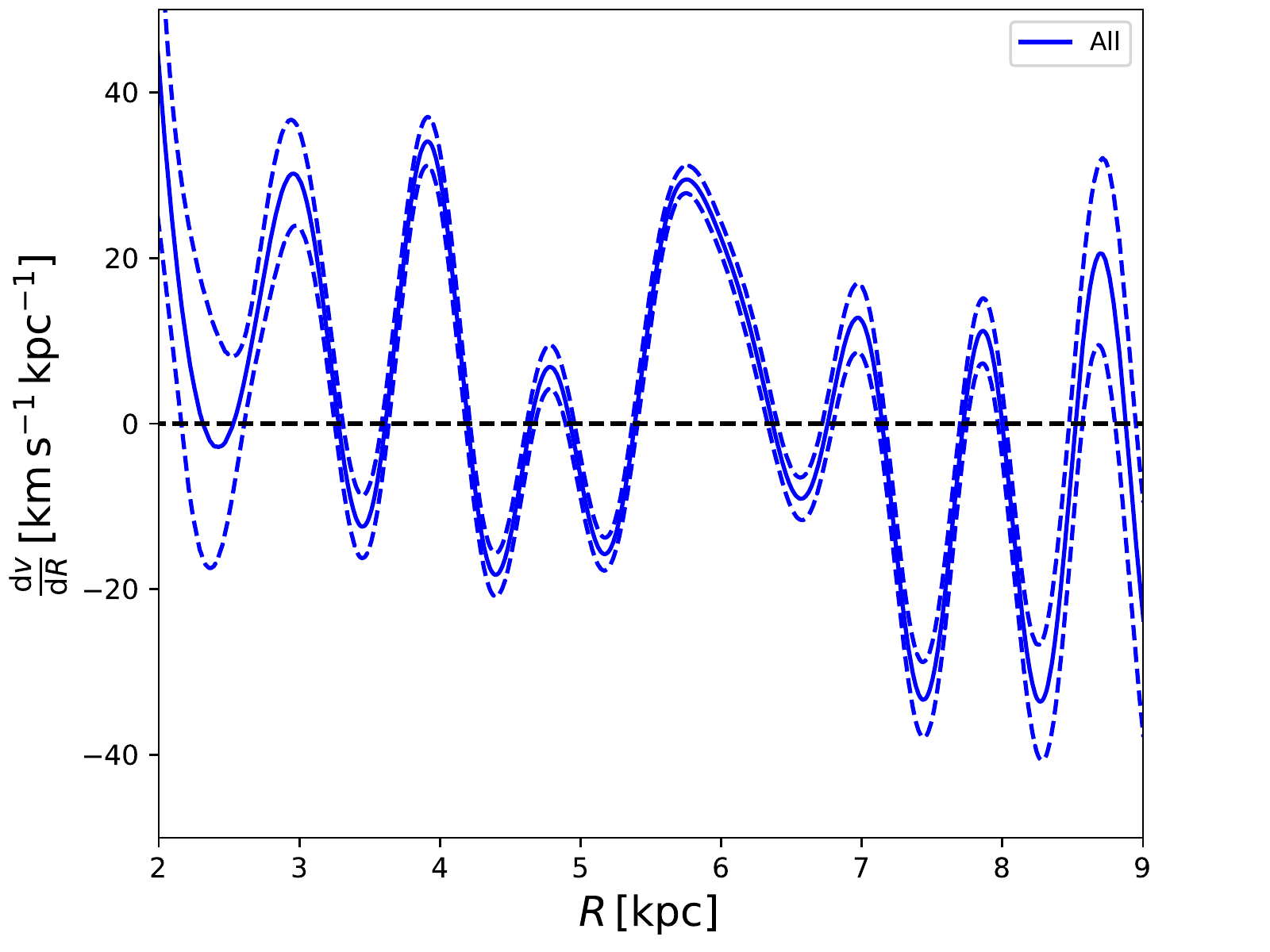}
    	\end{center}
    	\caption{The left panel shows the GP method circular velocity rotation curve slope $ dv/dR = \omega + Rd\omega/dR$ central value fits for the complete data set (in 52 bins), the positive azimuthal sector data, and the negative azimuthal sector data (both in 37 bins), as indicated in the left top linestyle legend, for the gaussian covariance function. The panel on the right shows the $dv/dR$ central value (solid line) and $\pm1\sigma$ uncertainty range (dashed line) for the complete data set.} 
    	\label{figure:dVdR}
    \end{figure*}
    \begin{figure*}
    	\centering
    	\begin{center}
    		\includegraphics[width=84mm]{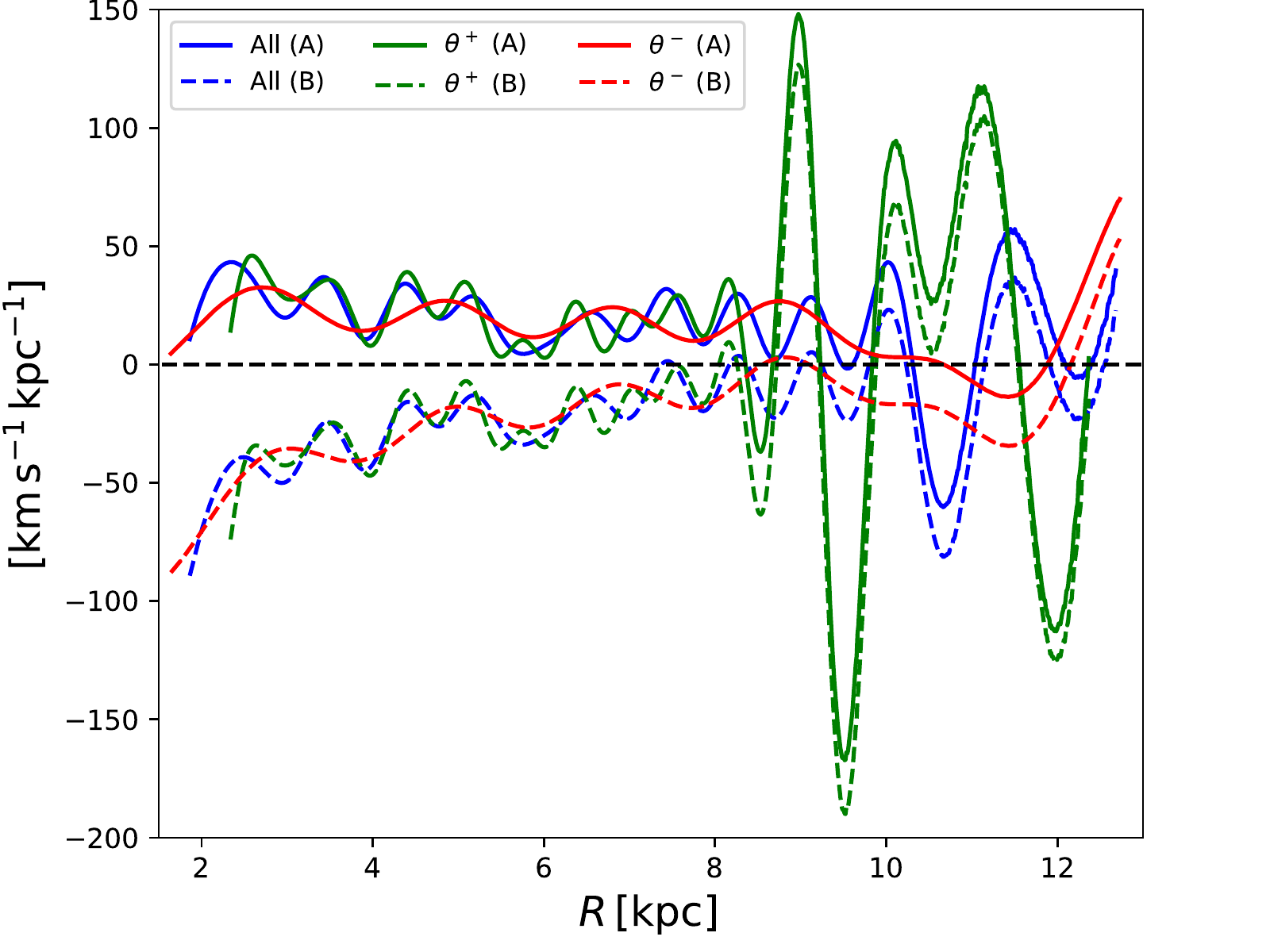}
    		\includegraphics[width=84mm]{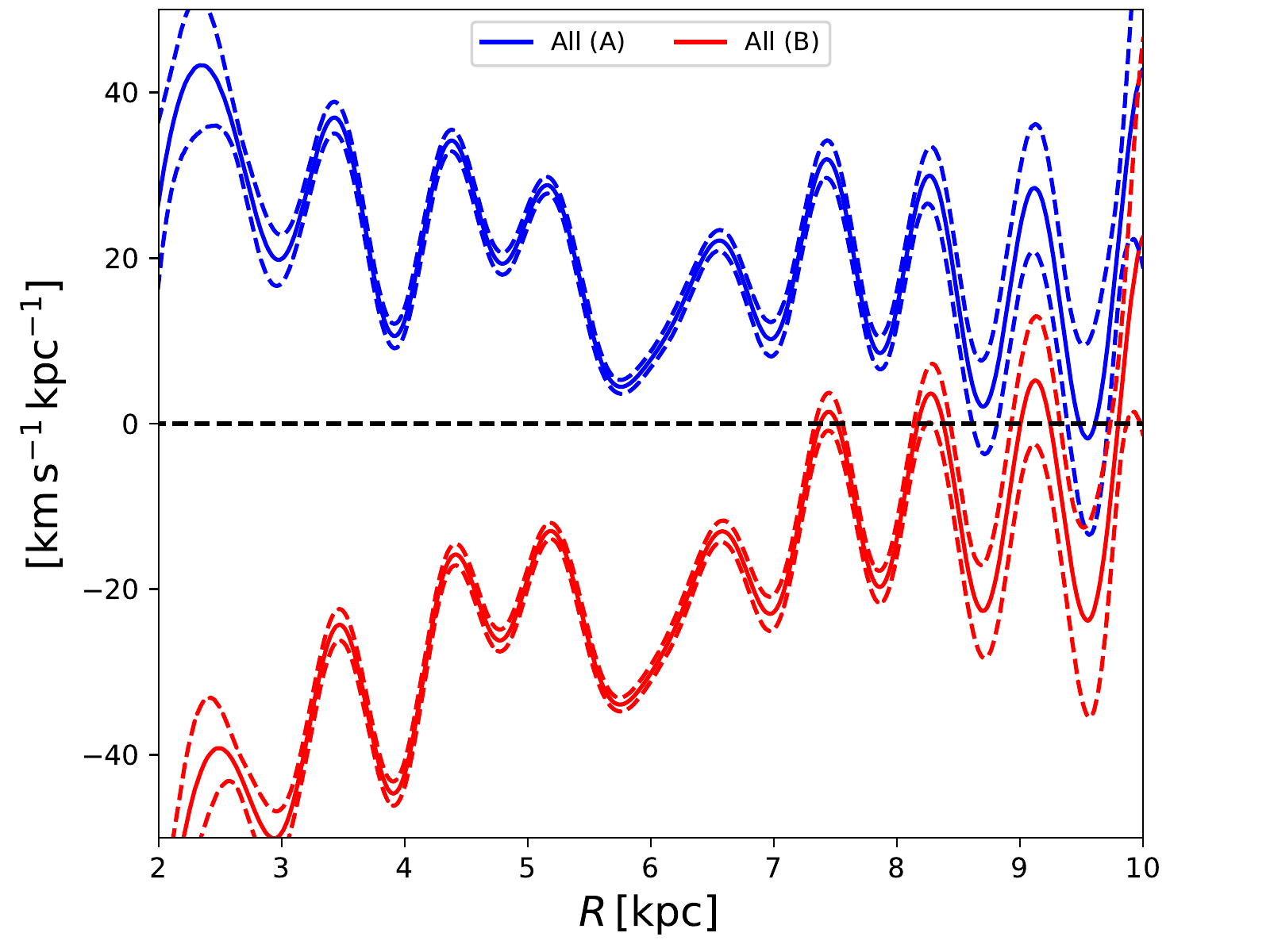}
    	\end{center}
    	\caption{The left panel shows the Oort functions, eq. \eqref{eq:Oort}, central values derived from the GP method fits for the complete data set (in 52 bins), the positive azimuthal sector data, and the negative azimuthal sector data (both in 37 bins), for the gaussian covariance function. We list the Oort constants, which are the values of the Oort functions at $R=R_0$, in Table \ref{table:Oort}. The right panel shows the Oort functions central values (solid lines) and $\pm1\sigma$ uncertainties (dashed lines) for the complete data set.} 
    	\label{figure:Oort}
    \end{figure*}
    \begin{figure*}
        \centering
    	\begin{center}
    	    \includegraphics[width=84mm]{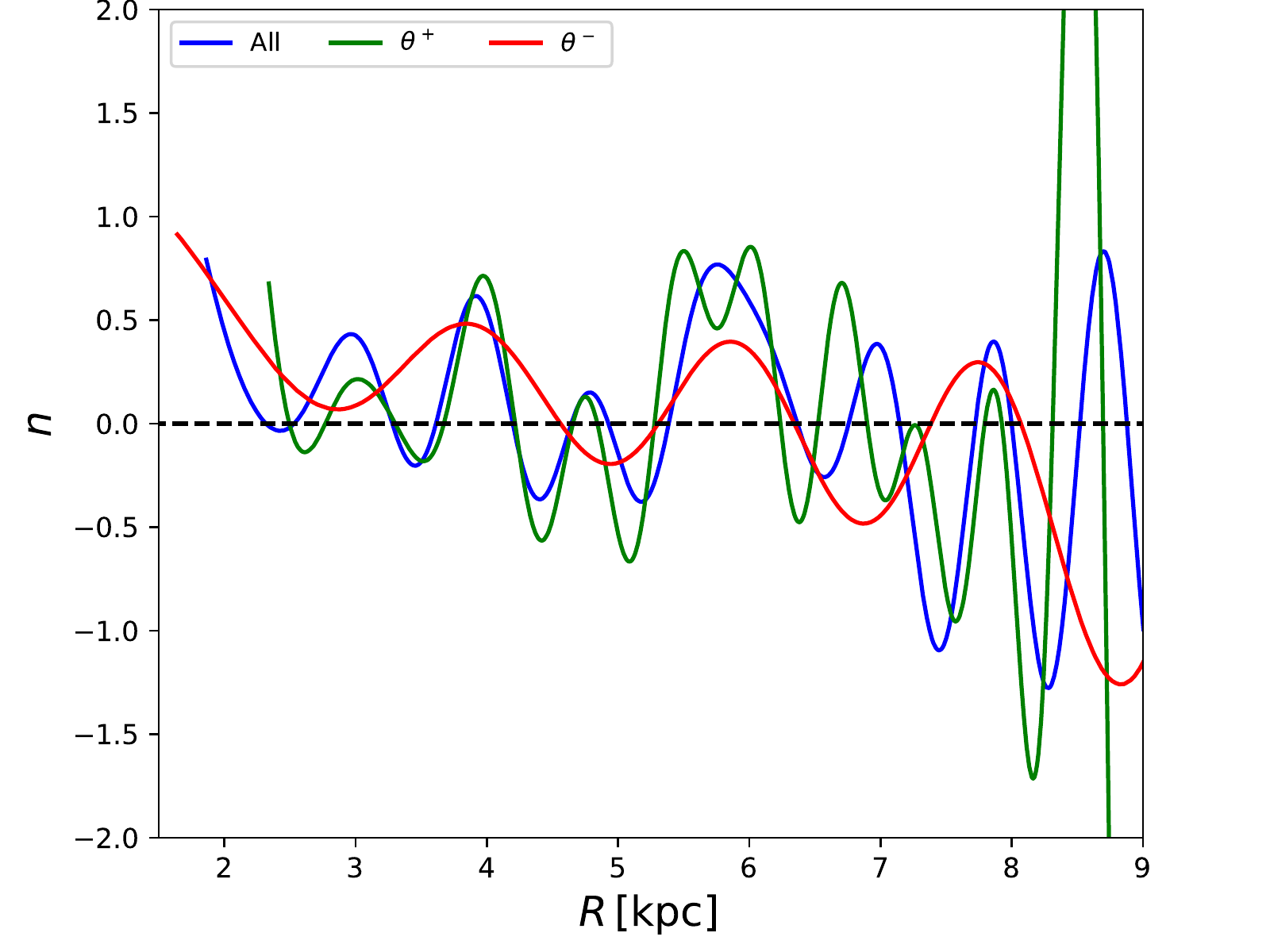}
    	    \includegraphics[width = 84mm]{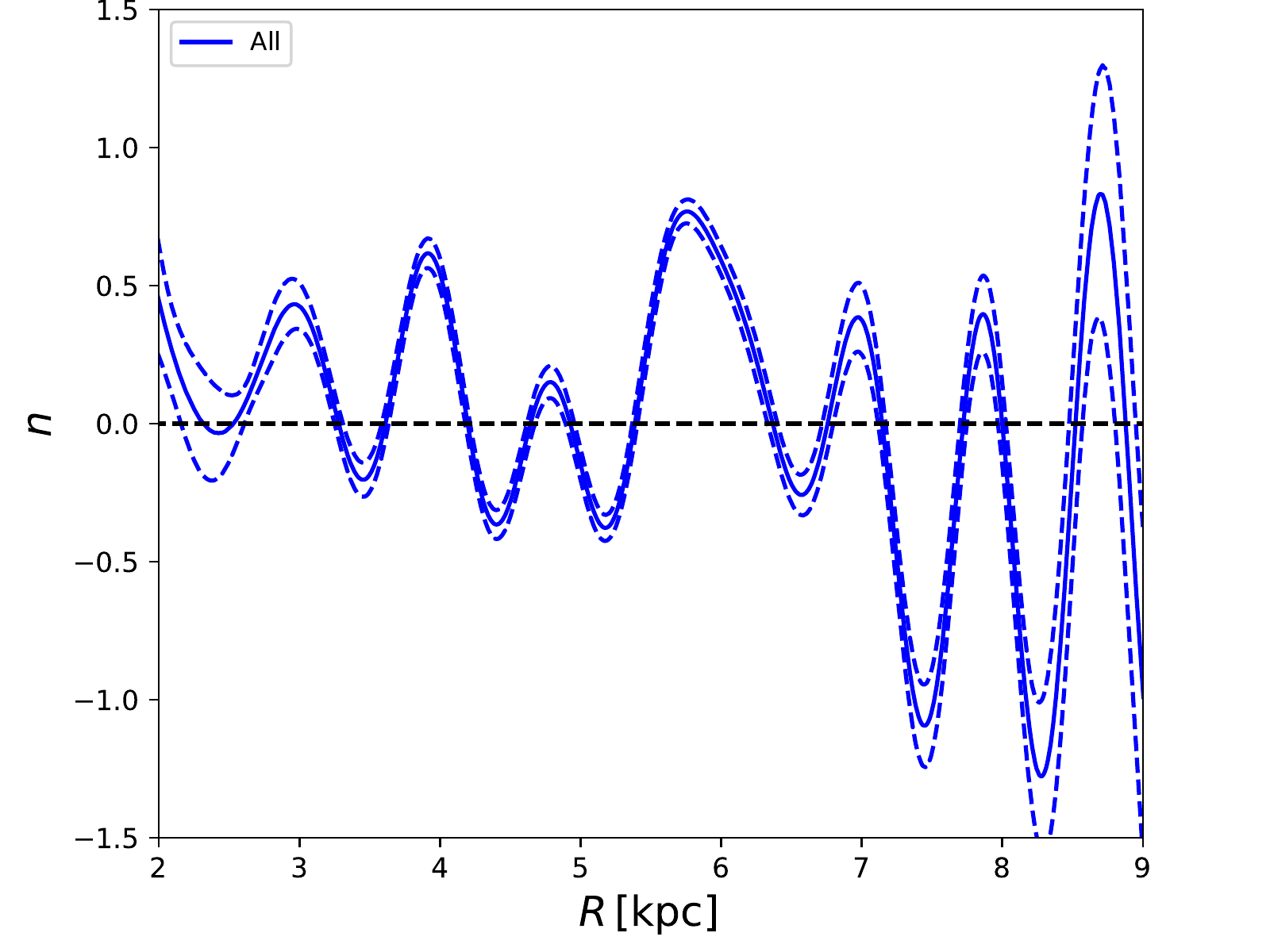}
    	\end{center}
    	\caption{The left panel show $n(R)$ central values computed using eq. \eqref{eq:Oort_n}, where $v(R)\propto R^n$ for the complete data set (in 52 bins), the positive azimuthal sector data, and the negative azimuthal sector data (both in 37 bins), for the gaussian covariance function. The right panel shows the $n(R)$ central value (solid line) and $\pm1\sigma$ uncertainties (dashed lines) for the complete data set. A flat rotation curve has $n= 0$ while $n = -1/2$ corresponds to Keplerian motion.} 
    	\label{fig:n(R)}
    \end{figure*}
    
	The GP method is a basic and useful regression method for determining a continuous function $f(x)$ that best represents a discrete set of measurements $f(x_i)\pm\sigma_i$ at $x_i$, where $i = 1,2,...,N$ and $\sigma_i$ are the 1$\sigma$ errors. A major advantage of the GP method is that it doesn't make an assumption about the form of the continuous function $f(x)$ and so is able to fit finer details in the discrete data than a fewer parameter simple function can.
	
	The GP method is based on the assumption that the value of the continuous function at position $x$ is a random variable that follows a gaussian distribution whose expectation and standard deviation, $\mu(x)$ and $\sigma(x)$, can be determined from the discrete data through a covariance (or kernel) function $k(x,x_i)$. The expected value and standard deviation are determined by
    \begin{equation}\label{eq:mu}
    \mu(x)=\sum_{i,j=1}^Nk(x,x_i)(M^{-1})_{ij}f(x_j),
    \end{equation}
    and
    \begin{equation}\label{eq:sigma}
    \sigma(x)=k(x,x)-\sum_{i,j=1}^Nk(x,x_i)(M^{-1})_{ij}k(x_j,x).
    \end{equation}
    Here the matrix $M_{ij}=k(x_i,x_j)+c_{ij}$ where $c_{ij}$ is the covariance matrix of the discrete measurements that are being fit. The matrix $c_{ij}$ is diagonal with elements $\sigma_i^2$ if the observed data are uncorrelated. Equations (\ref{eq:mu}) and (\ref{eq:sigma}) define the posterior distribution of the extrapolated points. The GP regression method has been widely used in cosmology \citep{Seikel2012JCAP...06..036S,Seikel2012PhRvD..86h3001S,Bilicki2012MNRAS.425.1664B,Cai2016PhRvD..93d3517C,Yu2016ApJ...828...85Y,Yu2017A&A...606A...3Y,Wang2017ApJ...847...45W,Yu2018ApJ...856....3Y}. Here we use it for the first time to determine the Milky Way rotation curve.
    
    Given a covariance function $k(x, x_i)$, eqns. \eqref{eq:mu} and \eqref{eq:sigma} determine the continuous function $f(x)$ from the discrete data. In practice, any covariance function that reasonably describes the relation among data will work. Here we consider the usual gaussian form
    \begin{equation}\label{eq:sqex}
    k(x,x_i)=\sigma_f^2\exp{\left[-\frac{(x-x_i)^2}{2l^2}\right]},
    \end{equation}
    where hyperparameters $\sigma_f$ and $l$ control the strength of the correlation of the function values and the coherence length of the correlation in $x$. The $\sigma_f$ and $l$ hyperparameters are optimized for the observed discrete measurements, $f(x_i)\pm\sigma_i$, by minimizing the log marginal likelihood \citep{Seikel2012JCAP...06..036S}
    \begin{equation}\label{likelihood}
    \begin{split}
        \ln\mathcal{L} = -\frac{1}{2}\sum_{i,j=1}^N[f(x_i)-\mu(x_i)]&(M^{-1})_{ij}[f(x_j)-\mu(x_j)]\\&-\frac{1}{2}
    \ln|M|-\frac{1}{2}N\ln{2\pi},
    \end{split}
    \end{equation}
    where $|M|$ is the determinant of the matrix $M_{ij}$. 
    
    Since it is a basic regression method, there are many GP method codes available. For our analyses here we use the Python based software \texttt{GaPP} \citep{Seikel2012JCAP...06..036S}.\footnote{For detailed information about \texttt{GaPP} see\\ \href{www.acgc.uct.ac.za/\~seikel/GAPP/Documentation/Documentation.html}{www.acgc.uct.ac.za/\textasciitilde seikel/GAPP/Documentation/Documentation.html}.}
    
    A major assumption of the GP method is the choice of the covariance (or kernel) function $k(x, x_i)$. It is important to see if the derived continuous function $f(x)$ depends significantly on this choice. To do this, in addition to the gaussian covariance function of eq. \eqref{eq:sqex}, we also consider Mat\'ern and Cauchy covariance functions,
    \begin{equation}\label{eq:Matern}
	k(x,x_i)=\sigma_f^2 \left[1+\frac{\sqrt{3}|x-x_i|}{l}\right]\exp{\left[-\frac{\sqrt{3}|x-x_i|}{l}\right]},
	\end{equation}
	and
	\begin{equation}\label{eq:Cauchyprocessing}
	k(x,x_i)=\sigma_f^2\frac{l}{(x-x_i)^2+l^2},
	\end{equation}
	and compare results derived by using each of these three covariance functions to determine which part of the results are insensitive to the choice of the covariance function.
	
	\subsection{Fitting results}\label{sec:FittingResults}
    
	\begin{table}
    	\caption{Oort constants and $dv/dR\ (R_0)$ determined from GP method fits using the gaussian covariance function.}
    	\label{table:Oort}
    	\centering
    	\begin{center}
    		\begin{tabular}{l@{\hspace{2\tabcolsep}}ccc}\hline\hline
    			{Data}&$A(R_0)$&$B(R_0)$&${\rm d}v/{\rm d}R\ (R_0)$\\\hline
				All	&$14.01\pm2.07$	&$-13.87\pm2.07$&$-0.14\pm4.14$\\
				N	&$21.81\pm5.00$	&$-6.06\pm5.00$	&$-15.75\pm10.00$\\
				S	&$12.31\pm1.42$&$-15.37\pm1.42$	&$3.06\pm2.83$\\\hline
    		\end{tabular}
    	\end{center}
    \end{table}
    \begin{table}
        \caption{Some Oort constants estimates since 2010.$^{\rm a}$}
        \centering
        \begin{tabular}{rrl}\hline\hline
            $A(R_0)^{\rm b}$&$B(R_0)^{\rm b}$&Reference\\\hline
            $17.8\pm0.8$&$-13.2\pm1.5$&\protect\cite{Bobylev2010a}\\
            $15.8\pm0.2$&$-10.9\pm0.2$&\protect\cite{Bobylev2010b}\\
            $17.9\pm0.5$&$-13.6\pm1.0$&\protect\cite{Bobylev2011}\\
            $17.1$&$-14.9$&\multirow{2}{*}{\cite{Klacka2012}}\\
            $15.0$&$-12.5$&\\
            $16.7\pm0.6$&$-12.0\pm1.0$&\protect\cite{Stepanishchev2013}\\
            $17.1\pm0.5$&$-11.6\pm0.8$&\protect\cite{Bobylev2016}\\
            $15.3 \pm 0.4$&$-11.9\pm0.4$&\protect\cite{Bovy2017}\\
            $16.5\pm0.5$&$-10.8\pm0.9$&\multirow{2}{*}{\protect\cite{Bobylev2017}}\\
            $17.8\pm0.5$&$-13.8\pm0.7$&\\
            $15.6\pm0.3$&$-10.7\pm0.5$&\protect\cite{Vityazev2017}\\
            $15.1\pm0.3$&$-12.2\pm0.4$&\protect\cite{Bobylev2018}\\\hline
            \multicolumn{3}{l}{$^{\rm a}$ These are not at the same $R_0$ and $v_0$. Two papers list values}\\
            \multicolumn{3}{l}{ from two different analyses. One paper does not provide error}\\
            \multicolumn{3}{l}{ bars.}\\
            \multicolumn{3}{l}{$^{\rm b}$ In units of $\rm km\ s^{-1}\ kpc^{-1}$.}\\
        \end{tabular}
        \label{tab:oort_external}
    \end{table}
    
    Again, as in the case for the $\chi^2$ fitting to simple functional forms in \S\ref{subsec:ffresults}, here we wish to determine what model independent conclusions can be drawn from the GP method fitting, so we derive rotation curves for a variety of data set and covariance function combinations.\footnote{The GP method assumes gaussianity of the discrete measurements and hence cannot be applied to the individual angular circular velocity measurements which are non-gaussian. Here we apply the GP method to the median statistics binned data.}
	
	Figure \ref{figure:gp1} shows the GP method linear circular velocity rotation curves derived using the gaussian, Mat\'ern, and Cauchy covariance functions, for the complete data set (52 bins, $\rm AeB_{52}$) and positive azimuthal and negative azimuthal data sets (37 bins, $\rm PeB_{37}$ and $\rm NeB_{37}$). There is no real difference between the rotation curves derived using the three different covariance functions for the $R$ ranges over which the GP method median statistics binned linear circular velocity is well determined with smaller error bars: $3\text{ kpc}\lesssim R\lesssim 8\text{ kpc}$ and perhaps even over $2\text{ kpc}\lesssim R\lesssim 9\text{ kpc}$ for the complete and negative azimuthal sector data sets. Beyond $R\gtrsim 8$ or 9 kpc, depending on the data set being analyzed, there are not enough data points to allow the GP method to reasonably approximate the rotation curve. The wild oscillations seen at larger $R$ are artifacts of overfitting due to this, and are not real. At smaller $R$ the GP method accurately captures the small-scale spatial structure in these data.
	
	Comparing the GP method rotation curves of Fig. \ref{figure:gp1} to those shown in \S\ref{subsec:ffresults} that were determined by fitting simple, few parameter, functions to the velocity data, we see that the GP method rotation curves capture much more of the small scale spatial information that is present in the binned velocity data. These panels also show that the complete data and the positive azimuthal sector data have more small scale spatial structure than do the negative azimuthal sector data.
	
	Figure \ref{figure:gp2} shows the effect on the GP method derived rotation curves resulting from changes in the median statistics binning. For the complete data set, the rotation curve in the range $3\text{ kpc}\lesssim R\lesssim 8\text{ kpc}$ is relatively insensitive to the binning. Figure \ref{figure:gp3} shows that this is also true for the $1/3\ R$ exchange, so the $R$ uncertainties cannot significantly affect the GP method derived rotation curves for the complete data set over the $3\text{ kpc}\lesssim R\lesssim 8\text{ kpc}$ range. 
	
	In \S \ref{subsec:ffresults}, we saw that the reduced $\chi^2$'s were quite large for the cases where we used the binned median statistics error bars. To reduce $\chi^2$ to unity, we had to introduce an additional error.\footnote{The large $\chi^2$'s are almost certainly more properly thought of as being mostly a consequence of the small scale spatial structure in the circular velocity field (that cannot be adequately described by the simple functions we used in \S\ref{subsec:ffresults}) rather than as being mostly a consequence of improperly estimated error bars.} We study this issue here by examining the consequence of introducing an additional overall error. Figure \ref{figure:gp4} shows the effects of adding an additional $\sigma_{\rm int} (R) = 2.0/R\rm\ km\ s^{-1}$ (large) error in quadrature to the median statistics binned angular circular velocity error bars prior to the GP method fitting procedure. This additional error results in a smoothing of the small scale spatial structure in the rotation curves. This effect is more prominent for the complete and positive azimuthal sector cases, which had more small scale spatial structure compared to the negative azimuthal sector data prior to the inclusion of $\sigma_{\rm int}$. 
	
	Figure \ref{figure:gp5} compares the rotation curves determined from the complete data set and the gas and star tracers data subsets. These rotation curves are largely mutually consistent, except, as discussed in \S\ref{subsubsec:tracertypes}, near $R\approx7$ kpc where the star tracers rotation curve lies a little below that for the gas tracers. This difference should be examined. In a more precise analysis it probably would be inappropriate to combine circular velocity data for star and gas tracers. However, this is probably not a significant issue for our less precise purposes here.
	
	Focusing on the complete data set presented in the top panels of Figs. \ref{figure:gp1} through \ref{figure:gp4}, we see that $v(R)$ increases from about $190\rm\ km\ s^{-1}$ at $R\approx2\rm \ kpc$ to about $230\rm\ km\ s^{-1}$ at $R\approx7\rm \ kpc$ and then declines to a little above $200\rm\ km\ s^{-1}$ at $R\approx10\rm \ kpc$. This trend is relatively independent of the binning used, the $1/3\ R$ error exchange, and the inclusion of a nonzero $\sigma_{\rm int}$. The Milky Way bar has a significant influence to perhaps $R\approx4$ kpc. Superposed on this overall rise and fall are finer scale spatial variations in $v(R)$.
	
	Figure \ref{figure:dVdR} shows the slope of the circular velocity rotation curve, $dv/dR$. We see, in the left panel, as is well-known, the complete data and the negative azimuthal sector data rotation curves are relatively flat over $3\text{ kpc}\lesssim R\lesssim 10$ kpc and the positive azimuthal sector curve is relatively flat over $3\text{ kpc}\lesssim R\lesssim 8$ kpc. Focusing more closely on the slope of the rotation curve for the complete data, shown in the right panel of Fig. \ref{figure:dVdR}, we see however that there are significant deviations from a flat rotation curve. We return to this point in the last paragraph of this subsection. The slopes at the solar distance $R = 8$ kpc are recorded in Table \ref{table:Oort}. The uncertainties plotted in Fig. \ref{figure:dVdR} and listed in Table \ref{table:Oort} were computed using Monte Carlo simulations with the covariance matrix obtained for the $d\omega/dR$ fits. We emphasize that the quantitative results depend on the median statistics binning used; we have checked that different binnings do not change the qualitative conclusions.
	
	The Oort constants $A(R_0)$ and $B(R_0)$ characterize the rotational properties of the Milky Way at the distance of the Sun from the Galactic center. For an azimuthally symmetric disk they characterize the rotational properties of the Milky Way at the position of the Sun. The Oort constants are evaluated from the Oort functions
	\begin{equation}\label{eq:Oort}
	    \begin{split}
	        A(R) &= \frac{1}{2}\left(\frac{v}{R} - \frac{dv}{dR}\right),\\
	        B(R) &= -\frac{1}{2}\left(\frac{v}{R} + \frac{dv}{dR}\right).
	    \end{split}
	\end{equation}
    We use the obtained GP method fits to the complete data set, and the positive azimuthal and negative azimuthal sector data subsets, to compute the respective Oort functions. The Oort functions are plotted in Fig. \ref{figure:Oort}, and the Oort constants are given in Table \ref{table:Oort}. The right panels of the figure shows the Oort functions central values for the complete data set, binned in 52 bins and computed using the gaussian covariance function, along with $1\sigma$ uncertainties. These uncertainties are computed using Monte Carlo simulations.
    
    Table \ref{tab:oort_external} lists some Oort constants values determined since 2010. These are from analyses that have different $R_0$ and $v_0$ values. Our estimates of $A(R_0)\text{ and }B(R_0)$ for the complete data set are mostly consistent with the values shown in this table. This is reassuring and indicates our median statistics GP method rotation curve is, at least qualitatively, a reasonable description of the Milky Way rotation curve. Our positive azimuthal sector data subset values are less inconsistent with the values in Table \ref{tab:oort_external}, but our negative azimuthal sector data subset values are mostly inconsistent with those of Table \ref{tab:oort_external}. A fair portion of the data is from measurements done before 2000. Therefore, it is worth comparing the Oort constant values with \textit{Hipparcos} astronomy measurements. \cite{Feast1997} find $A(R_0) = 14.82\pm0.84\text{ km s}^{-1}\text{ kpc}^{-1}\text{ and } B = -12.37\pm0.64\text{ km s}^{-1}\text{ kpc}^{-1}$, which is fairly consistent with our complete data and our negative azimuthal sector data derived Oort constants, but not our positive azimuthal sector Oort constants. The difference in Oort constant values for the two sectors is another indication that the circular velocity data are azimuthally anisotropic.
    
    Our Oort constant values are determined from our GP method fit Oort functions that are based on the median statistics binned data. The median statistics analyses ignore the error bars of the individual measurements and so are less constraining. In addition the GP method fit is to circular velocity measurements over a significant fraction of the Milky Way disk that are not necessarily azimuthally symmetric, which will introduce additional scatter. As a result our Oort constants error bars are mostly significantly larger than those shown in Table \ref{tab:oort_external}, which were determined largely from local measurements and make use of the error bars of the individual measurements.
    
	Assuming Newtonian gravity, the velocity of a tracer on a circular orbit of radius $R$ about the Galactic center depends on the mass interior to the orbit, $M(<R)$, through $v(R)\propto \sqrt{M(<R)/R}$. With $M(<R)\propto R^{2n + 1}$, we have $v(R)\propto R^{n}$, or 
	\begin{equation}\label{eq:Oort_n}
	    n(R) = \frac{R}{v}\frac{dv}{dR} = 1 + \frac{R}{\omega}\frac{d\omega}{dR}.
	\end{equation}
	Figure \ref{fig:n(R)} plots $n(R)$ for the complete data, and the positive azimuthal and negative azimuthal sector data subsets. For a rigid body, $n = 1$, Keplerian motion has $n = -1/2$, and $n=0$ corresponds to a flat rotation curve. Clearly, $n(R)$ shown in the right panel of Fig. \ref{fig:n(R)}, for the complete data set, does not bear much resemblance to $n = 0$ for a flat rotation curve. A similar conclusion can be drawn from the rotation curve slope plotted in the right panel of Fig. \ref{figure:dVdR}. The oscillatory structure in $R$ seen in the right panel of Fig. \ref{fig:n(R)} (as well as in the right panel of Fig. \ref{figure:dVdR}) is probably a consequence of the small-scale spatial variations in the circular velocity rotation curve, azimuthal anisotropy in the circular velocity field, and the inadequacy of the simple $v(R)\propto R^{n}$ model. It is also possible that the $1\sigma$ uncertainties shown in the right panels of Figs. \ref{figure:dVdR} and \ref{fig:n(R)} are underestimated, although the median statistics technique error bars are larger than the weighted mean ones, and the GP method fits are less constraining than model based ones \citep[e.g.,][]{Yu2018ApJ...856....3Y}.
	\begin{figure*}
		\begin{center}
			\includegraphics[width=\linewidth]{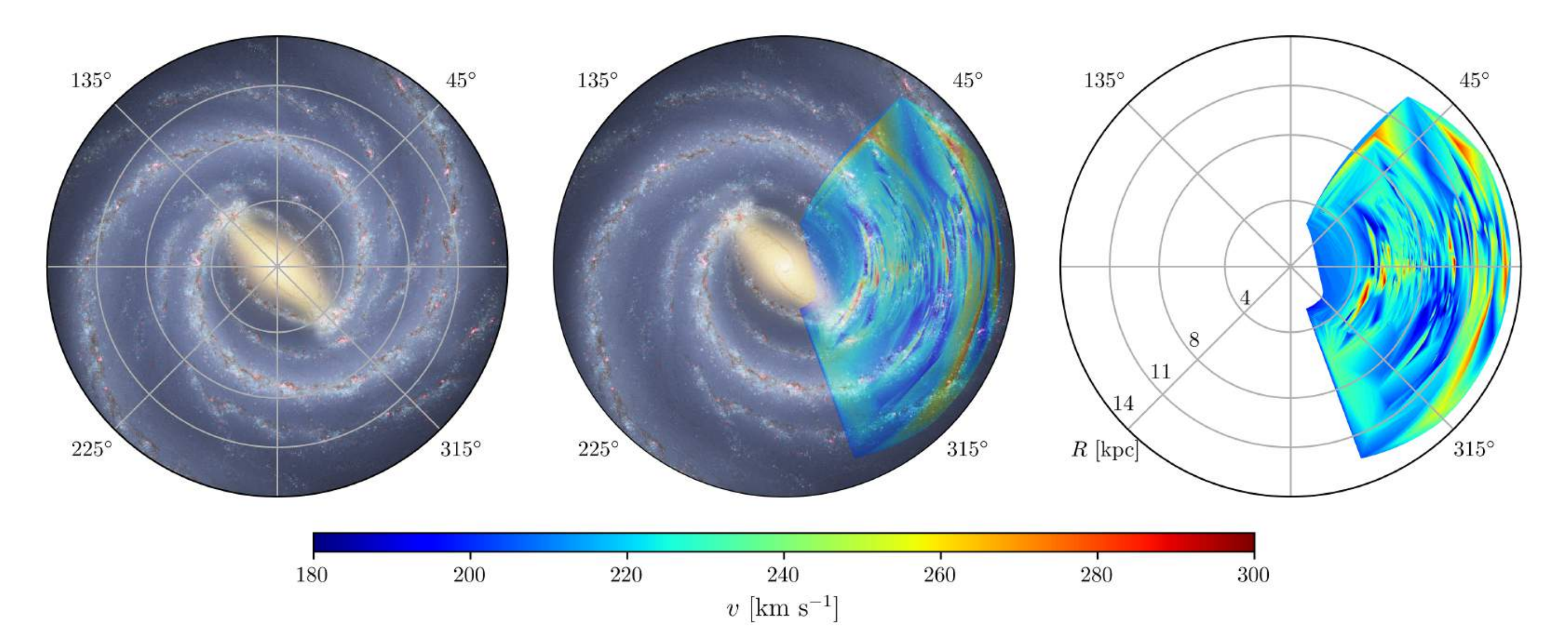}
		\end{center}
		\caption{The left panel shows an artist's rendition of the Milky Way galaxy [credits: NASA/JPL-Caltech/R. Hurt(SSC/Caltech)]. The image has been rotated and scaled to fit the location of the Sun and the Galactic center in our analyses ($R_0 = 8\rm\ kpc, \theta_{\rm Sun} = 0\degree$). The right panel shows circular velocity contours derived from the complete data limited to $290\degree \leq \theta\leq 70\degree$ (with 16 measurements with $v>300\rm\ km\ s^{-1}$ and 60 measurements with $v<180\rm\ km\ s^{-1}$ discarded to ensure that the circular velocity contour map color scale is not skewed). The middle panel is an overlap of the two figures.}
		\label{fig:spiral_arms}
	\end{figure*}
	\begin{figure}
		\begin{center}
			\centering
			\includegraphics[width=84mm]{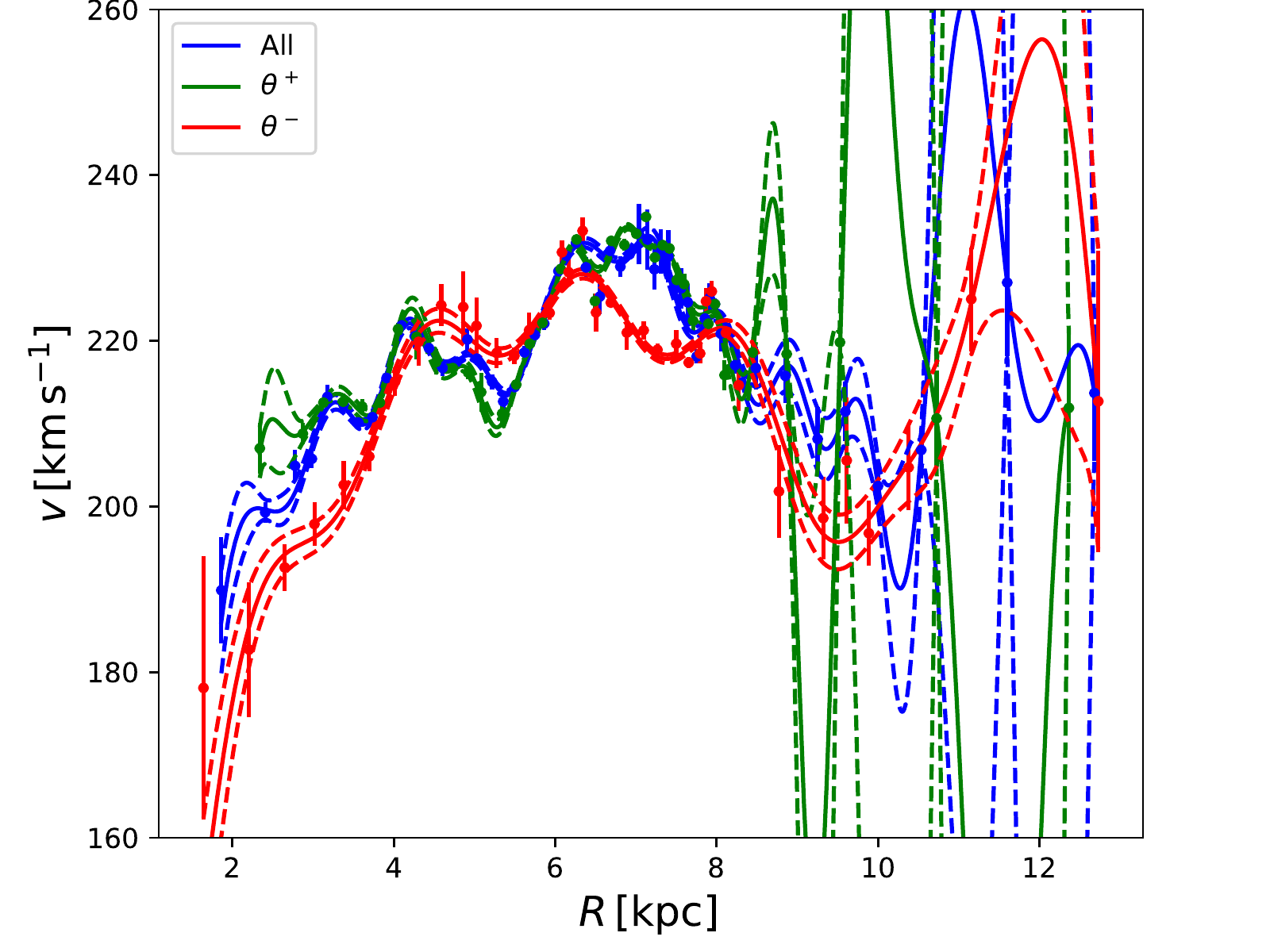}
		\end{center}
		\caption{The GP method $v(R)$ central values (solid lines) and $\pm1\sigma$ limits (dashed lines) for the complete data set (in 52 bins), the positive azimuthal sector data, and the negative azimuthal sector data (both in 37 bins) as indicated in the left top linestyle legend, for the gaussian covariance function.}
		\label{figure:gp6}
	\end{figure}
	\begin{figure}
		\centering
		\begin{center}
			\includegraphics[width=84mm]{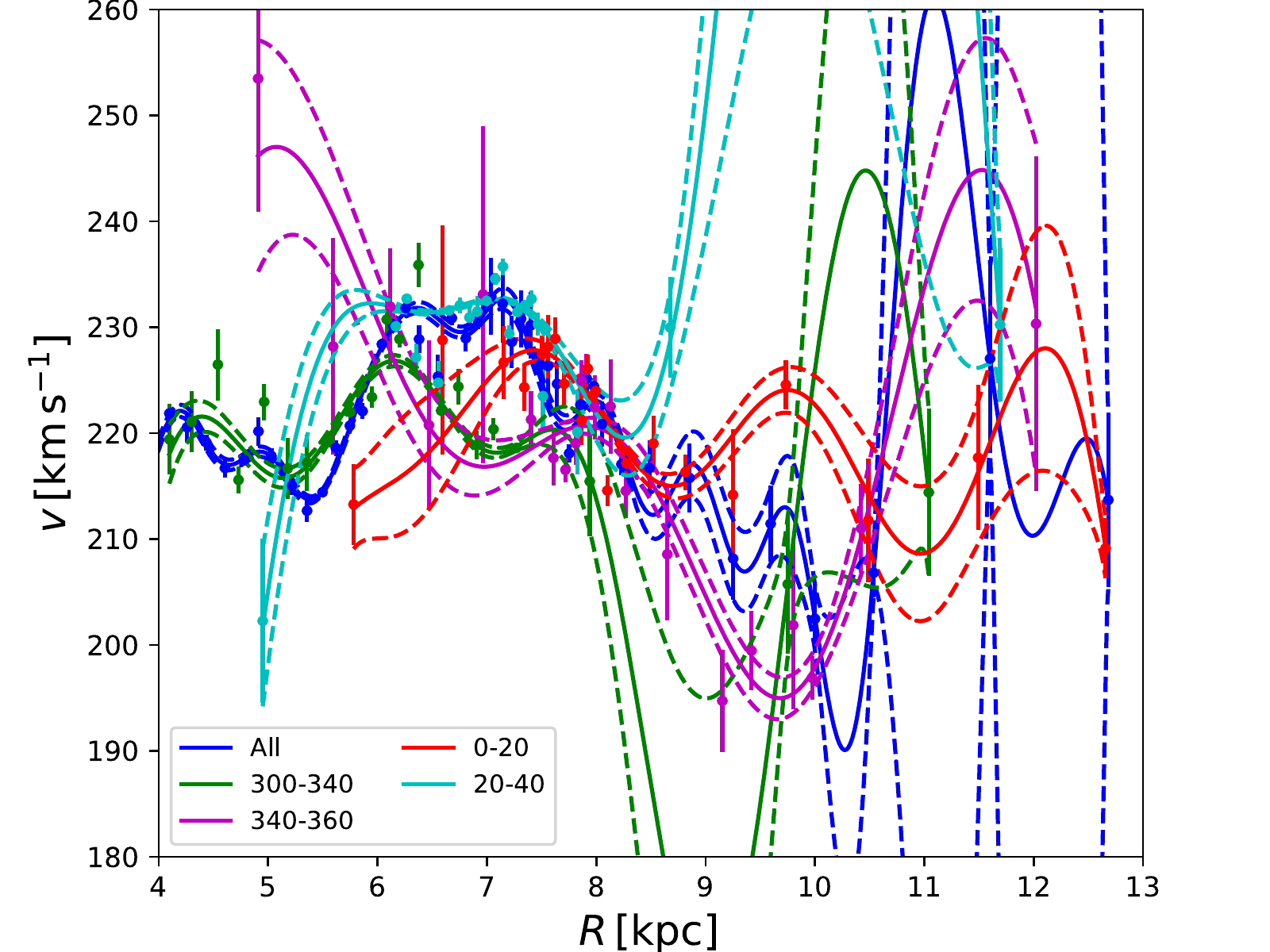}
		\end{center}
		\caption{The GP method $v(R)$ central values (solid lines) and $\pm1\sigma$ limits (dashed lines) for the complete data set (2706 measurements in 52 bins), and for data in four angular sectors (21 bins for each sector), 300\degree---340\degree\ (429 measurements), 340\degree---360\degree\ (432 measurements), 0\degree---20\degree\ (471 measurements), and 20\degree---40\degree\ (440 measurements), as indicated in the left bottom linestyle legends. These angular sectors do not contain a lot of measurements near the Galactic center, so we only show larger $R$($>4$ kpc) values in this figure. } \label{figure:4AngBin}
	\end{figure}
	
	\subsection{Azimuthal asymmetry}\label{subsec:azimuthal}
	\begin{figure*}
		\centering
		\begin{center}
			\includegraphics[width=84mm]{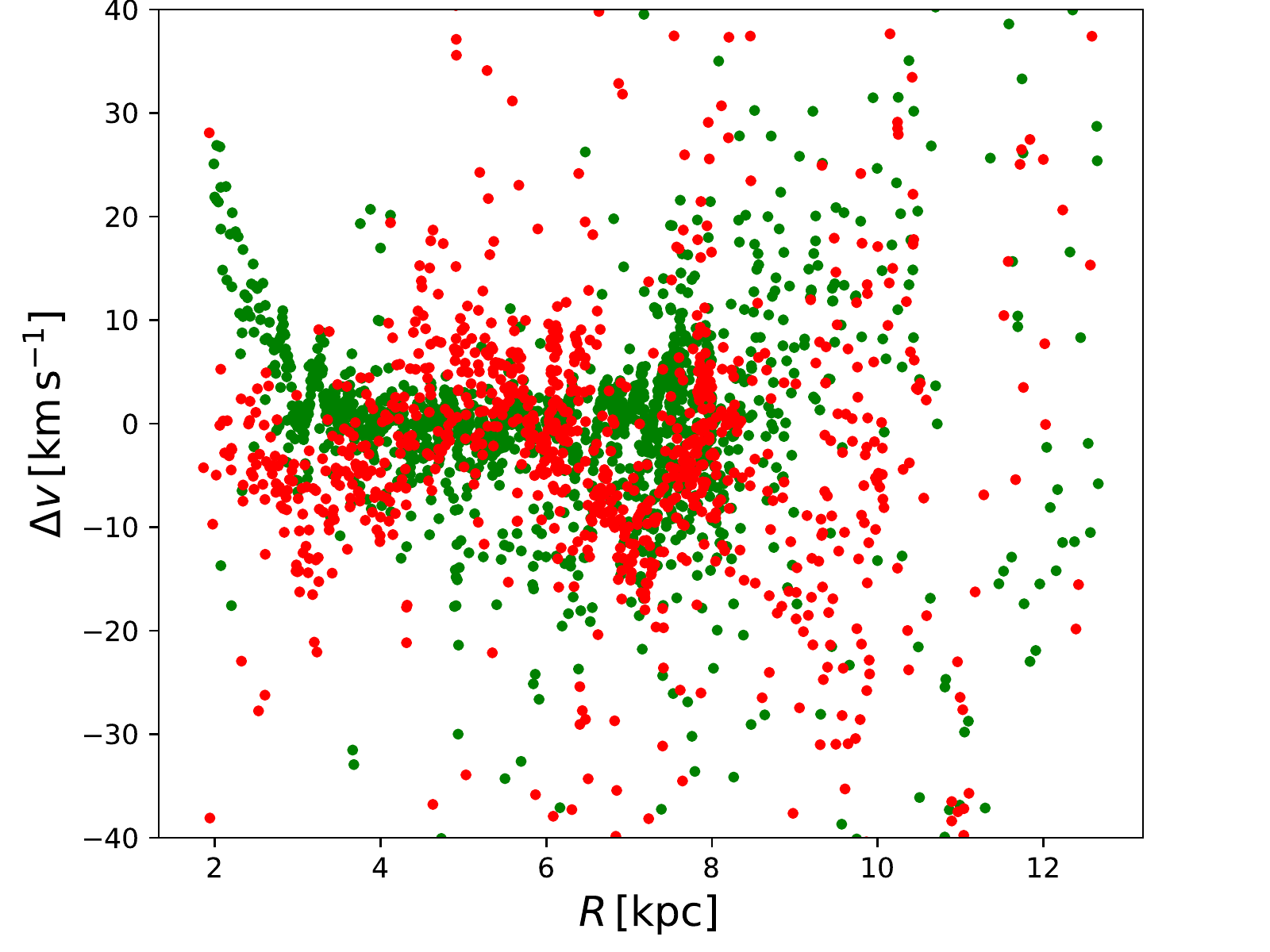}
			\includegraphics[width=84mm]{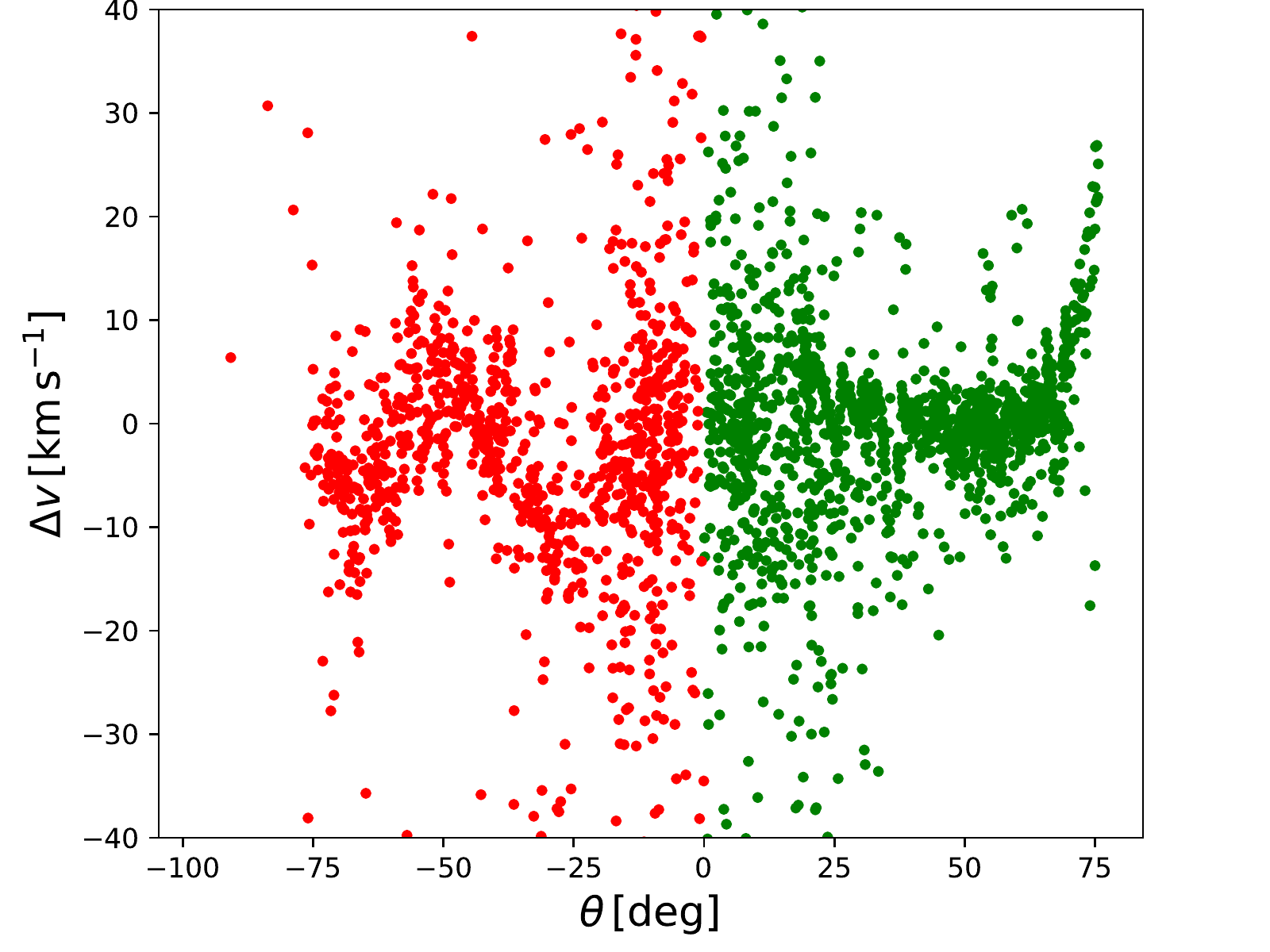}
		\end{center}
		\caption{Residual circular velocities relative to the GP method rotation curve derived from the complete data, binned in 52 bins, for the gaussian covariance function. For clarity we do not show error bars and some of the more deviant measurements. Green (red) points correspond to measurements in the positive (negative) azimuthal sector. We do not show 19 (36) measurements with $\Delta v > 40\rm\ km\ s^{-1}$ and 33 (39) measurements with $\Delta v < -40\rm\ km\ s^{-1}$ in the positive (negative) azimuthal. Left panel shows the residual velocities as a function of $R$ and the right panel shows them as a function of galactocentric longitude $\theta$.}
		\label{figure:res0}
	\end{figure*}
	\begin{figure}
	    \centering
	    \begin{center}
			\includegraphics[width=84mm]{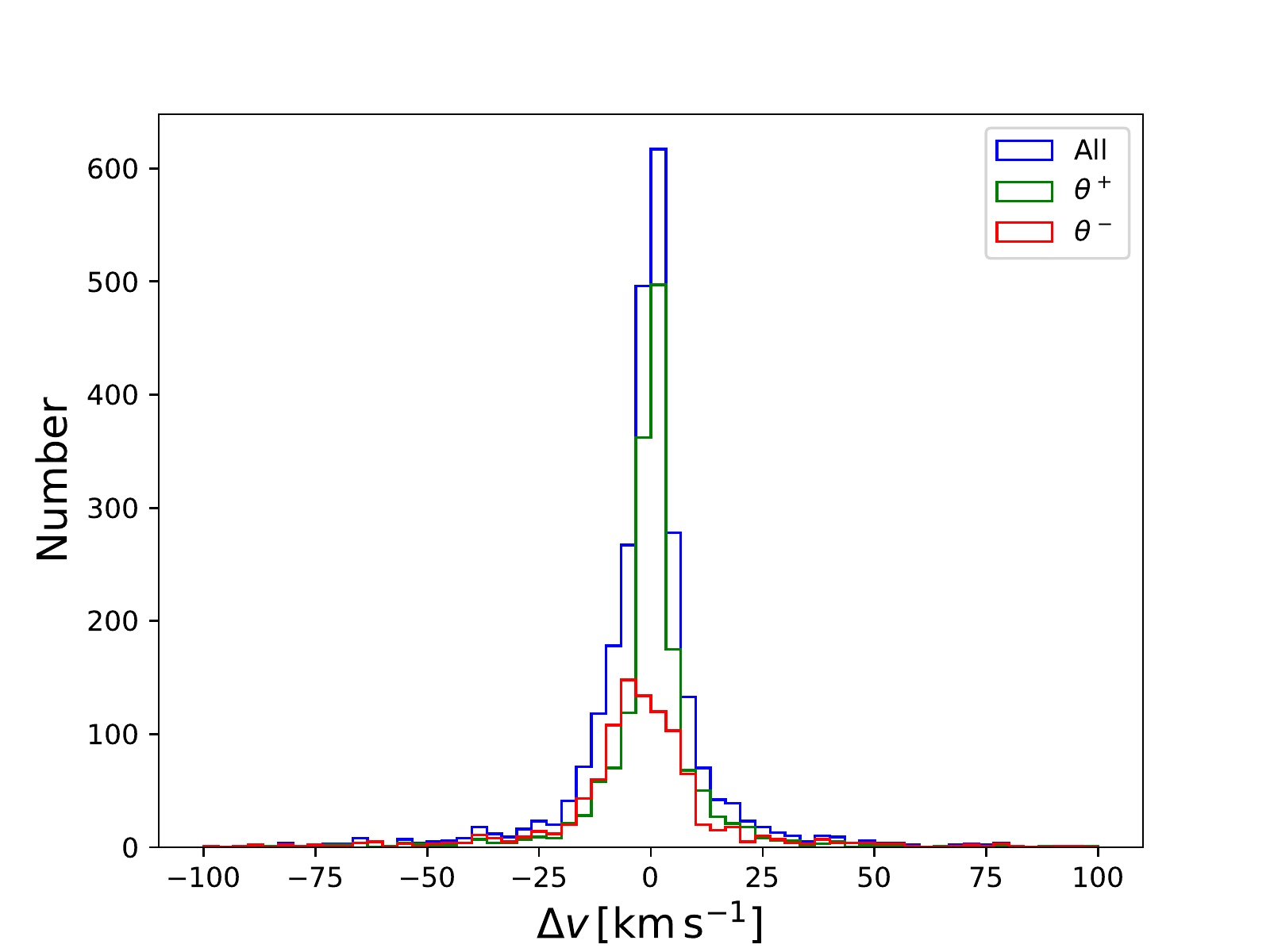}
	    \end{center}
	    \caption{Histograms of the residual velocities plotted in Fig. \ref{figure:res0}. The residual velocities are with respect to the GP method rotation curve derived from the complete data using the gaussian covariance function. To avoid compressing the histograms, we do not show 10 (9) velocity residues with $\Delta v < -100\rm\ km\ s^{-1}$ ($\Delta v > 100\rm\ km\ s^{-1}$). Of these 6 (2) correspond to the positive azimuthal sector data and 4 (7) correspond to the negative azimuthal sector data.}
	    \label{fig:res0_hist}
	\end{figure}
	\begin{figure*}
		\centering
		\begin{center}
			\includegraphics[width=\linewidth]{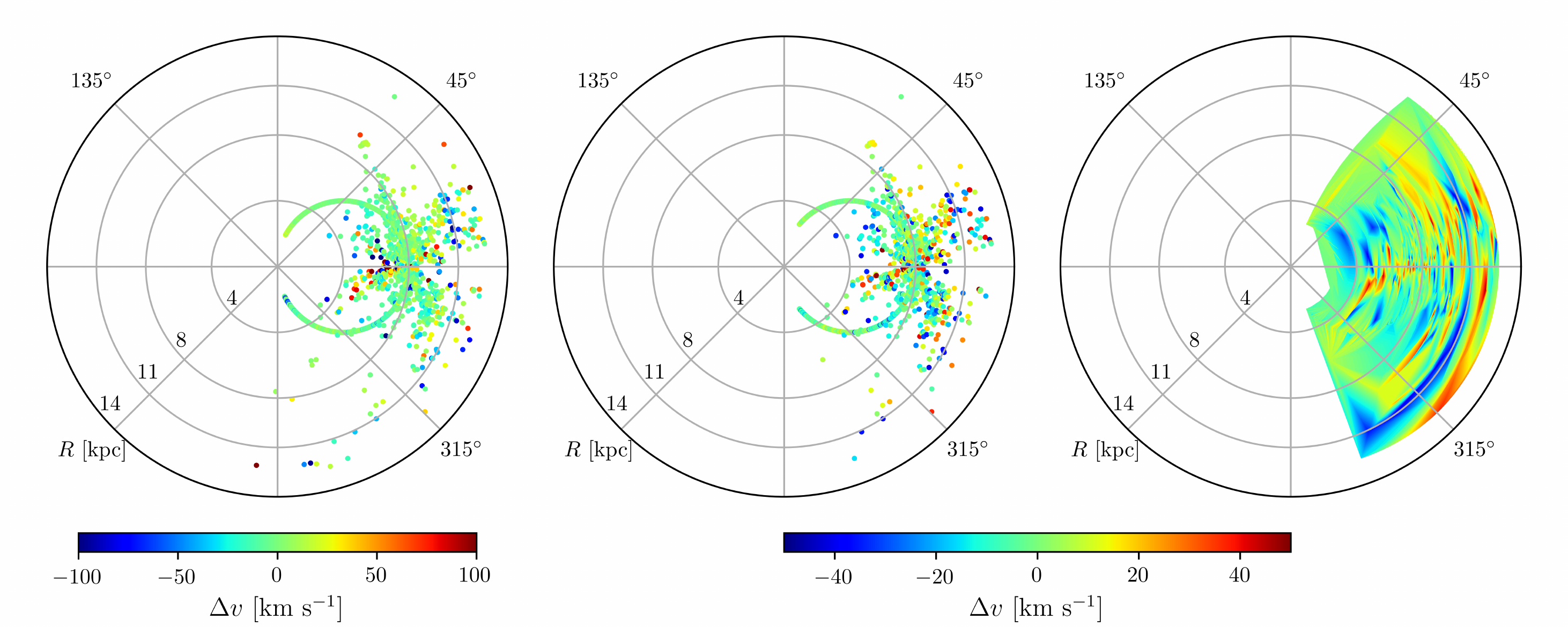}
		\end{center}
		\caption{The left panel shows the location and amplitude of all 2,654 velocity residues. To avoid skewing the color scale, we use dark blue (red) to show residual velocities $\Delta v \leq -100\rm\ km\ s^{-1}$ ($\Delta v \geq 100\rm\ km\ s^{-1}$). There are 10 (9) points with $\Delta v < -100\rm\ km\ s^{-1}$ ($\Delta v > 100\rm\ km\ s^{-1}$). The middle panel corresponds to the 2,467 measurements used to make the contour plot in the right panel and in Fig. \ref{fig:residuecontour}, restricted to $290\degree \leq \theta \leq 70\degree$ and $\left|\Delta v\right| \leq 50\rm\ km\ s^{-1}$.}
		\label{figure:res2_0}
	\end{figure*}
	\begin{figure*}
	    \centering
	    \begin{center}
	        \includegraphics[width=\linewidth]{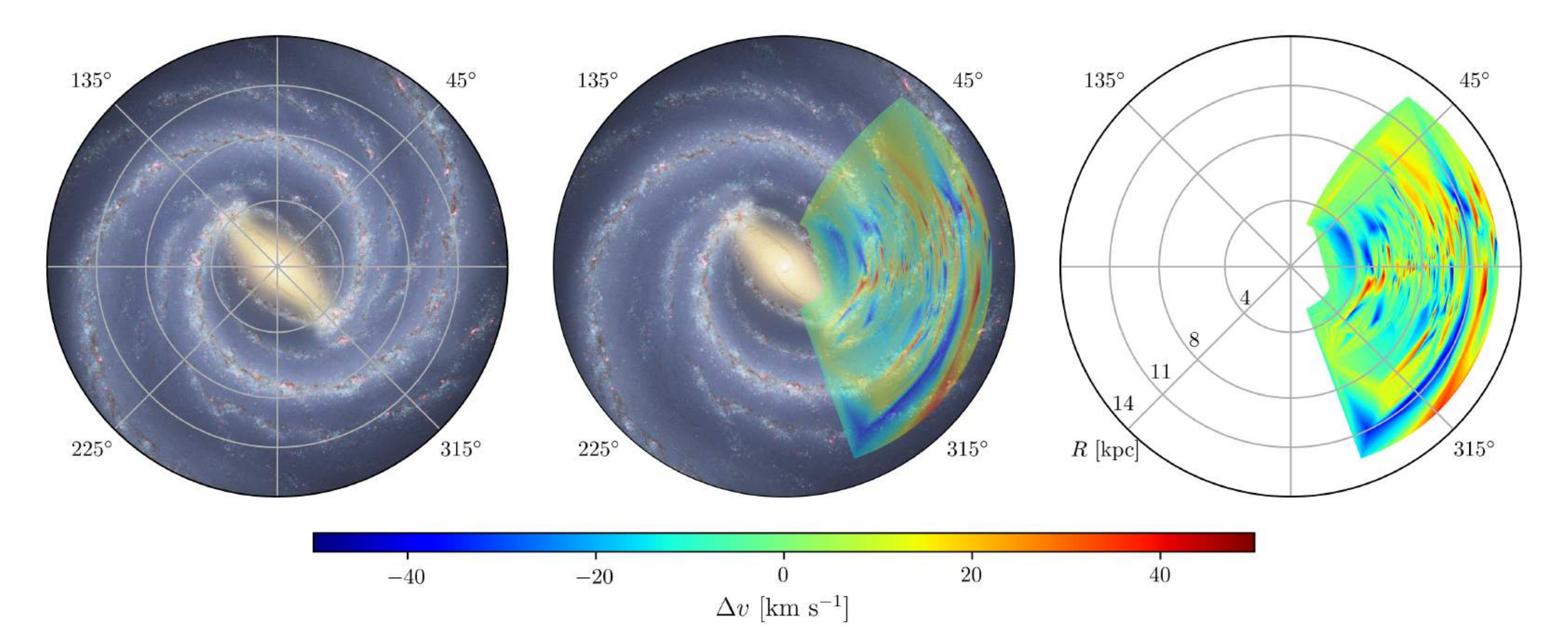}
	    \end{center}
	    \caption{The left panel shows an artist's rendition of the Milky Way galaxy [credits: NASA/JPL-Caltech/R. Hurt(SSC/Caltech)]. The image has been rotated and scaled to fit the location of the Sun and the Galactic center in our analyses ($R_0 = 8\rm\ kpc, \theta_{\rm Sun} = 0\degree$). The right panel shows residual circular velocity contours derived from the complete data limited to $290\degree \leq \theta\leq 70\degree$. The middle panel is an overlap of the two figures.}
	    \label{fig:residuecontour}
	\end{figure*}
    \begin{figure*}
		\centering
		\begin{center}
			\includegraphics[width=84mm]{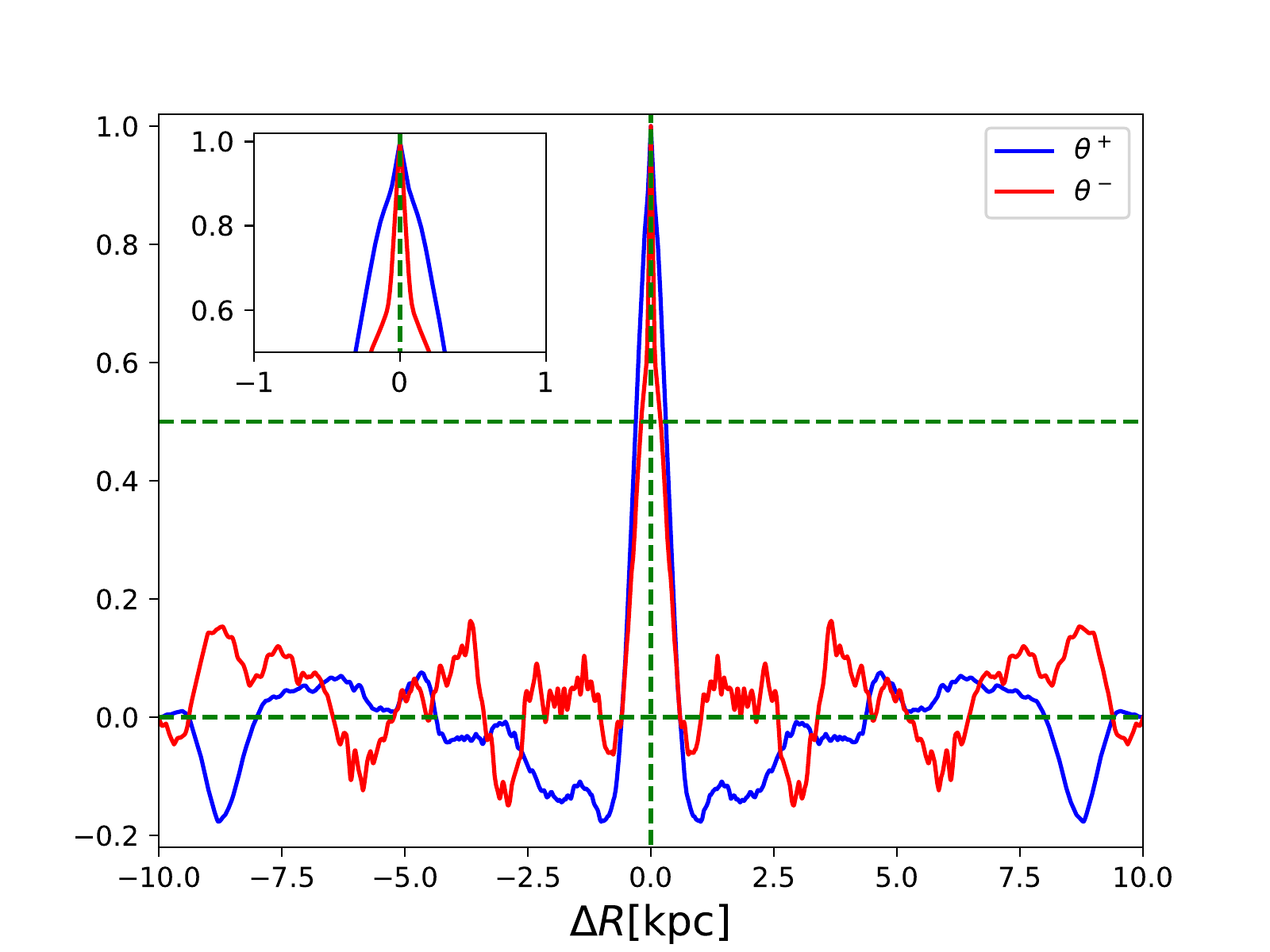}
			\includegraphics[width=84mm]{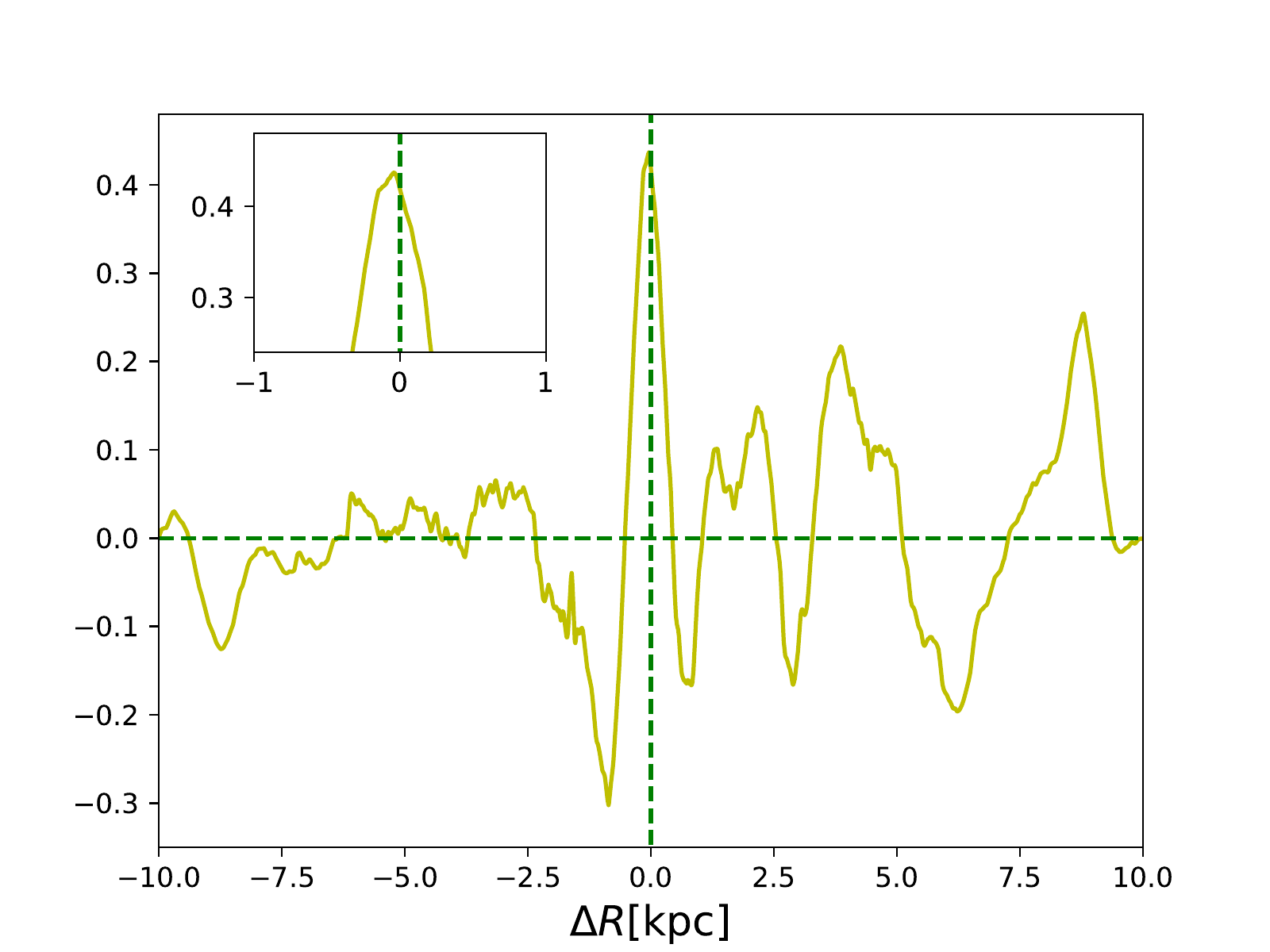}
		\end{center}
		\caption{The left planel shows the acfs and the right panel shows the ccf of positive azimuthal and negative azimuthal residual velocities data subsets. Vertical dashed lines indicate zero lag (see insets for details), each panel has a dashed horizontal line at zero amplitude, and the second dashed horizontal line in the left panel is at half the maximum amplitude.}
		\label{figure:acf2}
	\end{figure*}
	
	It has been known for quite a while that the Milky Way disk circular velocity field is not axisymmetric. See, e.g., \cite{Kerr1964}, \cite{Humphreys1970}, \cite{Georgelin1976}, \cite{Humphreys1976}, \cite{Blitz1991}, \cite{McClure-G07}, \cite{Bovy2015}, \cite{McClureGriffiths2016}, \cite{Bovy2017}, and references therein. In this subsection we use our GP method median statistics Milky Way rotation curves to study this azimuthal asymmetry. 
	
	Figure \ref{fig:spiral_arms} shows a contour map of the circular velocity field $v(R,\theta)$, where $\theta$ is galactocentric longitude, interpolated from the circular velocity data. We consider only the measurements in $\left|z\right| \leq 1.5$ kpc, $R\leq13.55$ kpc, and $290\degree \leq \theta \leq 70\degree$, so that the areal density of circular velocity measurement points is sufficiently large.\footnote{This map is over a smaller area than what is used in the $\chi^2$ and GP method analyses and we have discarded 141 of the 2,706 measurements used in those analyses. Additionally, so as to not skew the color scale, we have discarded 16 measurements with $v>300\rm\ km\ s^{-1}$ and 60 measurements with $v<180\rm\ km\ s^{-1}$. In all we use 2,489 of the 2,706 measurements to make this contour map.}
	
	To produce this contour map we used the \texttt{tricontourf} tool in the \texttt{matplotlib.pyplot} Python library. This tool uses Delaunay triangulation\footnote{The Delaunay triangulation method triangulates coordinates in space so that there is no other measurement coordinate in the circle circumscribing the triangles. This reduces the use of sliver triangles (triangles with one or more angles close to 0\degree) in its triangulation.} to triangulate the data and interpolates the data according to the triangulated and original data to find a possible value at any given position $(R, \theta)$ within the bounds of the data's positions.
	
	We overlap the contour circular velocity map with an artist's impression of the Milky Way disk, aligning the images using the position of the Sun and the Galactic center. Some regions with higher circular velocity seem to align with the spiral arms. In particular, there are higher circular rotation velocities at $\theta \sim 0\degree$ and $R \sim 5$ kpc (the Scutum-Centaurus spiral arm), $R \sim 8$ kpc (the Orion spur), and $R \sim 10\text{---}11$ kpc (the Perseus spiral arm). The circular velocity contour map clearly exhibits azimuthal anisotropy. Perhaps a better probe of azimuthal anisotropy is the residual velocity field, discussed later in this subsection.
	
	Figure \ref{figure:gp6} compares the rotation curves for the complete and positive azimuthal and negative azimuthal sector data sets. There are very significant differences between the positive azimuthal sector and negative azimuthal sector rotation curves, which further shows that azimuthal symmetry is broken. In the $3\text{ kpc}\lesssim R \lesssim 8\text{ kpc}$ range the complete data rotation curve is more consistent with the positive azimuthal sector rotation curve, probably because there are more measurements in that sector. 
	
	Figure \ref{figure:4AngBin} extends this comparison by analyzing narrower angular sectors. The GP method rotation curve fits for the four sectors are significantly different, further indicating azimuthal asymmetry.
	
	\begin{table}
	    \caption{Distribution of residual velocities plotted in Fig. \ref{figure:res0}, listed in terms of the residual velocity at different percentiles.}
	    \label{tab:residue}
	    \centering
	    \begin{tabular}{l@{\hspace{3\tabcolsep}}r@{\hspace{2\tabcolsep}}r@{\hspace{2\tabcolsep}}r@{\hspace{2\tabcolsep}}r@{\hspace{2\tabcolsep}}r}\hline\hline
	        \multirow{2}{*}{Data}&\multicolumn{5}{c}{Residual Velocity$^{\rm a}$ at}\\
	        	&2.5\%	&16\%	&50\%	&84\%	&97.5\%\\\hline
	        All	&$-41.49$&$-9.36$&$-0.04$&$6.49$&$35.05$\\ 
            P	&$-34.79$&$-6.18$&$0.47$&$5.72$	&$25.21$\\
            N	&$-53.78$&$-13.19$&$-2.49$&$7.65$&$47.86$\\\hline\hline
            \multicolumn{6}{l}{$^{\rm a}$ In units of km s$^{-1}$. \% here refers to percentile.}
	    \end{tabular}
	\end{table}
	
	Figure \ref{figure:res0} shows the residual circular velocities relative to the complete data GP method rotation curve (52 bins, gaussian covariance function). The residuals are color coded, with green (red) markers corresponding to the positive (negative) azimuthal sector data points.\footnote{The GP method rotation curve is only defined between the central radius of the lowest radius bin and the central radius of the highest radius bin. As a result we are unable to determine residual velocities for measurements that lie beyond the ends of the GP method rotation curve. Therefore we have to omit 52 out of 2,706 data points for these analyses, limiting the number of residual velocity data points to 2,654.} \cite{McClureGriffiths2016}, hereafter MD 2016, show a similar residual velocity plot in their Fig. 7. Theirs is for a smaller data compilation, and for $3\text{ kpc}\leq R\leq 8$ kpc, compared to the larger $R$ range shown in the left panel of Fig. \ref{figure:res0} here. The MD 2016 residuals are relative to a smooth linear rotation curve fit while ours are relative to the GP method fit that captures more of the small scale spatial structure present in the median statistics binned data. Our negative azimuthal sector (red) velocity residuals over $3\text{ kpc}\lesssim R \lesssim 8$ kpc show a qualitatively similar sawtooth spatial structure to the Quadrant IV (QIV) residuals in Fig. 7 of MD 2016. The sawtooth structure is much less prominent in our positive azimuthal sector residuals compared to that seen in the QI data of Fig. 7 of MD 2016. This is probably a consequence of two effects. As seen is Fig. \ref{figure:gp6}, and discussed earlier, over $3\text{ kpc}\lesssim R \lesssim 8$ kpc the rotation curves for the complete data set and for the positive azimuthal sector data subset are very similar, so the GP method complete data set rotation curve captures much of the small scale spatial variation present in the positive azimuthal sector velocity field. Thus the positive azimuthal sector residuals relative to the complete data rotation curve will have less small scale spatial structure than the negative aziuthal sector residual velocities, and this is evident in the left panel of Fig. \ref{figure:res0}. MD 2016 on the other hand fit a smooth rotation curve to their data, which will result in both QI and QIV residual velocities with small scale spatial structure.
	
	A similar conclusion may be drawn from the velocity residuals histograms shown in Fig. \ref{fig:res0_hist}. The negative azimuthal sector residual velocities are broader than the positive azimuthal sector and complete data residual velocities. See Table \ref{tab:residue} for numerical characteristics of the residual velocity distributions. We emphasize that the quantitative details depend on the median statistics binning used, but the qualitative conclusions don't.
	
	Figure \ref{figure:res2_0}, left panel, plots the location and amplitude of all 2,654 velocity residuals. To convert this to a velocity residuals contour map, we limit to an angular range of $290\degree\leq\theta\leq70\degree$ and omit 108 of the 2,654 measurements. To avoid skewing the color scale, we discard 35 measurements with $\Delta v > 50\rm\ km\ s^{-1}$ and 44 measurements with $\Delta v < - 50\rm\ km\ s^{-1}$. In all we use 2,467 velocity residuals, shown in the middle panel of Fig. \ref{figure:res2_0}, to make our velocity residuals contour map shown in the right panel of Fig. \ref{figure:res2_0} and in the central and right panels of Fig. \ref{fig:residuecontour}.
	
	This contour plot of residual velocities was made using the method used to produce Fig. \ref{fig:spiral_arms}. This residual velocity contour map is also azimuthally asymmetric, with the negative azimuthal sector exhibiting greater anisotropy. Somewhat prominent are anisotropies at $R\sim4$ kpc near where the Scutum-Centaurus spiral arm starts at the Galactic bar, at $R\sim9$ kpc at the Perseus spiral arm, and at $R\sim11$ kpc at the Outer spiral arm. They stand out more in the negative azimuthal sector, but some anisotropy is also evident in the positive azimuthal sector.
	
	It is of interest to examine the autocorrelation of the positive and negative azimuthal sector residual velocities, and the cross-correlation between the positive and the negative azimuthal secor residual velocities (MD 2016). Secondary peaks in the autocorrelation functions might be associated with physical processes in the Milky Way, and the scales at which these occur could help identify the relevant processes (MD 2016). A peak in the cross-correlation function at non-zero spatial displacement would signify a relative displacement between spatial structure in the positive and negative azimuthal sector residual velocities (MD 2016). The correlation function,
	\begin{equation}\label{eq:corr_cont}
	    C_{fg}(\Delta R) = K\int\Delta v_f(R + \Delta R)\times\overline{\Delta v_g(R)}dR,
	\end{equation}
	corresponds to the autocorrelation function (acf) for the case $f = g$, and the cross-correlation function (ccf) if $f \neq g$. Here $f\text{ and }g = \theta^+, \theta^-$, corresponding to the positive and negative azimuthal sector data subsets, and $K$ is a normalization constant, determined as follows. We normalize the positive and negative azimuthal sector residual velocity acfs by their respective global maxima and the ccf by the geometric mean of the global maxima of the positive and negative azimuthal sector acfs, $\sqrt{{\rm max}({\rm acf}(\theta^+))\times{\rm max}({\rm acf}(\theta^-))}$, where ${\rm max}({\rm acf}(f))$ stands for the maximum value of the acf of $v_f(R)$ obtained without normalization.
	
	The acf is useful when searching for periodicity in space, as manifested by secondary peaks in the acf. The ccf is used to look at projections of one function on the other at various spatial displacements. If $\Delta v_g(R) =\Delta v_f(R + a)$, we expect $C_{fg}(\Delta R)$ to form its peak at $\Delta R = a$.
	
	In the case at hand, for the discrete residual velocity function, eq. \eqref{eq:corr_cont} is replaced by
	\begin{equation}\label{eq:corr}
	    C_{fg}[\Delta R] = K\sum_{R} \Delta v_{f}[R + \Delta R]\times\overline{\Delta v_{g}[R]},
	\end{equation}
	where $f,\ g$ take on values in $\{\theta^+,\ \theta^-\}$ depending on what is being computed, and $\Delta v_{N}[R]\ (\Delta v_{S}[R])$ refers to the residual velocities of the positive (negative) azimuthal sector at distance $R$ from the Galactic center. To compute $\Delta v_{f}[R]$ at any $R$, we take the mean of the residual velocities of the ten nearest points from the $f$ data subset. We compute $\Delta v_{N}[R]\text{ and } \Delta v_{S}[R]$ over $2\text{ kpc} < R < 12\text{ kpc}$, at $dR = 5$ pc intervals to obtain a uniform sequence. To compute $C_{fg}(\Delta R)$ we use the \texttt{correlate} tool in the \texttt{numpy} Python library, which uses $\Delta v_{f}[R]$ and $\Delta v_{g}[R]$ as inputs.
	
	Here we focus on residual velocities with respect to the GP method fit for the complete data, in 52 bins and using a gaussian covariance function. The acf's and ccf depend on the median statistics binning used; we have checked that our qualitative conclusions do not depend significantly on the binning, although the binning does affect the quantitative results.
	
	The acf's are shown in the left panel of Fig. \ref{figure:acf2}. The half width at half maximum for the positive and negative azimuthal sector velocity residuals are 310 pc and 200 pc,\footnote{When computed for 43, 46, 49, and 52 bins, they range over 260---310 pc ($\theta^+$) and 190---280 pc ($\theta^-$); the variation is larger when more than 52 bins are allowed.} not far from the 290 pc and 210 pc QI and QIV values of MD 2016. The half width to first zero crossing are 590 pc in both the positive and negative azimuthal sectors\footnote{When computed for 43, 46, 49, and 52 bins, this ranges over 470---590 pc ($\theta^+$) and 500---700 pc ($\theta^-$).}; MD 2016 find 690 pc (QI) and 730 pc (QIV). MD 2016 emphasize that these length scales are significantly larger than the data resolution scale, and so argue that these are characteristic scales for dynamical effects.
	
	As MD 2016 did, we find a peak at close to zero lag in the ccf, Fig. \ref{figure:acf2} right panel. This peak has a normalized correlation of 0.44 at $-40$ pc shift, which means that the positive and negative azimuthal sector velocity residuals have a similar pattern with the positive azimuthal sector residuals shifted to a slightly smaller radius relative to the negative azimuthal sector residuals at the peak.\footnote{When computed for 43, 46, 49, and 52 bins, the correlation ranges between 0.37 and 0.82 and the shift between $-40$ pc and 0 pc.} These results are qualitatively consistent with those of MD 2016, who find a correlation of 0.27 and a shift of $-75$ pc between QI and QIV residuals. MD 2016 note that the shift is unexpectedly small.
	
	Compared to the acf's and ccf shown in Fig. 8 of MD 2016, our acf's and ccf in Fig. \ref{figure:acf2} have significantly more small-scale spatial structure and also do not have the well-aligned side-lobes at $\Delta R\sim \pm 2$ kpc in the two acf's and the ccf seen in Fig. 8 of MD 2016. We note however that our negative azimuthal sector acf has side lobes at $\Delta R\sim \pm 1\text{---}2$ kpc, $\Delta R\sim \pm 4$ kpc, and at $\Delta R\sim \pm 7.5\text{---}9$ kpc,\footnote{These $\Delta R$'s are close to the separations between the local overdensities in visible matter; there is a distance of $\sim2$ kpc between Perseus and Outer spiral arms, a distance of $\sim4$---5 kpc between the Scutum-Centaurus and Perseus spiral arms, and a distance of $\sim 7.5$ kpc between the Scutum-Centaurus and Outer spiral arms.} and the positive azimuthal sector acf has wider side lobes at $\Delta R\sim \pm 4.5\text{---}6$ kpc. These differences between our side lobes and those found by MD 2016 are probably a consequence of the different rotation curves used to determine the velocity residuals, the GP method median statistics rotation curve fit here and the simple functional form used in MD 2016.
	
	The results we find here are qualitatively consistent with those found by MD 2016. This indicates that the GP method median statistics rotation curve we derive here is, at least qualitatively, as reasonable an approximation to that of the Milky Way disk as the rotation curves derived using the conventional technique.
	
	Given that the small scale spatial structure in the rotation curve is anisotropic, e.g., we find that the positive and negative azimuthal sector data rotation curves are significantly different, it is probably more valuable now to extend beyond axisymmetric bulge, disk, and halo models \citep[e.g.,][and references therein]{Sofue2009,Sofue2013}, instead of trying to explain observed small scale spatial structures in the derived Milky Way rotation curves in terms of axisymmetric mass perturbations about a more uniform background. More importantly, the small scale spatial structure in the Milky Way rotation curve depends on the technique used to determine the rotation curve, and this effect must be accounted for when trying to relate such small scale spatial structure to models of the Milky Way mass distribution.
	
	\section{Conclusion}\label{sec:conclusions}
	
	\cite{Iocco2015} have recently compiled a very useful, large collection of Milky Way circular velocity measurements. We have studied and used their compilation in this paper. 
	
	Our main results, in summary, are:
	\begin{itemize}
	    \item We have found that the \cite{Iocco2015} circular velocity measurement error bars are non-gaussian.
	    \item To be able to make use of the \cite{Iocco2015} compilation, we have ignored the error bars on the individual measurements and used median statistics and the measured circular velocity values to determine new median statistics central values and error bars for binned circular velocity measurements.
	    \item We have provided tables of median statistics binned circular velocities for a variety of different tracer types and spatial distribution combinations. We emphasize that we only get good fits over $2\lesssim R\lesssim 10$ kpc (and over shorter ranges for the various data subsets) due to a low radial density of measurements outside the range.
	    \item We have found that simple, few parameter, functional descriptions of the Milky Way rotation curve are unable to adequately capture the small scale spatial variation in the circular velocity data, and so badly fit these data.
	    \item To capture this small scale spatial structure in the circular velocity field, we introduce and use the Gaussian Processes method to extract Milky Way rotation curves from these data. The GP method rotation curves we derive have significant small scale spatial structure. We believe that most of this small-scale spatial structure is physical and reinforces a revised picture of the Milky Way rotation curve.
	    \item Near $R \approx 7$ kpc the star tracers rotation curve lies a little below that derived from gas tracers. This possible discrepancy deserves careful study.
	    \item From the GP rotation curve derived using the median statistics binned measurements of the complete data, we measure the $A(R_0)$ and $B(R_0)$ Oort constants. Our measurements are consistent with, but less constraining than, many recent measurements of $A(R_0)$ and $B(R_0)$. It is reassuring that our new GP method median statistics Milky Way rotation curve passes this test.
	    \item By deriving GP method median statistics rotation curves for different angular sectors of the Milky Way disk, we show that the Milky Way circular velocity field is azimuthally anisotropic. We emphasize that the small scale spatial structure in GP method Milky Way rotation curves are azimuthally anisotropic.
	    \item We find that circular velocity residuals relative to the GP method rotation curve show evidence for characteristic spatial structures at length scales qualitatively consistent with those found by \cite{McClureGriffiths2016}. It is reassuring that our results are consistent with theirs; this validates the GP method median statistics rotation curve technique we have developed. As discussed by \cite{McClureGriffiths2016}, these measured spatial structure length scales may be used to elucidate dynamical Milky Way processes.
	    \item We have found that different tracers can result in slightly different Milky Way rotation curves. More importantly we have found that different techniques ($\chi^2$ fitting of simple functional forms versus the GP method) result in different Milky Way rotation curves. This probably means that determining more accurate Milky Way mass models necessitates avoiding the step of compressing circular rotation velocity data into a rotation curve.
	\end{itemize}
	
	While these results are interesting, they are based on the \cite{Iocco2015} data compilation of older, less precise data. It is of significant interest to apply the techniques we have developed and used here to more recent, better data.
	
	\section*{Acknowledgements}
	We are indebted to A. Quillen and S. Sharma for detailed comments that helped improve our paper. We thank P. Ghosh and D. Weinberg for useful discussions, and T. Camarillo for participating in the early stages of the gaussianity analyses. This work was partially supported by DOE grant DE--SC0019038 and NASA grant 12--EUCLID11--000. H. Yu also acknowledges support by the China Scholarship Council for studying abroad.

	\bibliographystyle{mnras}
	\bibliography{RCbib}
	
	\appendix
	\section{Gaussianity analyses tables}
	\begin{table}\tiny
		\caption{Angular and linear circular velocity median central values for the complete data.}
		\label{tab:cvall}
		\begin{center}
\begin{tabular}{l@{\hspace{2\tabcolsep}}c@{\hspace{2\tabcolsep}}c@{\hspace{1\tabcolsep}}c@{\hspace{1\tabcolsep}}c@{\hspace{1\tabcolsep}}c}\hline\hline
			Bin&Count&Bin Radii (kpc)&$\omega (\rm km\ s^{-1}\ kpc^{-1})$&Median $R \rm(kpc)$&$v \rm (km\ s^{-1})$\\\hline
			1&53&0.62-2.18&$102.0_{-3.0}^{+4.0}$&$1.86_{-0.05}^{+0.12}$&$190_{-11}^{+20}$\\\noalign{\vskip 0.5mm}
			2&53&2.2-2.6&$82.8_{-0.6}^{+0.4}$&$2.41_{-0.05}^{+0.03}$&$199_{-5}^{+4}$\\\noalign{\vskip 0.5mm}
			3&52&2.6-2.87&$73.8_{-0.5}^{+0.8}$&$2.78_{-0.03}^{+0.02}$&$205_{-4}^{+4}$\\\noalign{\vskip 0.5mm}
			4&52&2.87-3.08&$68.9_{-0.2}^{+0.5}$&$2.98_{-0.02}^{+0.01}$&$206_{-2}^{+2}$\\\noalign{\vskip 0.5mm}
			5&52&3.08-3.26&$67.1_{-0.4}^{+0.5}$&$3.18_{-0.02}^{+0.01}$&$213_{-3}^{+3}$\\\noalign{\vskip 0.5mm}
			6&52&3.27-3.46&$62.9_{-0.5}^{+0.2}$&$3.38_{-0.02}^{+0.01}$&$212_{-3}^{+1}$\\\noalign{\vskip 0.5mm}
			7&52&3.46-3.64&$59.0_{-0.1}^{+0.4}$&$3.56_{-0.02}^{+0.02}$&$210_{-2}^{+2}$\\
\hline\hline
		\end{tabular}
		\end{center}
	\end{table}
	\begin{table*} \tiny
		\caption{$N_\sigma$ analysis for the complete data, with (lower row for each bin) and without (upper row) $R$ error.}
		\label{tab:nsall}
		\begin{tabular}{l@{\hspace{1\tabcolsep}}c@{\hspace{1\tabcolsep}}c@{\hspace{3\tabcolsep}}c@{\hspace{1\tabcolsep}}c@{\hspace{1\tabcolsep}}c@{\hspace{1\tabcolsep}}c@{\hspace{1\tabcolsep}}c@{\hspace{1\tabcolsep}}c@{\hspace{3\tabcolsep}}c@{\hspace{1\tabcolsep}}c@{\hspace{1\tabcolsep}}c@{\hspace{1\tabcolsep}}c@{\hspace{1\tabcolsep}}c@{\hspace{1\tabcolsep}}c}\hline\hline
			&&&\multicolumn{6}{c}{Median}&\multicolumn{6}{c}{Weighted Mean}\\
			&&&$68.27\%$&$95.45\%$&$68.27\%$&$95.45\%$&EF&EF&$68.27\%$&$95.45\%$&$68.27\%$&$95.45\%$&EF&EF\\
			Bin&Count&Radii (kpc)&$N_{\sigma}$&$N_{\sigma}$&$|N_{\sigma}|$&$|N_{\sigma}|$&$|N_{\sigma}|$&$|N_{\sigma}|$&$N_{\sigma}$&$N_{\sigma}$&$|N_{\sigma}|$&$|N_{\sigma}|$&$|N_{\sigma}|$&$|N_{\sigma}|$\\\hline
			\noalign{\vskip 1mm}
			\multirow{2}{*}{1}&\multirow{2}{*}{53}&0.62-2.18&[-7.62,4.93]&[-15.3,12.9]&5.62&15.3&18.9\%&43.4\%&[-11.2,14.3]&[-30.9,20.6]&13.3&25.5&7.55\%&15.1\%\\
			&&0.62-2.33&[-6.34,8.49]&[-9.62,15.5]&8.22&13.9&18.9\%&34.0\%&[-12.2,17.1]&[-26.7,23.8]&14.8&25.7&13.2\%&22.6\%\\
			\multirow{2}{*}{2}&\multirow{2}{*}{53}&2.2-2.6&[-8.87,4.97]&[-19.2,10.2]&5.31&16.3&22.6\%&28.3\%&[-7.14,8.33]&[-18.5,13.7]&8.33&15.2&11.3\%&18.9\%\\
			&&2.02-2.73&[-2.63,3.49]&[-5.83,5.47]&2.8&5.77&22.6\%&47.2\%&[-2.91,14.5]&[-19.1,21.0]&11.6&21.0&13.2\%&18.9\%\\
			\multirow{2}{*}{3}&\multirow{2}{*}{52}&2.6-2.87&[-2.48,1.42]&[-5.61,3.71]&1.93&5.11&36.5\%&69.2\%&[-2.16,2.99]&[-7.61,6.82]&2.99&7.14&28.9\%&50.0\%\\
\hline\hline
		\end{tabular}
	\end{table*}	
	\begin{table}\tiny
		\begin{center}
			\caption{KS test results for $N_\sigma$ distributions for the complete data, with (lower rows) and without (upper rows) $R$ error.}
			\label{tab:ksall}
			\begin{tabular}{l@{\hspace{3\tabcolsep}}l@{\hspace{1\tabcolsep}}c@{\hspace{1\tabcolsep}}c@{\hspace{3\tabcolsep}}l@{\hspace{1\tabcolsep}}c@{\hspace{1\tabcolsep}}c}\hline\hline
				&\multicolumn{3}{c}{Median Statistics}&\multicolumn{3}{c}{Weighted Mean Statistics}\\
				Bin&Distribution&$S$&$p$&Distribution&$S$&$p$\\\hline
				\multirow{2}{*}{1}&Cauchy&2.56&0.96&Student's $t$, $\nu = 2$&9.04&1.0\\
				&Laplace&5.98&1.0&Student's $t$, $\nu = 2$&10.0&0.84\\
				\multirow{2}{*}{2}&Laplace&4.59&0.92&Gaussian&8.28&1.0\\
				&Gaussian&3.16&1.0&Laplace&8.78&1.0\\
				\multirow{2}{*}{3}&Student's $t$, $\nu = 2$&1.53&1.0&Student's $t$, $\nu = 3$&2.43&1.0\\
				&Laplace&4.0&1.0&Student's $t$, $\nu = 2$&4.06&1.0\\
\hline\hline
			\end{tabular}
		\end{center}
	\end{table}
	\begin{table}\tiny
		\caption{Angular and linear circular velocity median central values for gas tracers.}
		\label{tab:cvgas}
		\begin{center}
		\begin{tabular}{l@{\hspace{2\tabcolsep}}c@{\hspace{2\tabcolsep}}c@{\hspace{1\tabcolsep}}c@{\hspace{1\tabcolsep}}c@{\hspace{1\tabcolsep}}c}\hline\hline
			Bin&Count&Bin Radii (kpc)&$\omega (\rm km\ s^{-1}\ kpc^{-1})$&Median $R \rm(kpc)$&$v \rm (km\ s^{-1})$\\\hline\noalign{\vskip 0.5mm}
			1&47&0.62-2.1&$106.0_{-5.0}^{+2.0}$&$1.83_{-0.06}^{+0.11}$&$193_{-16}^{+15}$\\\noalign{\vskip 0.5mm}
			2&47&2.12-2.5&$87.3_{-4.0}^{+1.3}$&$2.34_{-0.02}^{+0.0}$&$204_{-11}^{+3}$\\\noalign{\vskip 0.5mm}
			3&47&2.51-2.81&$75.2_{-0.8}^{+0.2}$&$2.68_{-0.04}^{+0.03}$&$202_{-5}^{+3}$\\\noalign{\vskip 0.5mm}
			4&47&2.81-3.0&$70.1_{-0.4}^{+0.5}$&$2.9_{-0.02}^{+0.02}$&$203_{-2}^{+3}$\\\noalign{\vskip 0.5mm}
			5&47&3.0-3.17&$67.9_{-0.2}^{+0.1}$&$3.09_{-0.02}^{+0.01}$&$210_{-2}^{+1}$\\\noalign{\vskip 0.5mm}
			6&47&3.17-3.35&$66.0_{-1.3}^{+0.5}$&$3.26_{-0.01}^{+0.01}$&$215_{-5}^{+2}$\\\noalign{\vskip 0.5mm}
			7&47&3.36-3.51&$60.8_{-0.6}^{+0.9}$&$3.44_{-0.02}^{+0.02}$&$209_{-3}^{+4}$\\
\end{tabular}
		\end{center}
		\vspace{10mm}
	\end{table}
	\begin{table*} \tiny
		\caption{$N_\sigma$ analysis for gas tracers, with (lower row for each bin) and without (upper row) $R$ error.}
		\label{tab:nsgas}
		\begin{tabular}{l@{\hspace{1\tabcolsep}}c@{\hspace{1\tabcolsep}}c@{\hspace{3\tabcolsep}}c@{\hspace{1\tabcolsep}}c@{\hspace{1\tabcolsep}}c@{\hspace{1\tabcolsep}}c@{\hspace{1\tabcolsep}}c@{\hspace{1\tabcolsep}}c@{\hspace{3\tabcolsep}}c@{\hspace{1\tabcolsep}}c@{\hspace{1\tabcolsep}}c@{\hspace{1\tabcolsep}}c@{\hspace{1\tabcolsep}}c@{\hspace{1\tabcolsep}}c}\hline\hline
			&&&\multicolumn{6}{c}{Median}&\multicolumn{6}{c}{Weighted Mean}\\
			&&&$68.27\%$&$95.45\%$&$68.27\%$&$95.45\%$&EF&EF&$68.27\%$&$95.45\%$&$68.27\%$&$95.45\%$&EF&EF\\
			Bin&Count&Radii (kpc)&$N_{\sigma}$&$N_{\sigma}$&$|N_{\sigma}|$&$|N_{\sigma}|$&$|N_{\sigma}|$&$|N_{\sigma}|$&$N_{\sigma}$&$N_{\sigma}$&$|N_{\sigma}|$&$|N_{\sigma}|$&$|N_{\sigma}|$&$|N_{\sigma}|$\\\hline
			\noalign{\vskip 1mm}\multirow{2}{*}{1}&\multirow{2}{*}{47}&0.62-2.1&[-5.63,7.47]&[-9.61,14.5]&6.92&13.3&19.1\%&42.5\%&[-13.1,14.9]&[-31.1,20.5]&14.9&25.7&10.6\%&17.0\%\\
			&&0.62-2.25&[-8.08,7.88]&[-14.8,14.7]&8.08&14.8&21.3\%&23.4\%&[-11.2,16.4]&[-28.0,22.0]&14.5&27.3&2.13\%&12.8\%\\
			\multirow{2}{*}{2}&\multirow{2}{*}{47}&2.12-2.5&[-2.06,3.03]&[-6.13,6.61]&2.64&6.61&29.8\%&59.6\%&[-4.96,8.6]&[-17.6,13.1]&7.58&17.3&12.8\%&25.5\%\\
			&&1.99-2.61&[-6.82,4.01]&[-12.9,5.65]&4.67&11.4&21.3\%&44.7\%&[-13.3,14.4]&[-21.0,19.6]&14.2&21.0&8.51\%&21.3\%\\
			\multirow{2}{*}{3}&\multirow{2}{*}{47}&2.51-2.81&[-3.02,3.31]&[-5.47,6.52]&3.16&6.52&29.8\%&48.9\%&[-2.96,4.32]&[-9.26,8.73]&4.32&9.26&19.1\%&42.5\%\\
\hline\hline
		\end{tabular}
	\end{table*}
	\begin{table}\tiny
		\begin{center}
			\caption{KS test results for $N_\sigma$ distributions, with (lower rows) and without (upper rows) $R$ error, for gas tracers.}
			\label{tab:ksgas}
			\begin{tabular}{l@{\hspace{3\tabcolsep}}l@{\hspace{1\tabcolsep}}c@{\hspace{1\tabcolsep}}c@{\hspace{3\tabcolsep}}l@{\hspace{1\tabcolsep}}c@{\hspace{1\tabcolsep}}c}\hline\hline
				&\multicolumn{3}{c}{Median Statistics}&\multicolumn{3}{c}{Weighted Mean Statistics}\\
				Bin&Distribution&$S$&$p$&Distribution&$S$&$p$\\\hline
				\multirow{2}{*}{1}&Cauchy&2.88&0.96&Student's $t$, $\nu = 2$&9.9&1.0\\
				&Laplace&6.16&0.99&Student's $t$, $\nu = 2$&9.2&0.97\\
				\multirow{2}{*}{2}&Student's $t$, $\nu = 2$&2.23&1.0&Student's $t$, $\nu = 3$&7.14&1.0\\
				&Student's $t$, $\nu = 6$&3.7&0.99&Student's $t$, $\nu = 3$&9.79&1.0\\
				\multirow{2}{*}{3}&Laplace&2.68&1.0&Student's $t$, $\nu = 2$&3.17&1.0\\
				&Student's $t$, $\nu = 7$&2.8&1.0&Laplace&4.42&1.0\\
\hline\hline
			\end{tabular}
		\end{center}
	\end{table}
	\begin{table}\tiny
		\caption{Angular and linear circular linear velocity median central values for star tracers.}
		\label{tab:cvstar}
		\begin{center}
		\begin{tabular}{l@{\hspace{2\tabcolsep}}c@{\hspace{2\tabcolsep}}c@{\hspace{1\tabcolsep}}c@{\hspace{1\tabcolsep}}c@{\hspace{1\tabcolsep}}c}\hline\hline
			Bin&Count&Bin Radii (kpc)&$\omega (\rm km\ s^{-1}\ kpc^{-1})$&Median $R \rm(kpc)$&$v \rm (km\ s^{-1})$\\\hline\noalign{\vskip 0.5mm}
			1&22&3.62-5.86&$40.4_{-3.1}^{+6.6}$&$5.45_{-0.35}^{+0.13}$&$220_{-31}^{+41}$\\\noalign{\vskip 0.5mm}
			2&22&6.0-6.42&$32.4_{-1.0}^{+1.6}$&$6.32_{-0.05}^{+0.0}$&$205_{-8}^{+10}$\\\noalign{\vskip 0.5mm}
			3&22&6.42-6.83&$35.0_{-2.3}^{+1.5}$&$6.57_{-0.02}^{+0.04}$&$230_{-16}^{+11}$\\\noalign{\vskip 0.5mm}
			4&22&6.85-7.18&$32.2_{-0.6}^{+0.7}$&$7.0_{-0.06}^{+0.03}$&$226_{-7}^{+5}$\\\noalign{\vskip 0.5mm}
			5&22&7.18-7.31&$30.3_{-0.1}^{+1.4}$&$7.26_{-0.01}^{+0.01}$&$220_{-1}^{+10}$\\\noalign{\vskip 0.5mm}
			6&22&7.31-7.47&$29.9_{-0.5}^{+0.5}$&$7.4_{-0.02}^{+0.01}$&$222_{-4}^{+4}$\\\noalign{\vskip 0.5mm}
			7&22&7.5-7.58&$29.8_{-0.8}^{+0.4}$&$7.54_{-0.01}^{+0.01}$&$225_{-7}^{+3}$\\
\hline\hline
		\end{tabular}
		\end{center}
	\end{table}
	\begin{table*} \tiny
		\caption{$N_\sigma$ analysis for star tracers, with (lower row for each bin) and without (upper row) $R$ error.}
		\label{tab:nsstar}
		\begin{tabular}{l@{\hspace{1\tabcolsep}}c@{\hspace{1\tabcolsep}}c@{\hspace{3\tabcolsep}}c@{\hspace{1\tabcolsep}}c@{\hspace{1\tabcolsep}}c@{\hspace{1\tabcolsep}}c@{\hspace{1\tabcolsep}}c@{\hspace{1\tabcolsep}}c@{\hspace{3\tabcolsep}}c@{\hspace{1\tabcolsep}}c@{\hspace{1\tabcolsep}}c@{\hspace{1\tabcolsep}}c@{\hspace{1\tabcolsep}}c@{\hspace{1\tabcolsep}}c}\hline\hline
			&&&\multicolumn{6}{c}{Median}&\multicolumn{6}{c}{Weighted Mean}\\
			&&&$68.27\%$&$95.45\%$&$68.27\%$&$95.45\%$&EF&EF&$68.27\%$&$95.45\%$&$68.27\%$&$95.45\%$&EF&EF\\
			Bin&Count&Radii (kpc)&$N_{\sigma}$&$N_{\sigma}$&$|N_{\sigma}|$&$|N_{\sigma}|$&$|N_{\sigma}|$&$|N_{\sigma}|$&$N_{\sigma}$&$N_{\sigma}$&$|N_{\sigma}|$&$|N_{\sigma}|$&$|N_{\sigma}|$&$|N_{\sigma}|$\\\hline
			\noalign{\vskip 1mm}
			\multirow{2}{*}{1}&\multirow{2}{*}{22}&3.62-5.86&[-1.49,1.5]&[-2.98,2.03]&1.5&2.27&45.5\%&81.8\%&[-4.6,2.19]&[-7.09,9.4]&3.9&7.09&18.2\%&50.0\%\\
			&&3.62-6.19&[-1.59,2.8]&[-2.67,4.52]&2.01&3.51&31.8\%&68.2\%&[-5.13,2.38]&[-9.67,12.7]&4.37&9.67&27.3\%&40.9\%\\
			\multirow{2}{*}{2}&\multirow{2}{*}{22}&6.0-6.42&[-0.85,1.99]&[-7.23,4.69]&1.99&5.06&54.5\%&72.7\%&[-3.21,-0.34]&[-9.65,12.6]&2.67&9.65&9.09\%&54.5\%\\
			&&5.67-6.53&[-2.64,1.95]&[-5.3,4.75]&2.2&4.92&50.0\%&59.1\%&[-3.77,4.6]&[-6.97,5.89]&4.6&6.28&18.2\%&27.3\%\\
			\multirow{2}{*}{3}&\multirow{2}{*}{22}&6.42-6.83&[-1.55,0.86]&[-6.41,4.89]&1.1&4.89&63.6\%&81.8\%&[-1.83,0.81]&[-26.6,26.7]&1.83&26.6&50.0\%&72.7\%\\
\hline\hline
		\end{tabular}
	\end{table*}
	\begin{table*}\tiny
		\begin{center}
			\caption{KS test results for $N_\sigma$ distributions, with (lower rows) and without (upper rows) $R$ error, for star tracers.}
			\label{tab:ksstar}
			\begin{tabular}{l@{\hspace{3\tabcolsep}}l@{\hspace{1\tabcolsep}}c@{\hspace{1\tabcolsep}}c@{\hspace{3\tabcolsep}}l@{\hspace{1\tabcolsep}}c@{\hspace{1\tabcolsep}}c}\hline\hline
				&\multicolumn{3}{c}{Median Statistics}&\multicolumn{3}{c}{Weighted Mean Statistics}\\
				Bin&Distribution&$S$&$p$&Distribution&$S$&$p$\\\hline
				\multirow{2}{*}{1}&Gaussian&1.61&1.0&Student's $t$, $\nu = 2$&2.6&1.0\\
				&Gaussian&2.15&1.0&Student's $t$, $\nu = 12$&3.63&1.0\\
				\multirow{2}{*}{2}&Cauchy&0.98&1.0&Gaussian&3.34&0.95\\
				&Laplace&1.92&0.99&Gaussian&4.7&0.98\\
				\multirow{2}{*}{3}&Student's $t$, $\nu = 7$&1.19&1.0&Cauchy&1.08&1.0\\
				&Student's $t$, $\nu = 49$&0.91&0.95&Laplace&1.71&1.0\\
\hline\hline
			\end{tabular}
		\end{center}
	\end{table*}
	\begin{table}\tiny
		\caption{Angular and linear circular velocity median central values for data in 290\degree \ to 300\degree, 300\degree \ to 310\degree, 310\degree \ to 320\degree, and 320\degree \ to 330\degree\ sectors.}
		\label{tab:cvang1}
		\begin{center}
		\begin{tabular}{l@{\hspace{1\tabcolsep}}c@{\hspace{1\tabcolsep}}c@{\hspace{2\tabcolsep}}c@{\hspace{1\tabcolsep}}c@{\hspace{1\tabcolsep}}c@{\hspace{1\tabcolsep}}c}\hline\hline
			Angle (deg)&Bin&Count&Bin Radii (kpc)&$\omega (\rm km\ s^{-1}\ kpc^{-1})$&Median $R \rm(kpc)$&$v \rm (km\ s^{-1})$\\\hline
			\multirow{16.5}{*}{290-300}
			&1&11&2.74-2.86&$70.4_{-1.0}^{+0.6}$&$2.8_{-0.03}^{+0.02}$&$197_{-5}^{+3}$\\\noalign{\vskip 0.5mm}
			&2&11&2.87-3.0&$68.2_{-3.3}^{+0.4}$&$2.94_{-0.04}^{+0.04}$&$201_{-12}^{+4}$\\\noalign{\vskip 0.5mm}
			&3&11&3.01-3.13&$65.6_{-1.2}^{+0.2}$&$3.06_{-0.04}^{+0.03}$&$201_{-6}^{+2}$\\\noalign{\vskip 0.5mm}
			&4&11&3.14-3.25&$63.1_{-1.7}^{+0.9}$&$3.2_{-0.02}^{+0.02}$&$202_{-7}^{+4}$\\\noalign{\vskip 0.5mm}
			&5&11&3.25-3.42&$61.6_{-1.0}^{+1.0}$&$3.34_{-0.03}^{+0.04}$&$206_{-5}^{+6}$\\\noalign{\vskip 0.5mm}
			&6&11&3.42-3.52&$59.3_{-0.4}^{+0.2}$&$3.46_{-0.02}^{+0.02}$&$205_{-3}^{+2}$\\\noalign{\vskip 0.5mm}
			&7&11&3.54-3.65&$57.1_{-0.5}^{+0.6}$&$3.62_{-0.03}^{+0.0}$&$207_{-3}^{+2}$\\
\hline\hline
		\end{tabular}
		\end{center}
		\vspace{15mm}
	\end{table}
	\begin{table}\tiny
		\caption{Angular and linear circular velocity median central values for data in 340\degree \ to 350\degree, 350\degree \ to 0\degree, and 0\degree \ to 10\degree.}
		\label{tab:cvang2}
		\begin{center}
		\begin{tabular}{l@{\hspace{1\tabcolsep}}c@{\hspace{1\tabcolsep}}c@{\hspace{2\tabcolsep}}c@{\hspace{1\tabcolsep}}c@{\hspace{1\tabcolsep}}c@{\hspace{1\tabcolsep}}c}\hline\hline
			Angle (deg)&Bin&Count&Bin Radii (kpc)&$\omega (\rm km\ s^{-1}\ kpc^{-1})$&Median $R \rm(kpc)$&$v \rm (km\ s^{-1})$\\\hline
			\multirow{23}{*}{340-350}
			&1&15&4.28-4.91&$52.0_{-4.4}^{+0.7}$&$4.9_{-0.34}^{+0.01}$&$255_{-39}^{+4}$\\\noalign{\vskip 0.5mm}
			&2&15&4.92-6.24&$40.8_{-1.3}^{+1.8}$&$5.37_{-0.0}^{+0.29}$&$219_{-7}^{+21}$\\\noalign{\vskip 0.5mm}
			&3&15&6.33-7.31&$31.1_{-0.8}^{+0.3}$&$7.19_{-0.08}^{+0.04}$&$223_{-8}^{+3}$\\\noalign{\vskip 0.5mm}
			&4&15&7.41-7.6&$29.0_{-0.4}^{+0.2}$&$7.56_{-0.04}^{+0.01}$&$219_{-4}^{+2}$\\\noalign{\vskip 0.5mm}
			&5&15&7.61-7.73&$28.4_{-0.1}^{+0.0}$&$7.69_{-0.03}^{+0.01}$&$218_{-2}^{+1}$\\\noalign{\vskip 0.5mm}
			&6&15&7.73-7.81&$28.0_{-0.2}^{+0.1}$&$7.76_{-0.01}^{+0.01}$&$218_{-2}^{+1}$\\\noalign{\vskip 0.5mm}
			&7&15&7.81-7.86&$28.7_{-0.2}^{+0.2}$&$7.83_{-0.01}^{+0.01}$&$225_{-2}^{+2}$\\
\hline\hline
		\end{tabular}
		\end{center}
	\end{table}
	\begin{table}\tiny
		\caption{Angular and linear circular velocity median central values for data in 10\degree \ to 20\degree, 20\degree \ to 30\degree, and 30\degree \ to 40\degree\ sectors.}
		\label{tab:cvang3}
		\begin{center}
		\begin{tabular}{l@{\hspace{1\tabcolsep}}c@{\hspace{1\tabcolsep}}c@{\hspace{2\tabcolsep}}c@{\hspace{1\tabcolsep}}c@{\hspace{1\tabcolsep}}c@{\hspace{1\tabcolsep}}c}\hline\hline
			Angle (deg)&Bin&Count&Bin Radii (kpc)&$\omega (\rm km\ s^{-1}\ kpc^{-1})$&Median $R \rm(kpc)$&$v \rm (km\ s^{-1})$\\\hline
			\multirow{23}{*}{10-20}
			&1&15&4.9-6.22&$37.0_{-0.5}^{+0.2}$&$5.84_{-0.15}^{+0.16}$&$216_{-9}^{+7}$\\\noalign{\vskip 0.5mm}
			&2&15&6.27-7.2&$32.0_{-0.3}^{+0.8}$&$7.03_{-0.09}^{+0.06}$&$225_{-5}^{+8}$\\\noalign{\vskip 0.5mm}
			&3&15&7.22-7.5&$30.3_{-0.1}^{+0.4}$&$7.36_{-0.02}^{+0.02}$&$223_{-2}^{+4}$\\\noalign{\vskip 0.5mm}
			&4&15&7.5-7.56&$30.5_{-0.1}^{+0.1}$&$7.54_{-0.01}^{+0.01}$&$230_{-1}^{+1}$\\\noalign{\vskip 0.5mm}
			&5&15&7.56-7.62&$30.0_{-0.2}^{+0.2}$&$7.58_{-0.0}^{+0.03}$&$228_{-2}^{+2}$\\\noalign{\vskip 0.5mm}
			&6&15&7.62-7.66&$29.7_{-0.2}^{+0.4}$&$7.64_{-0.0}^{+0.0}$&$227_{-2}^{+3}$\\\noalign{\vskip 0.5mm}
			&7&15&7.67-7.8&$28.8_{-0.3}^{+0.3}$&$7.73_{-0.02}^{+0.01}$&$223_{-3}^{+3}$\\
\hline\hline
		\end{tabular}
	\end{center}
	\end{table}
	\begin{table}\tiny
		\caption{Angular and linear circular velocity median central values for data in 40\degree \ to 50\degree, 50\degree \ to 60\degree, and 60\degree \ to 70\degree\ sectors.}
		\label{tab:cvang4}
		\begin{center}
			\begin{tabular}{l@{\hspace{1\tabcolsep}}c@{\hspace{1\tabcolsep}}c@{\hspace{2\tabcolsep}}c@{\hspace{1\tabcolsep}}c@{\hspace{1\tabcolsep}}c@{\hspace{1\tabcolsep}}c}\hline\hline
				Angle (deg)&Bin&Count&Bin Radii (kpc)&$\omega (\rm km\ s^{-1}\ kpc^{-1})$&Median $R \rm(kpc)$&$v \rm (km\ s^{-1})$\\\hline
				\multirow{23.5}{*}{40-50}
				&1&15&5.15-5.24&$41.2_{-0.1}^{+0.1}$&$5.22_{-0.01}^{+0.01}$&$215_{-1}^{+1}$\\\noalign{\vskip 0.5mm}
				&2&15&5.25-5.31&$40.4_{-0.0}^{+0.2}$&$5.28_{-0.01}^{+0.01}$&$213_{-0}^{+1}$\\\noalign{\vskip 0.5mm}
				&3&15&5.32-5.38&$39.5_{-0.1}^{+0.0}$&$5.35_{-0.01}^{+0.0}$&$211_{-1}^{+0}$\\\noalign{\vskip 0.5mm}
				&4&15&5.38-5.44&$39.1_{-0.0}^{+0.1}$&$5.41_{-0.01}^{+0.01}$&$212_{-1}^{+1}$\\\noalign{\vskip 0.5mm}
				&5&15&5.45-5.51&$38.9_{-0.0}^{+0.0}$&$5.47_{-0.01}^{+0.01}$&$213_{-0}^{+1}$\\\noalign{\vskip 0.5mm}
				&6&15&5.51-5.56&$39.0_{-0.0}^{+0.0}$&$5.54_{-0.01}^{+0.01}$&$216_{-1}^{+0}$\\\noalign{\vskip 0.5mm}
				&7&15&5.57-5.63&$38.8_{-0.1}^{+0.1}$&$5.6_{-0.01}^{+0.01}$&$218_{-1}^{+1}$\\
\hline\hline
			\end{tabular}
		\end{center}
	\end{table}
	\begin{table*} \tiny
		\caption{$N_\sigma$ analysis for data within 290\degree \ to 300\degree, 300\degree \ to 310\degree, 310\degree \ to 320\degree, and 320\degree \ to 330\degree\ sectors, with (lower row for each bin) and without (upper row) $R$ error.}
		\label{tab:nsang1}
		\begin{tabular}{l@{\hspace{3\tabcolsep}}c@{\hspace{1\tabcolsep}}c@{\hspace{1\tabcolsep}}c@{\hspace{3\tabcolsep}}c@{\hspace{1\tabcolsep}}c@{\hspace{1\tabcolsep}}c@{\hspace{1\tabcolsep}}c@{\hspace{1\tabcolsep}}c@{\hspace{1\tabcolsep}}c@{\hspace{3\tabcolsep}}c@{\hspace{1\tabcolsep}}c@{\hspace{1\tabcolsep}}c@{\hspace{1\tabcolsep}}c@{\hspace{1\tabcolsep}}c@{\hspace{1\tabcolsep}}c}\hline\hline
			&&&&\multicolumn{6}{c}{Median}&\multicolumn{6}{c}{Weighted Mean}\\
			&&&&$68.27\%$&$95.45\%$&$68.27\%$&$95.45\%$&EF&EF&$68.27\%$&$95.45\%$&$68.27\%$&$95.45\%$&EF&EF\\
			Angle (deg)&Bin&Count&Radii (kpc)&$N_{\sigma}$&$N_{\sigma}$&$|N_{\sigma}|$&$|N_{\sigma}|$&$|N_{\sigma}|$&$|N_{\sigma}|$&$N_{\sigma}$&$N_{\sigma}$&$|N_{\sigma}|$&$|N_{\sigma}|$&$|N_{\sigma}|$&$|N_{\sigma}|$\\\hline\noalign{\vskip 1mm}\multirow{20}{*}{290-300}
			&\multirow{2}{*}{1}&\multirow{2}{*}{11}&2.74-2.86&[-1.42,1.2]&[-1.83,2.56]&1.2&2.56&54.5\%&90.9\%&[-2.91,2.54]&[-3.21,3.0]&2.54&3.21&54.5\%&63.6\%\\
			&&&2.74-2.88&[-1.21,1.2]&[-10.7,2.56]&1.2&10.7&54.5\%&81.8\%&[-1.59,3.38]&[-16.1,3.78]&1.79&16.1&36.4\%&72.7\%\\
			&\multirow{2}{*}{2}&\multirow{2}{*}{11}&2.87-3.0&[-1.04,0.97]&[-3.59,2.2]&1.01&3.59&63.6\%&81.8\%&[-6.31,3.6]&[-12.9,4.65]&3.6&12.9&18.2\%&54.5\%\\
			&&&2.83-3.02&[-1.81,0.73]&[-2.33,0.98]&0.98&2.33&72.7\%&90.9\%&[-5.81,3.09]&[-6.07,4.52]&3.09&6.07&18.2\%&54.5\%\\
			&\multirow{2}{*}{3}&\multirow{2}{*}{11}&3.01-3.13&[-1.32,1.34]&[-1.41,1.41]&1.32&1.41&45.5\%&100.0\%&[-1.92,1.85]&[-4.1,1.94]&1.85&4.1&27.3\%&90.9\%\\
\hline\hline
		\end{tabular}
		\vspace{15mm}
	\end{table*}
	\begin{table*} \tiny
		\caption{$N_\sigma$ analysis for data within 340\degree \ to 350\degree, 350\degree \ to 0\degree, and 0\degree \ to 10\degree\ sectors, with (lower row for each bin) and without (upper row) $R$ error.}
		\label{tab:nsang2}
		\begin{tabular}{l@{\hspace{3\tabcolsep}}c@{\hspace{1\tabcolsep}}c@{\hspace{1\tabcolsep}}c@{\hspace{3\tabcolsep}}c@{\hspace{1\tabcolsep}}c@{\hspace{1\tabcolsep}}c@{\hspace{1\tabcolsep}}c@{\hspace{1\tabcolsep}}c@{\hspace{1\tabcolsep}}c@{\hspace{3\tabcolsep}}c@{\hspace{1\tabcolsep}}c@{\hspace{1\tabcolsep}}c@{\hspace{1\tabcolsep}}c@{\hspace{1\tabcolsep}}c@{\hspace{1\tabcolsep}}c}\hline\hline
			&&&&\multicolumn{6}{c}{Median}&\multicolumn{6}{c}{Weighted Mean}\\
			&&&&$68.27\%$&$95.45\%$&$68.27\%$&$95.45\%$&EF&EF&$68.27\%$&$95.45\%$&$68.27\%$&$95.45\%$&EF&EF\\
			Angle (deg)&Bin&Count&Radii (kpc)&$N_{\sigma}$&$N_{\sigma}$&$|N_{\sigma}|$&$|N_{\sigma}|$&$|N_{\sigma}|$&$|N_{\sigma}|$&$N_{\sigma}$&$N_{\sigma}$&$|N_{\sigma}|$&$|N_{\sigma}|$&$|N_{\sigma}|$&$|N_{\sigma}|$\\\hline\noalign{\vskip 1mm}\multirow{28}{*}{340-350}
			&\multirow{2}{*}{1}&\multirow{2}{*}{15}&4.28-4.91&[-1.11,1.21]&[-1.28,2.42]&1.21&2.42&40.0\%&93.3\%&[-2.38,3.72]&[-5.7,5.12]&3.0&5.7&13.3\%&40.0\%\\
			&&&4.28-5.22&[-1.12,1.17]&[-2.45,2.31]&1.17&2.45&40.0\%&86.7\%&[-2.13,3.34]&[-5.08,5.42]&3.34&5.42&13.3\%&46.7\%\\
			&\multirow{2}{*}{2}&\multirow{2}{*}{15}&4.92-6.24&[-1.85,5.14]&[-2.64,5.69]&2.64&5.69&40.0\%&60.0\%&[-2.98,8.97]&[-7.91,11.3]&7.13&11.3&13.3\%&40.0\%\\
			&&&4.91-6.53&[-1.17,6.9]&[-1.78,7.78]&2.28&7.78&40.0\%&66.7\%&[-5.24,10.2]&[-8.68,11.6]&6.0&11.6&13.3\%&26.7\%\\
			&\multirow{2}{*}{3}&\multirow{2}{*}{15}&6.33-7.31&[-1.29,2.52]&[-1.31,7.3]&1.31&7.3&46.7\%&80.0\%&[-2.65,2.6]&[-2.67,8.84]&2.65&8.84&40.0\%&53.3\%\\
\hline\hline
		\end{tabular}
	\end{table*}
	\begin{table*} \tiny
		\caption{$N_\sigma$ analysis for data within 10\degree \ to 20\degree, 20\degree \ to 30\degree, and 30\degree \ to 40\degree\ sectors, with (lower row for each bin) and without (upper row) $R$ error.}
		\label{tab:nsang3}
		\begin{tabular}{l@{\hspace{3\tabcolsep}}c@{\hspace{1\tabcolsep}}c@{\hspace{1\tabcolsep}}c@{\hspace{3\tabcolsep}}c@{\hspace{1\tabcolsep}}c@{\hspace{1\tabcolsep}}c@{\hspace{1\tabcolsep}}c@{\hspace{1\tabcolsep}}c@{\hspace{1\tabcolsep}}c@{\hspace{3\tabcolsep}}c@{\hspace{1\tabcolsep}}c@{\hspace{1\tabcolsep}}c@{\hspace{1\tabcolsep}}c@{\hspace{1\tabcolsep}}c@{\hspace{1\tabcolsep}}c}\hline\hline
			&&&&\multicolumn{6}{c}{Median}&\multicolumn{6}{c}{Weighted Mean}\\
			&&&&$68.27\%$&$95.45\%$&$68.27\%$&$95.45\%$&EF&EF&$68.27\%$&$95.45\%$&$68.27\%$&$95.45\%$&EF&EF\\
			Angle (deg)&Bin&Count&Radii (kpc)&$N_{\sigma}$&$N_{\sigma}$&$|N_{\sigma}|$&$|N_{\sigma}|$&$|N_{\sigma}|$&$|N_{\sigma}|$&$N_{\sigma}$&$N_{\sigma}$&$|N_{\sigma}|$&$|N_{\sigma}|$&$|N_{\sigma}|$&$|N_{\sigma}|$\\\hline\noalign{\vskip 1mm}\multirow{28}{*}{10-20}
			&\multirow{2}{*}{1}&\multirow{2}{*}{15}&4.9-6.22&[-1.08,3.56]&[-1.56,3.93]&1.56&3.93&46.7\%&73.3\%&[-1.35,3.43]&[-2.58,3.82]&2.58&3.82&46.7\%&60.0\%\\
			&&&4.9-6.48&[-0.92,3.44]&[-6.99,3.82]&3.2&6.99&53.3\%&66.7\%&[-2.18,4.05]&[-7.53,4.36]&3.23&7.53&40.0\%&53.3\%\\
			&\multirow{2}{*}{2}&\multirow{2}{*}{15}&6.27-7.2&[-1.65,2.14]&[-5.28,3.86]&2.14&5.28&46.7\%&60.0\%&[-3.52,2.07]&[-5.6,6.79]&3.52&6.79&33.3\%&46.7\%\\
			&&&6.21-7.29&[-1.15,3.89]&[-2.51,4.48]&2.51&4.48&46.7\%&60.0\%&[-2.22,5.61]&[-5.2,6.66]&4.62&6.66&13.3\%&46.7\%\\
			&\multirow{2}{*}{3}&\multirow{2}{*}{15}&7.22-7.5&[-0.88,1.15]&[-1.75,1.77]&1.15&1.77&66.7\%&100.0\%&[-0.89,1.54]&[-1.93,1.89]&1.54&1.93&60.0\%&100.0\%\\
\hline\hline
		\end{tabular}
		\vspace{7.5mm}
	\end{table*}
	\begin{table*} \tiny
		\caption{$N_\sigma$ analysis for data within 40\degree \ to 50\degree, 50\degree \ to 60\degree, and 60\degree \ to 70\degree\ sectors, with (lower row for each bin) and without (upper row) $R$ error.}
		\label{tab:nsang4}
		\begin{tabular}{l@{\hspace{3\tabcolsep}}c@{\hspace{1\tabcolsep}}c@{\hspace{1\tabcolsep}}c@{\hspace{3\tabcolsep}}c@{\hspace{1\tabcolsep}}c@{\hspace{1\tabcolsep}}c@{\hspace{1\tabcolsep}}c@{\hspace{1\tabcolsep}}c@{\hspace{1\tabcolsep}}c@{\hspace{3\tabcolsep}}c@{\hspace{1\tabcolsep}}c@{\hspace{1\tabcolsep}}c@{\hspace{1\tabcolsep}}c@{\hspace{1\tabcolsep}}c@{\hspace{1\tabcolsep}}c}\hline\hline
			&&&&\multicolumn{6}{c}{Median}&\multicolumn{6}{c}{Weighted Mean}\\
			&&&&$68.27\%$&$95.45\%$&$68.27\%$&$95.45\%$&EF&EF&$68.27\%$&$95.45\%$&$68.27\%$&$95.45\%$&EF&EF\\
			Angle (deg)&Bin&Count&Radii (kpc)&$N_{\sigma}$&$N_{\sigma}$&$|N_{\sigma}|$&$|N_{\sigma}|$&$|N_{\sigma}|$&$|N_{\sigma}|$&$N_{\sigma}$&$N_{\sigma}$&$|N_{\sigma}|$&$|N_{\sigma}|$&$|N_{\sigma}|$&$|N_{\sigma}|$\\\hline\noalign{\vskip 1mm}\multirow{28}{*}{40-50}
			&\multirow{2}{*}{1}&\multirow{2}{*}{15}&5.15-5.24&[-0.69,1.02]&[-1.53,2.48]&1.02&2.48&66.7\%&93.3\%&[-1.12,0.94]&[-1.58,2.44]&1.12&2.44&66.7\%&93.3\%\\
			&&&5.15-5.26&[-1.22,1.11]&[-1.58,2.52]&1.22&2.52&60.0\%&93.3\%&[-1.38,1.76]&[-2.62,2.7]&1.76&2.7&40.0\%&73.3\%\\
			&\multirow{2}{*}{2}&\multirow{2}{*}{15}&5.25-5.31&[-0.52,1.13]&[-1.68,2.0]&1.08&2.0&60.0\%&93.3\%&[-0.65,1.17]&[-2.17,2.39]&1.17&2.39&60.0\%&86.7\%\\
			&&&5.23-5.33&[-1.4,1.07]&[-2.31,4.14]&1.4&4.14&46.7\%&73.3\%&[-2.51,1.17]&[-4.16,4.85]&2.51&4.85&40.0\%&66.7\%\\
			&\multirow{2}{*}{3}&\multirow{2}{*}{15}&5.32-5.38&[-0.54,1.12]&[-0.76,2.19]&0.69&2.19&80.0\%&86.7\%&[-0.89,0.95]&[-1.16,2.1]&0.92&2.1&80.0\%&86.7\%\\
\hline\hline
		\end{tabular}
	\end{table*}
	\begin{table}\tiny
		\caption{KS rest results for error distributions in 290\degree \ to 300\degree, 300\degree \ to 310\degree, 310\degree \ to 320\degree, and 320\degree \ to 330\degree\ sectors, with (lower rows) and without (upper rows) $R$ error.}
		\label{tab:ksang1}
		\begin{center}
			\begin{tabular}{l@{\hspace{1\tabcolsep}}l@{\hspace{3\tabcolsep}}l@{\hspace{1\tabcolsep}}c@{\hspace{1\tabcolsep}}c@{\hspace{3\tabcolsep}}l@{\hspace{1\tabcolsep}}c@{\hspace{1\tabcolsep}}c}\hline\hline
				&&\multicolumn{3}{c}{Median Statistics}&\multicolumn{3}{c}{Weighted Mean Statistics}\\
				Angle(deg)&Bin&Distribution&$S$&$p$&Distribution&$S$&$p$\\\hline
				\noalign{\vskip 1mm}\multirow{20}{*}{290-300}
				&\multirow{2}{*}{1}&Student's $t$, $\nu = 95$&1.23&1.0&Student's $t$, $\nu = 2$&1.41&1.0\\
				&&Student's $t$, $\nu = 20$&1.19&1.0&Student's $t$, $\nu = 2$&1.65&1.0\\
				&\multirow{2}{*}{2}&Laplace&0.82&1.0&Gaussian&3.51&1.0\\
				&&Student's $t$, $\nu = 4$&0.99&1.0&Gaussian&3.43&1.0\\
				&\multirow{2}{*}{3}&Gaussian&1.12&0.96&Gaussian&1.92&0.94\\
				&&Laplace&1.03&1.0&Cauchy&1.41&0.96\\
\hline\hline
			\end{tabular}
		\end{center}
		\vspace{15mm}
	\end{table}
	\begin{table}\tiny
		\caption{KS rest results for error distributions in 340\degree \ to 350\degree, 350\degree \ to 0\degree, and 0\degree \ to 10\degree\ sectors, with (lower rows) and without (upper rows) $R$ error.}
		\label{tab:ksang2}
		\begin{center}
			\begin{tabular}{l@{\hspace{1\tabcolsep}}l@{\hspace{3\tabcolsep}}l@{\hspace{1\tabcolsep}}c@{\hspace{1\tabcolsep}}c@{\hspace{3\tabcolsep}}l@{\hspace{1\tabcolsep}}c@{\hspace{1\tabcolsep}}c}\hline\hline
				&&\multicolumn{3}{c}{Median Statistics}&\multicolumn{3}{c}{Weighted Mean Statistics}\\
				Angle(deg)&Bin&Distribution&$S$&$p$&Distribution&$S$&$p$\\\hline
				\noalign{\vskip 1mm}\multirow{28}{*}{340-350}
				&\multirow{2}{*}{1}&Gaussian&1.24&0.78&Gaussian&3.65&0.98\\
				&&Gaussian&1.33&0.92&Gaussian&3.82&0.99\\
				&\multirow{2}{*}{2}&Cauchy&0.97&0.97&Cauchy&2.36&1.0\\
				&&Student's $t$, $\nu = 2$&1.97&1.0&Student's $t$, $\nu = 5$&5.52&1.0\\
				&\multirow{2}{*}{3}&Gaussian&1.48&1.0&Laplace&2.19&0.99\\
				&&Laplace&1.39&1.0&Laplace&2.7&1.0\\
\hline\hline
			\end{tabular}
		\end{center}
	\end{table}
	\begin{table}\tiny
		\caption{KS rest results for error distributions in 10\degree \ to 20\degree, 20\degree \ to 30\degree, and 30\degree \ to 40\degree\ sectors, with (lower rows) and without (upper rows) $R$ error.}
		\label{tab:ksang3}
		\begin{center}
			\begin{tabular}{l@{\hspace{1\tabcolsep}}l@{\hspace{3\tabcolsep}}l@{\hspace{1\tabcolsep}}c@{\hspace{1\tabcolsep}}c@{\hspace{3\tabcolsep}}l@{\hspace{1\tabcolsep}}c@{\hspace{1\tabcolsep}}c}\hline\hline
				&&\multicolumn{3}{c}{Median Statistics}&\multicolumn{3}{c}{Weighted Mean Statistics}\\
				Angle(deg)&Bin&Distribution&$S$&$p$&Distribution&$S$&$p$\\\hline
				\noalign{\vskip 1mm}\multirow{28}{*}{10-20}
				&\multirow{2}{*}{1}&Laplace&1.62&1.0&Student's $t$, $\nu = 4$&1.89&1.0\\
				&&Cauchy&1.08&1.0&Laplace&2.3&1.0\\
				&\multirow{2}{*}{2}&Laplace&1.75&1.0&Student's $t$, $\nu = 3$&2.56&1.0\\
				&&Cauchy&0.9&0.98&Student's $t$, $\nu = 3$&3.23&1.0\\
				&\multirow{2}{*}{3}&Student's $t$, $\nu = 3$&0.8&1.0&Student's $t$, $\nu = 5$&1.06&1.0\\
				&&Gaussian&1.51&1.0&Gaussian&1.78&1.0\\
\hline\hline
			\end{tabular}
		\end{center}
		\vspace{7.5mm}
	\end{table}
	\begin{table}\tiny
		\caption{KS rest results for error distributions in 40\degree \ to 50\degree, 50\degree \ to 60\degree, and 60\degree \ to 70\degree\ sectors, with (lower rows) and without (upper rows) $R$ error.}
		\label{tab:ksang4}
		\begin{center}
			\begin{tabular}{l@{\hspace{1\tabcolsep}}l@{\hspace{3\tabcolsep}}l@{\hspace{1\tabcolsep}}c@{\hspace{1\tabcolsep}}c@{\hspace{3\tabcolsep}}l@{\hspace{1\tabcolsep}}c@{\hspace{1\tabcolsep}}c}\hline\hline
				&&\multicolumn{3}{c}{Median Statistics}&\multicolumn{3}{c}{Weighted Mean Statistics}\\
				Angle(deg)&Bin&Distribution&$S$&$p$&Distribution&$S$&$p$\\\hline
				\noalign{\vskip 1mm}\multirow{28}{*}{40-50}
				&\multirow{2}{*}{1}&Student's $t$, $\nu = 6$&0.91&1.0&Gaussian&1.21&1.0\\
				&&Gaussian&1.16&1.0&Student's $t$, $\nu = 9$&1.5&1.0\\
				&\multirow{2}{*}{2}&Cauchy&0.41&0.97&Gaussian&1.02&1.0\\
				&&Laplace&1.33&1.0&Cauchy&0.96&1.0\\
				&\multirow{2}{*}{3}&Student's $t$, $\nu = 3$&0.6&1.0&Student's $t$, $\nu = 10$&0.81&1.0\\
				&&Laplace&1.45&0.97&Gaussian&2.49&1.0\\
\hline\hline
			\end{tabular}
		\end{center}
	\end{table}

	\bsp
	\label{lastpage}
	
\end{document}